\documentclass[onecolumn,amsmath,amssymb,nofootinbib,12pt]{article}
\pdfoutput=1
\usepackage{jheppub}

\usepackage{graphicx}
\usepackage{dcolumn}
\usepackage{bm,psfrag}

\usepackage{subfigure}
\usepackage{float}
\usepackage{color}
\usepackage{booktabs}
\usepackage{multirow}

\newcommand{\ben}{\begin{equation}}
\newcommand{\een}{\end{equation}}
\newcommand{\be}{\begin{equation}}
\newcommand{\ee}{\end{equation}}
\newcommand{\bea}{\begin{eqnarray}}
\newcommand{\eea}{\end{eqnarray}}
\newcommand{\ba}{\begin{eqnarray}}
\newcommand{\ea}{\end{eqnarray}}

\newcommand{\beq}{\begin{equation}}
\newcommand{\eeq}{\end{equation}}
\newcommand{\beqa}{\begin{eqnarray}}
\newcommand{\eeqa}{\end{eqnarray}}
\newcommand{\beqar}{\begin{eqnarray*}}
\newcommand{\eeqar}{\end{eqnarray*}}
\newcommand{\e}{\epsilon}
\newcommand{\reef}[1]{(\ref{#1})}

\newcommand{\eg}{{\it e.g.,}\ }
\newcommand{\ie}{{\it i.e.,}\ }

\newcommand{\labell}[1]{\label{#1}} 

\newcommand{\hd}{{\hat d}}
\newcommand{\hchi}{{\hat \chi}}

\def\t6 {T_\mt{D6}}

\newcommand{\mt}[1]{\textrm{\tiny #1}}

\newcommand{\vk}{{\vec{k}}}
\newcommand{\vx}{{\vec{x}}}
\newcommand{\hJ}{{\hat{J}}}

\def\cale         {{\cal E}}

\def\cO        {{\cal O}}

\def\ee           {{\rm e}}

\def\sqr#1#2{{\vcenter{\vbox{\hrule height.#2pt
 \hbox{\vrule width.#2pt height#1pt \kern#1pt
 \vrule width.#2pt}\hrule height.#2pt}}}}

\def\IZ{\mathbb{Z}}
\def\a{\alpha}

\def\e{\epsilon}

\def\ee{\cale}

\def\aa1{\phi}
\def\cc1{\psi}

\def\vev#1{\langle #1 \rangle}

\def\hi{\hat{\bf i}}
\def\hj{\hat{\bf j}}
\def\vM{\vec{M}}
\def\vn{{\vec{n}}}
\def\vk{{\vec{k}}}

\def\bg{{\bar{g}}}
\def\bk{{\bar{k}}}
\def\vp{{\vec{p}}}

\def\nd{{ \vphantom{\dagger}}}

\newcommand{\dt}{\delta t}

\newcommand{\op}{\langle \bar\chi \chi \rangle}

\begin{document}

\preprint{arXiv:1706.nnnnn [hep-th]}

\title{An exactly solvable quench protocol for integrable spin models}

\author{Diptarka Das,$^{1}$ Sumit R. Das,$^{2}$ Dami\'an A. Galante,$^{3,4}$ Robert C. Myers$^4$ and Krishnendu Sengupta$^5$}
\affiliation{$^1$\,Department of Physics, University of California at San Diego,\\
\vphantom{k}\ \ La Jolla, CA 92093, USA}
\affiliation{$^2$\,Department of Physics and Astronomy, University of Kentucky,\\
\vphantom{k}\ \ Lexington, KY 40506, USA}
\affiliation{$^3$\, Institute for Theoretical Physics Amsterdam and Delta Institute for Theoretical Physics, \\
\vphantom{k}\ \ University of Amsterdam, Science Park 904, 1090 GL
Amsterdam, The Netherlands} \affiliation{$^4$\,Perimeter Institute
for Theoretical Physics, Waterloo, ON N2L 2Y5, Canada}
\affiliation{$^5$\, Theoretical Physics Department, Indian
Association for the Cultivation of Science, \\
Jadavpur, Kolkata-700032, India}

\emailAdd{diptarka@physics.ucsd.edu}
\emailAdd{das@pa.uky.edu}
\emailAdd{d.a.galante@uva.nl}
\emailAdd{rmyers@perimeterinstitute.ca}
\emailAdd{tpks@iacs.res.in}

\date{\today}

\abstract{Quantum quenches in continuum field theory across critical
points are known to display different scaling behaviours in different
regimes of the quench rate. We extend these results to integrable
lattice models such as the transverse field Ising model on a
one-dimensional chain and the Kitaev model on a two-dimensional
honeycomb lattice using  a nonlinear quench protocol which allows for
exact analytical solutions of the dynamics. Our quench protocol
starts with a finite mass gap at early times and crosses a critical
point or a critical region, and we study the behaviour of one point
functions of the quenched operator at the critical point or in the
critical region as a function of the quench rate. For quench rates
slow compared to the initial mass gap, we find the expected
Kibble-Zurek scaling. In contrast, for rates fast compared to the
mass gap, but slow compared to the inverse lattice spacing, we find
scaling behaviour similar to smooth fast continuum quenches. For
quench rates of the same order of the lattice scale, the one point
function saturates as a function of the rate, approaching the
results of an abrupt quench. The presence of an extended critical
surface in the Kitaev model leads to a variety of scaling exponents
depending on the starting point and on the time where the operator
is measured. We discuss the role of the amplitude of the quench in
determining the extent of the slow (Kibble-Zurek) and fast
quench regimes, and the onset of the saturation.}

\maketitle


\section{Introduction}

The dynamics of a closed quantum system following a smooth quench
involving a critical point is expected to carry universal signatures
of the critical theory. Of course, the best known of such behaviours
would be Kibble-Zurek (KZ) scaling \cite{kibble,zurek}, which has
received considerable attention in the past several years
\cite{more,qcritkz}. One might characterize the corresponding
quenches as ``slow'' since they involve a protocol where the
evolution remains adiabatic until the system approaches very close
to the critical point. However, recent holographic studies
\cite{numer,fastQ} also revealed interesting new scaling behaviour
when critical points were probed by ``fast'' quench protocols. Later
examination showed that this fast quench scaling is a universal
behaviour for quantum field theories flowing from a UV fixed point,
which is described by the conformal field theory
\cite{dgm1,dgm2,dgm3}. Hence various renormalized observables in
such continuum systems can exhibit a variety of scaling behaviours
for different regimes of the quench rate. In fact, it was explicitly
shown in \cite{dgm4} that scaling behaviour of various (renormalized)
observables smoothly interpolates between the Kibble-Zurek scaling
and the fast quench scaling with appropriate quench protocols in
free scalar and fermion field theories.

However, the above results beg the question: how do these scaling
regimes manifest in the presence of a finite UV cutoff? This
question is particularly important if we wish to make any contact
with real experimental systems which always have a finite lattice
spacing (\eg cold atom systems in optical lattices). The observables
of interest are now bare quantities, and while we can vary the
quench rate over a broad range of scales, the cutoff scale (\eg the
inverse lattice spacing) will always place limitations on fast
quench protocols. However, we might still expect that  when the
quench rate is taken far below the cutoff scale, the results would
match those for renormalized quantities for the corresponding
continuum fixed point theory.

In this paper, we will explore quench dynamics in spin systems with
a smooth nonlinear quench protocol for which the quantum dynamics
can be solved exactly for any value of the quench rate. We will
focus on two such models: the transverse field Ising model in one
space dimension and the Kitaev honeycomb model \cite{kitaev} in two
space dimensions. The ability to solve the dynamics stems from the
fact that both these models can be written in terms of free fermions
in momentum space. In both cases, we will measure the expectation
value of the quenched operator, which in the fermionic language is
${\bar{\psi}}\psi$, where $\psi$ is the fermionic field used to
represent the integrable spin models via Jordan-Wigner
transformation.

For quenches where the couplings vary linearly in time, the
response of the (1+1)-dimensional Ising model has been examined in
\cite{dziarmaga}, while the Kitaev model has been studied in
\cite{sengupta,hikichi}. Both the Ising \cite{powell,cc2} and the Kitaev \cite{kquench1,kquench2} model have also been studied for
instantaneous quenches. In contrast, we will study
the quench protocols in which the couplings vary smoothly over a (finite) time interval (characterized by the duration $\dt$)
and saturate to
constant values at early and late times. Among other things, this
allows us to investigate the dependence on the quench rate and the
amplitude of the quench separately, and to scan the entire range of
quench rates, including the new fast scaling regime
\cite{dgm1,dgm2,dgm3, dgm4},  as well as the regime where
the quench rate becomes of the order of the UV cutoff. In the latter
regime the response saturates as a function of the quench rate, as
one expects for an instantaneous quench. However we find that the
$\dt$ at which this saturation happens is proportional to the
amplitude for large amplitudes, while it is independent of the
amplitude for small amplitudes.

While the Ising model has an isolated
critical point separating two gapped phases, the Kitaev model is
particularly interesting since there is a whole critical region in
the space of couplings. This allows us to study a new class of
quenches in which the couplings are varied entirely within this
critical region.

The remainder of the paper is organized as follows: The remainder of this section
provides a more detailed review of the various quench regimes, which we introduced in
the introductory discussion above. In section \ref{sec-2}, we review the two lattice models and in
particular, we derive the relevant continuum theory describing the
corresponding critical points. We then introduce our specific quench
profiles for the couplings and discuss the exact time-dependent
solutions in section \ref{sec-3}. Our various results for the
response of these lattice models in different quench regimes are
presented in section \ref{sec-4}. Section \ref{sec-5} contains a
brief discussion of our results. In appendix \ref{app-kubo}, we
derive the fast scaling behaviour using linear response theory around
a CFT in arbitrary dimensions. In appendix \ref{app-sat}, we provide some technical details required to understand the amplitude dependence of the value of $\dt$ at which the response saturates. The final appendix \ref{app-cluster}
discusses the Cluster Ising model on a one-dimensional closed chain,
which can also be studied in the same manner as the Ising model.

\subsection{Quench regimes: Slow, Fast and Instantaneous}

\noindent {\underbar{Slow Quench}}: For a quench protocol where the system starts with a finite mass gap and then crosses or approaches a critical point, at a rate which is slow compared to the initial gap, many systems show
Kibble-Zurek (KZ) scaling \cite{kibble,zurek,more,qcritkz}. Consider examining a relativistic system (\ie the dynamical critical exponent is $z=1$) with the simple power-law protocol
\ben
g(t) - g_c \sim g_0\   \left( {t}/{\dt} \right)^{r} \,.
\label{0-1}
\een
It follows that the instantaneous energy gap $E_g(t)$  is given by
\ben
E_{g} (t) = \kappa\,  | g(t) - g_c |^{\nu} =E_0\ |t/\dt|^{r\nu}\,,
\label{1-1-3}
\een
where $\nu$ is the correlation length exponent for the critical point. Now the initial
adiabatic evolution breaks down at $t=-t_{KZ}$, as determined by the Landau criterion
\ben
\frac{1}{E_g(t)^2}\frac{dE_g(t)}{dt} \bigg|_{t= -t_{KZ}} \sim 1\,.
\label{1-1-2}
\een
Then the early-time one-point function of an
operator $\cO_{\Delta}$ with conformal dimension $\Delta$ is
expected to exhibit Kibble-Zurek scaling:
\ben
\langle{\cal O}_\Delta (t)\rangle \sim
(t_{KZ})^{-\Delta}~F( t/t_{KZ}) \,.
\label{1-1}
\een
There are similar KZ scaling relationships for correlation functions.
Given the profile \reef{1-1-3} above, one finds that $t_{KZ}$
is related to the inverse quench rate $\dt$ by the scaling relation
\ben t_{KZ} \sim
\left(\frac{\dt^{r\nu}}{E_0}\right)^{\frac{1}{r\nu+1}}\,.
\label{1-2}
\een

We will be considering more general quench protocols where the
instantaneous energy gap takes the form \ben E_g (t) = E_0\, f
(t/\dt) \,, \label{1-1-1} \een where the function $f(x) \rightarrow
1$ at $x \rightarrow -\infty$. Only in the regime $x\ll1$ does the
profile $f(x)$ approach zero as $f(x) \sim x^{r\nu}$. Now imagine
that the system starts in its ground state at $t = -\infty$ and
initially evolves adiabatically since $E_g (t)$ is only changing
very slowly. However, as described above, this initial adiabatic
evolution breaks down when the Landau criterion \reef{1-1-2} is
satisfied. At this point, the Kibble-Zurek scaling \reef{1-1}
described above can appear if two conditions are satisfied: ({\it
i}) $t_{KZ}$ is such that the function $f(x)$ can be approximated by
the power law $x^{r\nu}$, \ie we must demand that $t_{KZ} \ll \dt$;
and ({\it ii}) at $t=-t_{KZ}$, the instantaneous gap is much smaller
than the UV cutoff scale $\Lambda_\mt{UV}$, \ie we require
$E_{KZ}\equiv E_g(t=-t_{KZ})\ll\Lambda_\mt{UV}$. Combining
Eq.~(\ref{1-2}) with the first condition yields \ben \dt \gg
1/E_0\,. \label{1-1-4} \een Similarly, the second condition can be
expressed as \ben \dt \gg \frac1{E_0}\left(
\frac{E_0}{\Lambda_\mt{UV}} \right)^{1+\frac{1}{r\nu}}\,.
\label{1-1-5} \een When $E_0 \ll \Lambda_\mt{UV}$, the condition
(\ref{1-1-4}) implies the condition (\ref{1-1-5}). However, when
$E_0$ is of the same order or larger than $\Lambda_\mt{UV}$, the
condition (\ref{1-1-5}) is the stronger restriction. Note that we
are considering protocols where $E_0 > E_{KZ}$ \footnote{For the
quench protocols and the lattice models studied in the following, we
will have $\nu=1$ and $r=1$. Further, the UV cutoff scale is simply
the inverse lattice spacing, \ie $\Lambda_\mt{UV}=1/a$, and so
Eq.~(\ref{1-1-5}) can be written as $\dt \gg E_0\, a^2$.}.

This approach is closely related to quench protocols used more
commonly in discussions of KZ behaviour in the condensed matter
literature, where the behaviour \reef{1-1-3} is often considered to
hold for {\em all} times. In particular, if we define $K\equiv
E_0/\dt^{r\nu}$, the expression \reef{1-1-3} for the instantaneous
gap becomes $E_g(t) = K\, t^{r\nu}$. Note that in this case $E_0$
and $\delta t$ can not be separately varied. The system is prepared
in the ground state of the theory at some (finite) initial time $t =
t_i$. The KZ time and energy gap are then given by $t_{KZ} =
K^{-\frac{1}{r\nu+1}}$ and $E_{KZ} = K^{\frac{1}{r\nu+1}}$,
respectively. Now the first condition above is replaced by $|t_i|\ge
t_{KZ}$, which can be equivalently written as \ben K < E(t_i)^{r\nu
+ 1}\,. \label{1-1-11} \een The second condition above, \ie $E_{KZ}
\ll \Lambda_\mt{UV}$, can be expressed as \ben K \ll
\Lambda_\mt{UV}^{r\nu+1}\,. \label{1-1-10} \een Hence if $E(t_i) <
\Lambda_\mt{UV}$, the first condition implies the first, while if
$E(t_i) \gtrsim \Lambda_\mt{UV}$, the second constraint is the
stronger one. Often one actually considers the situation where $t_i
\to -\infty$. In this case, the only constraint is
Eq.~(\ref{1-1-10}), since $E(t_i)$ diverges rendering the
inequality \reef{1-1-11} trivial.
\\ \\
\noindent {\underbar{Fast Quench}}: As noted above, recently a new
scaling behaviour was discovered for smooth but fast quenches. In
particular, consider a generic action \ben S = S_\mt{CFT} + \int dt
\int d^{d-1}x~\lambda (t)\, \cO_\Delta(x,t) \,, \label{1-4a} \een
where $S_\mt{CFT}$ is the conformal field theory action describing
the UV fixed point, and $\cO_\Delta$ is a relevant operator with
conformal dimension $\Delta$. The quench profile for coupling
$\lambda(t)$ starts from some initial value $\lambda_\mt{init}$ and
smoothly changes to the final value $\lambda_\mt{fin}$ over a time
scale $\dt$. If this time scale is {\it fast} compared to all
{physical} scales, but {\em slow} compared to the scale of the UV
cutoff, \ie \ben
 \Lambda_\mt{UV}^{-1}\ll \dt \ll \lambda_\mt{init}^{1/(\Delta-d)}\,, \lambda_\mt{fin}^{1/(\Delta-d)}\,,
|\lambda_\mt{fin}-\lambda_\mt{init}|^{1/(\Delta-d)} \,, \label{1-5}
\een then the response of various {\em renormalized} quantities
exhibit scaling at early times \cite{dgm1,dgm2,dgm3,dgm4}. For
example, during the quench process, the renormalized expectation
value $\vev{\cO_\Delta}_{ren}$ behaves as \ben
\vev{\cO_\Delta}_{ren} \sim \frac{\delta\lambda}{\dt^{2\Delta-d}}
\,. \label{1-6} \een where
$\delta\lambda=\lambda_\mt{fin}-\lambda_\mt{init}$. Similarly, the
energy density scales as \ben {\cal{E}}_{ren} \sim  \frac{\delta
\lambda^{\,2}}{ \dt^{2\Delta-d}} \,. \label{1-7} \een This scaling
behaviour was originally discovered in holographic computations
\cite{numer,fastQ} but then it was shown to hold in free field
theories, and further argued to be true for general interacting
theories \cite{dgm1,dgm2}.

It may seem mysterious that the underlying QFT has been regulated
and renormalized and yet the above expressions are divergent in the
limit $\dt\to0$, when $\Delta>d/2$. These divergences arise because
as $\dt$ shrinks, the quench is exciting a growing number of short
wavelength modes.  Further there is an infinite ``reservoir'' of
such modes available as long as they are arranged as excitations of
the UV fixed-point CFT.  Implicitly the latter holds for {\em
renormalized} quantities as in Eqs.~(\ref{1-6}) and \reef{1-7},
which are defined in a procedure which involves taking the limit
$\Lambda_\mt{UV}\to\infty$. Of course, if the cutoff scale
$\Lambda_\mt{UV}$ is held fixed while $\dt$ continues to shrink,
eventually we will encounter $\dt\sim 1/\Lambda_\mt{UV}$ and the
above scaling behaviour will no longer be applicable.
\\ \\
\noindent {\underbar{Instantaneous Quench}}: The approach, which is most commonly discussed in the quantum quench literature, \eg \cite{powell,cc2,cc3,more}, involves preparing a system in the ground state of an initial Hamiltonian and then time evolving this state with a new or final Hamiltonian. There are some scaling results known to hold in the situation where the ground state of the initial Hamiltonian is gapped and the final Hamiltonian corresponds to a critical phase \cite{cc2,cc3}, in particular, the latter is a (1+1)-dimensional CFT. One can imagine that this describes an instantaneous quench, where initially the couplings of Hamiltonian are held constant, then at a single moment of time, the couplings are {\it instantaneously} changed to produce the final Hamiltonian and subsequently the couplings are fixed at their new values. Further this interpretation naively suggests that this protocol corresponds to the $\dt\to0$ limit of the smooth fast quenches described above. However, as emphasized in \cite{dgm2,dgm3}, this limit does not reproduce the instantaneous quench\footnote{While the divergences discussed above are not encountered, one can still expect that the late-time long-distance quantities for a smooth fast quench should agree with those of an instantaneous quench. The comparison of UV finite quantities was examined in detail in \cite{dgm3} for exactly solvable quenches in free field theory.} because implicitly the former assumes that quench rate is always small compared to the UV cutoff, \ie $\Lambda_\mt{UV}\gg 1/\dt$. Instead, the instantaneous quench implicitly assumes that $\dt\to0$ while $\Lambda_\mt{UV}$ remains fixed.\footnote{The papers \cite{cc2,cc3} argue that the state which results from this kind of quench can be well approximated by a state of the form $e^{-\beta H_{CFT}}|B\rangle$ where $|B\rangle$ is a boundary state of the final CFT and $H_{CFT}$ is the final Hamiltonian. This approximation is expected to hold for IR quantities, however, some subtleties have been discussed recently in \cite{mandal,mezei}.}

While the above discussion applies quite generally, implicitly we
are assuming $2\Delta>d$ in which case Eqs.~\reef{1-6} and
\reef{1-7} would produce divergences in the limit $\dt\to0$.
However, we can also consider the situation where $2\Delta< d$ in
which case the expressions in these formulae would vanish when
$\dt\to0$. In fact, these expressions no longer capture the leading
contributions in this situation and instead the quench will produce
finite results for $\vev{\cO_\Delta}_{ren}$ and ${\cal{E}}_{ren}$.
In this case, the results in free field theories indicate that these
final answers do not depend on UV details and the responses for the
instantaneous and the smooth quenches will agree \cite{dgm3}. This behaviour is also manifest in the excess energy above the adiabatic value at late times, which is UV finite for any finite $\dt$. Here again free field studies \cite{dgm3} show that  for $2\Delta < d$, this quantity remains finite in the $\dt \rightarrow 0$ limit and
reproduces instantaneous quench results. Instead it is a subleading term which displays scaling behaviour analogous to that in Eq.~\reef{1-7}. For $2\Delta > d$ the smooth quench answers diverge in this $\dt \rightarrow 0$ limit displaying scaling, while the instantaneous quench answer has a UV divergence.

Having introduced a finite UV cutoff in the present paper, we are
certainly able to study the new regime where
$\Lambda_\mt{UV}\lesssim 1/\dt$, which we will refer to as the
instantaneous quench regime. As noted above, the fast quench scaling
does not apply in this regime. In particular, the divergences which
the $\dt\to0$ limit would produce in Eq.~\reef{1-6} are avoided and
instead we will see that the response saturates when we enter the
instantaneous quench regime.\footnote{As anticipated in \cite{dgm3},
this behaviour is similar to that of correlation functions at finite
spatial separation $\delta\vec{x}$ in renormalized quantum field
theories. In particular, when $\dt$ is small (as specified in
Eq.~\reef{1-5}) and $\dt>|\delta\vec{x}|$, the correlation
functions exhibit fast quench scaling analogous to Eq.~\reef{1-6}
but when $\dt<|\delta\vec{x}|$, the correlation function saturates
so that a finite $\dt\to0$ limit exists. \label{footy78}}

An interesting feature of this regime is that for protocols of the form (\ref{1-1-1}), the value of $\dt$ where this saturation occurs is independent of the amplitude $E_0$  for small amplitudes, while it becomes proportional to $E_0$ for $E_0 \sim O(\Lambda_\mt{UV})$ or larger. We provide a physical explanation of this behaviour in terms of the value of $\dt$ at which the largest momentum mode $k_{max}$ (in lattice units) contributing to the observable departs from adiabatic behaviour. A detailed study of the Ising model reveals that for large amplitudes $k_{max}$ is at the UV cutoff, \ie $k_{max} \sim \pi$. The Landau criterion for this mode leads to the above result. In other words, saturation happens when all possible modes get excited. On the other hand, we find that
for small amplitudes $k_{max} \propto E_0$, \ie all modes do not contribute significantly to the observable we calculate. In this case, the Landau criterion shows that the saturation value is independent of the amplitude.

\section{The models}\label{sec-2}

In this section, we will describe the lattice models of interest and set our notation.

\subsection{Transverse Field Ising Model}
We write the Hamiltonian for the transverse field Ising model as
\ben H_\mt{Ising} = - \sum_{n} \left[ h(t)\, \tau^{(3)}(n) +
J\,\tau^{(1)}(n)\,\tau^{(1)}(n+1) \right]\,, \label{2-1} \een where
$\tau^{(i)}$ denote the Pauli spin operators, and $n$ denotes the
site indices of the one-dimensional chain. The time dependent
coupling $h(t)$ denotes the transverse magnetic field and $J$ is the
interaction strength between the nearest-neighbor spins. Both
couplings have dimensions of energy here. We will follow the
conventions of \cite{kogut}. Using the well-known Jordan-Wigner
transformation, the Hamiltonian given by Eq.~\reef{2-1} can be
rewritten as the theory of a one-component fermion $c(n)$ at each
site, whose Fourier components will be denoted by $d(q)$, \ben c(n)
= \frac{e^{-\frac{i\pi}{4}}}{\sqrt{2N+1}} \sum_{m=-N}^N e^{-iqn}
d(q)\qquad {\rm with}\ \ \ q = \frac{2\pi m}{2N+1}\,. \label{2-2}
\een The latter have the usual anti-commutation relations \ben \{
d(q) , d^\dagger (q^\prime)\} = \delta_{m m^\prime}\ ,\qquad\{ d(q),
d(q^\prime)\} = \{ d^\dagger (q) , d^\dagger (q^\prime) \} =0\,.
\label{relate} \een Note that $q$ is periodic from Eq.~\reef{2-2},
\ie the expression is invariant under $q\to q+2\pi$. However, it is
convenient to shift the momenta \ben q \rightarrow k = q - \pi \,,
\een and to introduce a two-component Majorana fermion \beq \chi(k)
= \left(
\begin{array}{c}
d(k+\pi) \\
d^\dagger (-k-\pi)
\end{array} \right) \,.
\label{fermi} \eeq The Hamiltonian then becomes \ben H_\mt{Ising}
=2J \sum_{k >0} \chi^\dagger (k) \left[ (\cos k - g(t) ) \,\sigma_3
+ \sin k\,\sigma_1 \right] \chi(k) \,, \label{2-5} \een where
$\sigma_i$ denote Pauli matrices in the particle-hole space of
fermions and we have introduced the (dimensionless) coupling $g(t)=
h(t)/J$. Note that the momentum sum runs over half the Brillouin
zone \cite{kogut}. Now it is convenient to consider the limit $N\to
\infty$, in which we are considering an infinite chain of spins.
This will certainly remove the possibility of having our quench
results infected by any finite size effects. In this limit, the
momentum $k$ becomes a continuous variable on the range $[-\pi,\pi]$
--- although as noted above, $k\sim k+2\pi$. We would have the more or
less standard replacements: \ben \frac{1}{2N+1}\sum_{m=-N}^N\
\longrightarrow\ \frac{1}{2\pi}\int_{-\pi}^\pi dk \quad{\rm
and}\quad (2N+1)\,\delta_{mm'}\ \longrightarrow\
2\pi\,\delta(k-k')\,. \label{replace} \een Hence it is convenient to
rescale operators $\hat d(k)=\sqrt{2N+1}\,d(k)$ so that in the
continuous limit, the anti-commutation \reef{relate} become \ben \{
\hat d(k) , \hat d^\dagger (k^\prime)\} = 2\pi\,\delta(k-k') \qquad
, \qquad \{ \hd(k), \hd(k^\prime)\} = \{ \hd^\dagger (k) ,
\hd^\dagger (k^\prime) \} =0 \,. \label{relate2} \een Now
constructing the fermion $\hchi(q)$ from these rescaled operators as
in Eq.~\reef{fermi}, the Hamiltonian \reef{2-5} becomes \ben
H_\mt{Ising} = 2J\int_{0}^\pi \frac{dk}{2\pi}\ \hchi^\dagger (k)
\left[ (\cos k - g(t) )\, \sigma_3 +  \sin k\,\sigma_1 \right]
\hchi(k)\,. \label{ham2} \een For a given momentum $k$, the
instantaneous energy eigenvalues are given by \ben E = \pm 2J \sqrt{
(\cos k-g(t))^2 + \sin^2 k} \,. \label{green} \een This dispersion
relation makes clear that $g = 1$ corresponds to a critical point
where the gapless mode is $k=0$.\footnote{A second critical point
occurs at $g=-1$, for which $k=\pi$ becomes the gapless mode.
\label{footy}}

The critical point at $g = 1$ separates two massive phases (the
paramagnetic and the ferromagnetic phases of the Ising model), and
the continuum limit around this critical point is a massive Majorana
fermion. This may be seen as usual by first expanding around the
critical coupling with
\beq g = 1 -\epsilon (t)\,,
\label{work}
\eeq
and then explicitly introducing the lattice spacing $a$ to define the
following dimensionful quantities:
 \ben
p=k/a\,, \qquad m(t)=\epsilon(t)/a\,,\quad{\rm and}\quad
\psi(p)=a^{1/2}\,\hchi(a\,p) \,.
\label{2-5a}
 \een
Finally we also define the dimensionless spin coupling: $\hJ=J\,a$.\footnote{Up to the dimensionless factor of $\hJ$, we can think that the interaction strength defines the inverse lattice spacing, \ie $J=\hJ/a$. \label{footy3}} We then take the continuum limit with $a \rightarrow 0$ holding $\hJ$, $p$, $m(t)$, and $\psi(p)$ fixed. In this limit, all of the terms in $H_\mt{Ising}$ which are higher order in $a$ vanish  and we are left with the continuum Hamiltonian,
\ben
H_\mt{Ising}^\mt{cont} =
2\hJ\int_0^\infty \frac{dp}{2\pi}\ \psi^\dagger (p)  \left[ m(t)\,
\sigma_3 +p\, \sigma_1 \right] \psi (p)\,,
\label{2-5b}
\een
which corresponds to the theory of a massive Majorana fermion with a (time dependent) mass $m(t)$.

Before ending this subsection, we note that another related
integrable model which shows similar behaviour is the Cluster-Ising
model on a one-dimensional closed chain. This model is reviewed in appendix
\ref{app-cluster}, where it is shown that the corresponding Hamiltonian can be reduced to three copies of the continuum Ising
Hamiltonian \reef{2-5b}. Thus the exactly solvable quench protocols, which
we will discuss below, also apply to the Cluster-Ising model.

\subsection{Kitaev Honeycomb Model} \label{honey}

\begin{figure}[h!]
\setlength{\abovecaptionskip}{0 pt}
\centering
\includegraphics[scale=0.6]{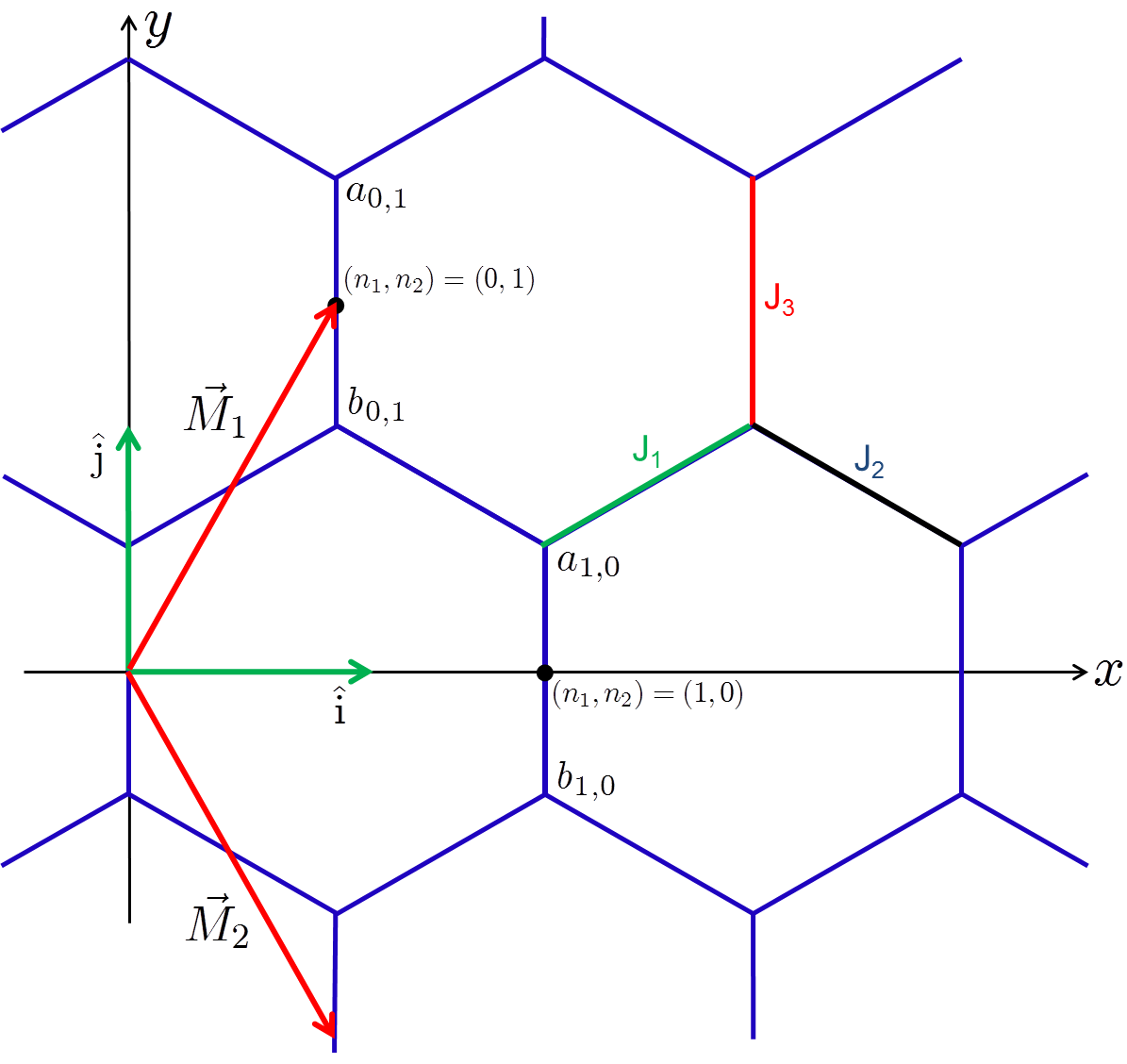}
\caption{The honeycomb lattice on which the Kitaev model \reef{2-9}
is defined. $J_1$ and $J_2$ correspond to the interaction strength
on the horizontal bonds tilted upward and downward, respectively,
while $J_3$ defines the interaction strength for the vertical bonds.
The fermions $a_{\vn}$ and $b_{\vn}$ live on the sites at the top
and bottom of the vertical bond labeled by ${\vn}$. }
\label{lattice}
\end{figure}

This model in $2+1$ dimensions is defined on a (spatial) honeycomb
lattice. The Hamiltonian can be written as
\ben
H_\mt{Kitaev} =
\sum_{j+l ={\rm even}} \left[ J_1\, \tau^{(1)}_{j,l}
\tau^{(1)}_{j+1,l} + J_2\, \tau^{(2)}_{j,l} \tau^{(2)}_{j-1,l} +
J_3\, \tau^{(3)}_{j,l} \tau^{(3)}_{j,l+1} \right]\,.
\label{2-6}
\een
where $(j,l)$ denote the column and row indices of a site on the lattice.
Typically, $(2+1)$-dimensional models are not solvable,
however, remarkably this model can be solved exactly for constant couplings
$J_i$ \cite{kitaev}, by rewriting it as a fermionic theory. We
will use the fermionic theory, which results from the Jordan-Wigner
transformation given in \cite{nussinov,feng}. The latter introduces two sets
of real fermionic fields obeying the standard anti-commutation
relations. To express the Hamiltonian in terms of these fermionic
fields, we first denote the unit vectors in the $x$ and $y$
directions by $\hi$ and $\hj$, respectively --- see figure \ref{lattice}. Then the vectors
\ben
\vM_1 =
\frac{\sqrt{3}}{2}\, \hi + \frac{3}{2}\, \hj\,,\qquad \vM_2 =
\frac{\sqrt{3}}{2}\, \hi - \frac{3}{2}\, \hj \,,
\label{2-7}
\een
span the reciprocal lattice. We also define the
vectors,
\ben
\vn = \sqrt{3}\, n_1\, \hi +
n_2\left(\frac{\sqrt{3}}{2}\, \hi + \frac{3}{2}\, \hj\right) \,,
\label{nnn}
\een
where $n_1$ and $n_2$ are integers. The vectors $\vn$ denote
the midpoints of the vertical bonds in the honeycomb lattice, \ie the lattice sites are positioned at $\vn\pm \hj/2$.
The Hamiltonian \reef{2-6} can now be written as \ben H_\mt{Kitaev}
= i \sum_{\vn} [ J_1\,b_{\vn} a_{\vn -\vM_1} + J_2\, b_{\vn} a_{\vn
+\vM_2} +J_3\, D_{\vn}\, b_{\vn} a_{\vn} ]  \,, \label{2-9} \een
where the fermion $a_{\vn}$ lives on the site at the top of the
vertical bond labeled by ${\vn}$ while the fermion $b_{\vn}$ lives
on the bottom site --- see figure \ref{lattice}.\footnote{We wish to
emphasize that the labeling in figure \ref{lattice} refers to the
fermionized version of the Kitaev model, given in Eq.~\reef{2-9}.
In particular, the bonds associated with $J_1$ and $J_2$ are
reversed in the original spin Hamiltonian \reef{2-6}. This reversal
of roles is perhaps not so surprising since the Jordan-Wigner
transformation used to produce the fermionic description is
nonlocal.} The quantity $D_{\vn}$ is an operator which is bilinear
in the fermions, can take values $\pm 1$ on each link, and commutes
with the Hamiltonian. This allows us to think of $D_{\vn}$ as
representing a static $Z_2$ gauge field living on the links of the
honeycomb lattice, which is coupled to the fermions. The key point
which makes the Kitaev model integrable is that $D_{\vec n}$ is
conserved leading to an infinite number of conserved quantities; the
ground state sector of the model corresponds to choice of $D_{\vec
n}=1$ on each link \cite{kitaev}.

In the following, we will study quantum quench from the ground state
in this model with protocols where $J_1$ and $J_2$ are held constant
but $J_3$ is time dependent. Since the $D_{\vn}$ commute with
$H_{\rm Kitaev}$, they remain unity throughout the full dynamics. In
this case, we can set $D_{\vn}=1$ in Eq.~(\ref{2-9}) making the
time dependent Hamiltonian quadratic in the fermions.

Now let us first define the Fourier modes $a_\vk$ and $b_\vk$ \ben
\left( \begin{array}{c} a_\vn \\ b_\vn \end{array} \right) =
\sqrt{\frac{4}{N}}\sum_\vk \left[ \left( \begin{array}{c} a_\vk \\
b_\vk \end{array} \right) e^{i \vk \cdot \vn} + \left(
\begin{array}{c} a^\dagger_\vk \\ b^\dagger_\vk \end{array} \right)
e^{-i \vk \cdot \vn} \right] \,, \label{2-10} \een where $N$ is the
total number of sites (assumed to be even) and $\vk$ extends over
half the Brillouin zone. We then define a two-component spinor
\cite{sengupta} \ben \chi_\vk = \frac{1}{\sqrt{2}} \left(
\begin{array}{c}  a_\vk +i b_\vk
\\ i(a_\vk - i b_\vk) \end{array} \right)\,.
\label{2-11}
\een
The Hamiltonian \reef{2-9} with a time dependent $J_3$ then becomes
\bea
H_\mt{Kitaev} &=& 2\int \frac{d^2k}{4\pi^2}\,  \chi^\dagger(\vk) \Big[(J_3(t) +J_1 \cos k_1+ J_2 \cos k_2)\, \sigma_3
\nonumber\\
&&\qquad\qquad\qquad\qquad +  (J_1 \sin k_1 - J_2 \sin k_2)\,
\sigma_1 \Big] \chi(\vk) \,, \labell{2-12} \eea where
 $\sigma_i$ denote Pauli matrices in the space of fermions and we have defined
 \ben k_1 \equiv \vk \cdot \vM_1\,,\qquad k_2 \equiv
\vk \cdot \vM_2\,. \label{2-13} \een Implicitly, we are again
considering the limit of an infinite lattice size, which has
resulted in replacing the momentum sum with an integral in
Eq.~\reef{2-13}. The measure is defined as $d^2k\equiv dk_1 dk_2$,
where the range of the integral is $0 \leq k_1,k_2 \leq \pi$
\cite{sengupta}.

For constant $J_3$, the model is critical over a two-dimensional region in the
$(J_1,J_2,J_3)$ hyperplane. Indeed the gap vanishes when
\bea
\cos k_1  =  - \frac{J_3^2+J_1^2-J_2^2}{2J_3 J_1}\,,\qquad&&\quad \cos
k_2  =  - \frac{J_3^2+J_2^2-J_1^2}{2J_3 J_2}\,, \label{2-14}\\
J_1 \sin k_1 &=& J_2 \sin k_2\,. \nonumber
 \eea
The solutions to these equations (\ref{2-14}) can be represented by a triangle with
sides of lengths $J_1,J_2,J_3$ and the angle between the $(J_1,J_3)$
sides being $k_1$ while that between the $(J_2,J_3)$ sides being
$k_2$. These conditions result in two bands of critical couplings satisfying $|J_1-J_2| \leq |J_3| \leq |J_1+J_2|$.

The continuum limit of the model depends on whether the limit is
constructed around a point in the interior of one of these critical
bands, or around a point on one of the edges. To simplify our
discussion of quenches in the following, we will set  $J_1 = J_2 =
J> 0$ and define $J_3 \equiv -2J\,g(t)$ for which the Hamiltonian
\reef{2-12} simplifies to \ben H_\mt{Kitaev} = 2J\int
\frac{d^2k}{4\pi^2}\,  \chi^\dagger(\vk) \Big[(\cos k_1+  \cos
k_2-2\,g(t) )\, \sigma_3 +  ( \sin k_1 -  \sin k_2)\, \sigma_1 \Big]
\chi(\vk) \,. \label{2-12a} \een Within this two-dimensional space
of couplings, the gapless constraints \reef{2-14} reduce to \ben k_1
= k_2 = \bk\,,~~~~~~~~~\cos \bk =g(t)\,, \label{2-15} \een and the
critical region becomes $|g(t)| \leq 1$. If we think of $g$ as a
constant for a moment, there are three distinct classes of critical
models corresponding to: 1) interior points with $1<|g|<0$; 2) the
edge points with $g=\pm1$; and 3) the ``interior" edge points with
$g=0$. Although the latter lies in the interior of the critical
region $-1 \leq g \leq 1$, it corresponds to the point where the
edges of the otherwise two distinct bands described previously merge
together with the choice $J_1 = J_2$. We now consider the continuum
limit associated with each of these critical models.

\subsection*{1) Interior points with $1<|g|<0$}

We expect the continuum theory for such interior points to be
a massless theory, as may be seen as follows: As before, we introduce a lattice spacing $a$ and then expand around the critical
point with
\ben
g = \bg + a \,m\quad{\rm and}\quad
k_i = \bk + \frac{a\,m}{\sin\! \bk}+  a\,p_i \,, ~~~i=1,2
\label{2-16}
\een
where $\cos\bk=\bg$. 
Further, we rescale the fields as \ben \chi(\vk) =  \psi (\vp)/a\,,
\een and define the dimensionless coupling $\hJ=3\sqrt{3}\,Ja/2$.
Now in the continuum limit $a\rightarrow 0$, the coupling $\hJ$, the
momenta $p_i$, the mass scale $m$ and the field $\psi(\vp)$ are held
fixed. With this limit, Eq.~\reef{2-12} yields the continuum
Hamiltonian \ben H^\mt{cont(1)}_\mt{Kitaev} = -2\hJ \int \frac{d^2
p}{(2\pi)^2}\, \psi^\dagger (\vp) \left[ 3\cos\! \bk \,p_y\,
\sigma_1 + \sqrt{3} \sin\!\bk \,p_x\, \sigma_3 \right] \psi (\vp)
\,, \label{2-19} \een where we used Eq.~(\ref{2-7}) to relate
$(p_1,p_2)$ to the momenta along the $x$ and $y$ axes, \ie \ben
\label{pxy} p_1 + p_2 = \sqrt{3} p_x\,,~~~~~~~p_1-p_2 = 3 p_y\,.
\een Note that implicitly we have introduced the standard measure
$d^2p=dp_xdp_y$ and these momentum integrals have infinite range.
Further, the Jacobian arising in transforming from $dp_1dp_2$ has
been absorbed in $\hJ$. Eq.~\reef{2-19} is the Hamiltonian of a
massless Dirac fermion in $2+1$ dimensions.  Of course, to obtain
the standard form of the Dirac Hamiltonian, one has to rescale the
spatial coordinates by a factor depending on $\bk$, \ie on $\bg$.

Let us make a few additional comments: First, the (energy) scaling
dimension of the momentum space field $\psi (\vp)$ is $-1$. Therefore
the corresponding scaling dimension of the position space field is
$+1$, and the operator ${\bar \psi}(\vx) \psi(\vx)$  has scaling
dimension $2$, as expected in a ($2+1$)-dimensional relativistic
theory.

Second, we note that in the following, our quenches result from
introducing a time dependence in the coupling $J_3(t)$, or in the
mass scale $m(t)$ above. Given that the latter scale does not appear
in the continuum Hamiltonian \reef{2-19}, it may naively appear that
such quenches do not effect the continuum limit. However, note that
$m(t)$ also appears in the definition of the momenta $p_i$ in
Eq.~\reef{2-16}. If we make a more conventional expansion removing
the latter shift, \ie we use \ben g = \bg + a \,m\quad{\rm and}\quad
k_i = \bk +  a\,\tilde p_i \,, ~~~i=1,2\,, \label{2-16a} \een the
continuum Hamiltonian becomes \ben H^\mt{cont(1)}_\mt{Kitaev} =
-2\hJ \int \frac{d^2 \tilde p}{(2\pi)^2}\, \psi^\dagger (\vp) \left[
3\cos\! \bk \,\tilde p_y\, \sigma_1 +
\left(\sqrt{3}\sin\!\bk\,\tilde p_x+2\,m(t)\right) \sigma_3 \right]
\psi (\vp) \,. \label{2-19a} \een Hence the dispersion relation
becomes \ben E^2 = 4\hJ^2 \left[ 9\cos^2\! \bk\,\tilde p_y^2 +
\left(\sqrt{3}\sin\!\bk\,\tilde p_x+2\,m(t)\right)^2\right] \,,
\label{wild} \een and we can see that a time dependent $m(t)$ shifts
the zero of this dispersion relation for the low-energy modes. Hence
a time varying $m(t)$ will produce a nontrivial quench. Of course,
from the perspective of a relativistic field theory, $m(t)$ appears
here as an unconventional coupling for the continuum Hamiltonian. In
fact, this coupling begins to reveal the anisotropic nature of the
underlying lattice model.

\subsection*{2) Edge points with $g=\pm1$}

For simplicity, let us focus on the lower edge of the critical
region where $g=-1$ (\ie $J_3=2J$) and hence where $\cos \bk =- 1$
(and $\sin \bk = 0$). Expanding around the critical point with \ben
g = -1 + \epsilon\quad{\rm and}\quad k_i =  \pi+q_i\,, 
\een
the Hamiltonian \reef{2-12} becomes to lowest order
\ben
H_\mt{Kitaev} =
2 J \int \frac{d^2 q}{4\pi^2}\,  \chi^\dagger (\vec q) \left[  (q_2 -
  q_1)\, \sigma_1 + \left( \frac{1}{2} \left(q_1^2 +q_2^2\right)-2\,\epsilon \right) \sigma_3 \right] \chi (\vec q)\,.
\label{2-21} \een Now we introduce a lattice spacing $a$ and define
\ben p_y = \frac{q_1 - q_2}{3\,a}\,,\quad p_x = \frac{q_1 +
q_2}{\sqrt{3\,a}}\,, \quad m = \frac{\epsilon}{a}\,\ \ {\rm and}\ \
\psi(\vec p)=a^{3/4}\,\chi (\vec q)\,, \label{2-23-2} \een as well
as $\hJ=3\sqrt{3}\,Ja/2$. Then continuum limit is obtained by taking
$a \rightarrow 0$ while keeping $\hJ,\, p_x,\, p_y,\, m$ and
$\psi(\vec p)$ finite. The resulting continuum Hamiltonian
then takes the form:
\ben
H^\mt{cont{(2)}}_\mt{Kitaev} = -2\hJ\int \frac{d^2 p}{4\pi^2}\, \psi^\dagger (\vp)
\left[ 3\, p_y \,\sigma_1 + \left(2\,m - \frac{3}{4}\, p_x^2\right)\sigma_3 \right] \psi (\vp)\,.
\label{2-23-3}
\een
At precisely $m =0$, this is the well-known semi-Dirac point.

This anisotropic theory has the dispersion relation \ben E^2 =
4\hJ^2 \left[ 9\,p_y^2 + \left(2\,m- \frac{3}{4}\,
p_x^2\right)^2\right]\,. \label{wild2} \een As in the previous case,
it is clear that a time dependent $m(t)$ will produce an interesting
quench. In contrast to Eq.~\reef{wild}, this mass parameter can not
be absorbed by a shift in the momentum of the low energy modes.
However, for $m\ge0$, the theory is still gapless with $E=0$ for
$(p_x,p_y)=(\pm2\sqrt{2m/3},0)$. For $m<0$, the theory has a gap
with $E_\mt{gap}=\pm 4J|m|$ at $(p_x,p_y)=(0,0)$. Further in the
latter case, for very low-lying modes, we might write the dispersion
relation \reef{wild2} as \ben E^2 \simeq 4\hJ^2 \left[ 9\,p_y^2 +
{3}\,|m|\, p_x^2 +4\,m^2+ \cdots \right]\,, \label{wild2a} \een
which has a form closer to the familiar relativistic Klein-Gordon
relation. However,  a quench with $m(t)$ would differ from the
familiar mass quench, \eg \cite{dgm1,dgm2} since the anisotropy
between the $x$ and $y$ directions is also changed as $m$ varies
with time.

In position space, the Hamiltonian \reef{2-23-3} would take the form
\ben H^\mt{cont{(2)}}_\mt{Kitaev} = 2\hJ\int dx \, dy\, \psi^\dagger
(x,y) \left[ 3i\, \sigma_1\, \partial_y - \sigma_3\left(2\,m
+\frac{3}{4}\, \partial_x^2\right) \right] \psi (x,y)\,.
\label{2-23-4} \een Note that for this anisotropic critical point,
the (energy) scaling dimension of the $y$ coordinate is $(-1)$ as
usual, however, the scaling dimension of $x$ is $(-1/2)$. Further,
the scaling dimension of the momentum space field $\psi (\vp)$ is
$-3/4$, and consequently the scaling dimension of the position space
field $\psi (x,y)$ is $+3/4$. The operator ${\bar \psi}(x,y)
\psi(x,y)$ then has a scaling dimension $3/2$.

\subsection*{{\it iii}) ``Interior" edge points with $g=0$}

There is a distinct critical point at $g = 0$, \ie where $J_3=0=\cos
\bk$. This corresponds to $\sin \bk = +1$, \ie $\bk=\pi/2$,
in the following. Hence expanding around the critical point with
\ben
g= \epsilon\qquad {\rm and}\qquad k_i = \frac{\pi}2 + q_i\,, 
\een
the Hamiltonian \reef{2-12} becomes to lowest order
\ben
H_\mt{Kitaev} =
-2 J \int \frac{d^2 q}{4\pi^2}\,  \chi^\dagger (\vec q) \left[  \frac12\,(q_1^2 -
  q_2^2)\, \sigma_1 + \left(q_1 +q_2+2\,\epsilon\right) \sigma_3 \right] \chi (\vec q)\,.
\label{2-21x} \een Now we introduce a lattice spacing $a$ and define
\ben p_y = \frac{q_1 - q_2}{3}\,,\quad p_x = \frac{q_1 +
q_2}{\sqrt{3}\,a}\,, \quad m = \frac{\epsilon}{a}\,,\quad \psi(\vec
p)=a^{1/2}\,\chi (\vec q)\,, \label{2-23-2x} \een and, as before,
$\hJ=3\sqrt{3}\,Ja/2$. Then continuum limit is obtained by   taking
$a \rightarrow 0$ while keeping $\hJ,\, p_x,\, p_y,\, m$ and
$\psi(\vec p)$ finite. The resulting continuum Hamiltonian  takes
the form: \ben H^\mt{cont{(3)}}_\mt{Kitaev} = -2\hJ\int \frac{d^2
p}{4\pi^2}\, \psi^\dagger (\vp) \left[ \frac{3\sqrt{3}}2\, p_y\,p_x
\,\sigma_1 +\left(\sqrt{3}\, p_x+2\,m\right)\sigma_3 \right] \psi
(\vp)\,. \label{2-23-3x} \een

Note that from the scaling in Eq.~\reef{2-23-2x}, $p_x$ has the
standard (energy) scaling dimension of (+1) and further can be made
as large as we like in the continuum theory. On the other hand,
$p_y$ is dimensionless and implicitly, the above results are only
valid of $p_y\ll1$. We can remove the latter restriction by
retaining the full nonlinearity of $p_y$ in the `continuum' theory.
This approach yields \ben H^\mt{`cont'(3)}_\mt{Kitaev} = -2\hJ\int
\frac{d^2 p}{4\pi^2}\, \psi^\dagger (\vp) \left[ \sqrt{3}\,
\sin\!\left(\frac{3\,p_y}{2}\right)p_x \,\sigma_1 +\left(\sqrt{3}\,
\cos\!\left(\frac{3\,p_y}{2}\right)p_x+2\,m\right)\sigma_3 \right]
\psi (\vp)\,, \label{2-23-3y} \een where $p_y$ can take finite
values above. However, we see that this momentum is periodic with
period $p_y\sim p_y+4\pi/3$. In some sense, this scaling limit has
only produced a continuum theory in the $x$ direction and the $y$
direction remains discrete. That is, we could interpret
Eq.~\reef{2-23-3y} as the Hamiltonian of (coupled) fermions living
on a family of one-dimensional defects, \ie the position space field
would take the form $\psi(x,n_y)$, where $x$ labels the position
along the defects and $n_y$ labels on which defect the fermion
resides. This behaviour is not unexpected since with $J_3=0$, the
Kitaev Hamiltonian \reef{2-6} reduces to a family of uncoupled
one-dimensional spin chains stretching in the $x$ direction. Here
$m$ introduces a small coupling between these chains. This unusual
anisotropic theory \reef{2-23-3y} has the dispersion relation \ben
E^2 = 4\hJ^2 \left[ 3\,p_x^2 + 4\,m^2
+4\sqrt{3}\,m\,\cos\!\left(\frac{3\,p_y}{2}\right) p_x\right]\,.
\een Note that with $m=0$ (vanishing coupling between the
one-dimensional defects), this dispersion relation becomes
independent of $p_y$.

For this anisotropic critical theory, the (energy) scaling
dimensions of the $x$ coordinate is $(-1)$ as usual. The scaling
dimension of the momentum space field $\psi (\vp)$ is $(-1/2)$ and
the scaling dimension of the position space field $\psi (x,n_y)$ is
$(+1/2)$, as appropriate for a one-dimensional fermion. The operator
${\bar \psi}(x,n_y) \psi(x,n_y)$ then has the scaling dimension $1$,
again as in a one-dimensional theory.

\section{Quantization}\label{sec-3}

The two lattice Hamiltonians given in Eqs.~\reef{ham2} and
\reef{2-12a} are both of the form \ben H= \int
\frac{d^{D}k}{(2\pi)^{D}}\ \chi^\dagger (\vk)
\left[-m(\vk,t)\,\sigma_3 + G(\vk)\, \sigma_1 \right] \chi (\vk) \,,
\label{3-1} \een where $D$ is the number of (spatial) dimensions.
The functions $m(\vk,t)$ and $G(\vk)$ are given in Table
\ref{table1} for the two models. Note that $G(\vk)$ is an odd
function of the momentum, \ie $G(-\vk) = - G(\vk)$. The
two-component spinor $\chi (\vk,t)$ is written as
\begin{eqnarray}
\chi(\vk) = \left( \begin{array}{c}
\chi_1 (\vk) \\
\chi_2 (\vk)
\end{array} \right) \,.
\label{3-1-1}
\end{eqnarray}

For the Ising model, there is an additional Majorana condition
\ben
\chi_2 (\vk) = \chi^\dagger_1 (-\vk) \,.
\label{major}
\een
\begin{table}[h!]
\begin{center}
  \begin{tabular}{ | c | c | c | c|}
    \hline
     & $D$ & $m(k,t)$ & $G(k)$ \\ \hline
    {\rm Ising} & 1 & $-2J(\cos k-g(t))$ & $2J\sin k$ \\ \hline
          {\rm Kitaev} & 2 & $-2J( \cos k_1 +\cos k_2-2\,g(t))$ & $2J( \sin k_1 - \sin k_2)$ \\
    \hline
  \end{tabular}
\end{center}
\caption{Couplings for lattice models}
\label{table1}
\end{table}

Now consider the Heisenberg equation of motion for the above
Hamiltonian,
\ben i \partial_t \chi (\vk,t) = \left[-
m(\vk,t)\, \sigma_3 + G(\vk)\, \sigma_1 \right] \,\chi (\vk,t) \,.
\label{3-4}
\een
The two independent solutions may be expressed in the form
\begin{eqnarray}
U(\vk,t) = \left( \begin{array}{c}
-i\partial_t + m(\vk,t) \\
-G(\vk)
\end{array} \right) \phi(\vk,t) \label{3-4a} \,, \\
V(\vk,t) = \left( \begin{array}{c}
G(\vk) \\
i\partial_t + m(\vk,t)
\end{array} \right) \phi^\star (\vk,t) \,,
\label{3-4b}
\end{eqnarray}
where the scalar function $\phi (\vk,t)$ satisfies the equation
\ben
\partial_t^2 \phi + i \partial_t m(\vk,t)\,\phi + [ G(\vk)^2 + m(\vk,t)^2 ] \phi = 0\,.
\label{3-5}
\een

Our aim is to quantize this theory with a time dependence of the
couplings which saturate to constant values in the past and the
future. In particular, we choose the profile
\ben
g(t)= a + b \tanh (t/\dt)\,,
\label{31-1}
\een
which means that $m(k,t)$ is of the form
\ben
m(k,t) = A(\vk) + B \tanh (t/\dt)\,,
\label{3-2}
\een
where the function $A(\vk)$ and the constant $B$ are given in the table \ref{table2x}.
\begin{table}[h!]
\begin{center}
  \begin{tabular}{ | c | c | c | }
    \hline
     & $A(\vk)$ & $B$ \\ \hline
    {\rm Ising} & $-2J( \cos k-a)$ & $2J\,b$ \\ \hline
          {\rm Kitaev} & $-2J(\cos k_1 +\cos k_2-2\,a)$ & $4J\,b$ \\
    \hline
  \end{tabular}
\end{center}
\caption{Parameters for lattice models}
\label{table2x}
\end{table}

The profile \reef{31-1} was chosen because Eq.~(\ref{3-5}) can be
exactly solved for $m(\vk,t)$  of the form given in
Eq.~(\ref{3-2}). The ``in" positive energy solution, \ie the
solution which behaves as a pure positive frequency mode at $t
\rightarrow \infty$, is given by \cite{Duncan,dgm2} \bea
\label{3-33}
\phi_{in}(\vk,t) & = & \frac{1}{|G(k)|}\sqrt{\frac{\omega_{in} + m_{in}}{2 \omega_{in}}} {\rm exp}[- i\omega_+ (\vk)  t - i \omega_-(\vk)  \dt \log ( 2 \cosh (t/\dt))]\\
& & _2 F_1 [ 1+ i \omega_-(\vk) \dt + i B \dt, i \omega_-(\vk) \dt - i B \dt; 1 - \omega_{in}(\vk) \dt; \frac{1}{2} (1 + \tanh (t/\dt) )] \,,
\nonumber
\eea
where we have defined
\bea
\omega_{in} & = & \sqrt{ G(\vk)^2 + (A(\vk) - B)^2} \,, \nonumber \\
\omega_{out} & = & \sqrt{ G(\vk)^2 + (A(\vk) + B)^2 } \,, \label{3-7}  \\
\omega_\pm & = &\frac{1}{2} ( \omega_{out} \pm \omega_{in} ) \,.
\nonumber \eea Substituting $\phi_{in}$ into Eqs.~\reef{3-4a} and
\reef{3-4b}, we get the solutions $U_{in} (\vk,t)$ and
$V_{in}(\vk,t)$ which are positive and negative frequency
respectively in terms of the ``in" modes. The field can be now
expanded in terms of the ``in" oscillators \ben \chi (\vk,t) =  a
(\vk) U_{in} (\vk,t) + b^\dagger (-\vk) V_{in}(-\vk,t)  \,,
\label{3-8} \een where the usual anti-commutation relations hold,
\ie \bea
&\{ a(\vk),a(\vk^\prime) \} =  \{ b(\vk),b(\vk^\prime) \}= \{ a(\vk),b(\vk^\prime) \} = \{ a(\vk),b^\dagger (\vk^\prime) \} = 0& \nonumber \,, \\
&\{ a(\vk),a^\dagger(\vk^\prime) \} = \{
b(\vk),b^\dagger(\vk^\prime) \}= \delta^{d} (\vk - \vk^\prime) \,. &
\label{3-9}
\eea
Further, the Majorana condition \reef{major}   requires $a(\vk) = b(\vk) $ for the Ising model. One can similarly introduce the ``out" modes, or for that matter, any
Bogoliubov transform of these modes.

In studying these quenches, we will begin the system in the ground state
of the Hamiltonian at $t \rightarrow -\infty$. The Heisenberg picture
state is then the ``in" vacuum
\ben
a_{in} (\vk) |0\rangle_{in} =
b_{in} (\vk) |0\rangle_{in} = 0 \,.
\label{3-10}
\een
We will examine the quenches by following the expectation value of local bilinears of
the fermionic operators in this  ``in" vacuum, \ie $c_n$ for the Ising model and $(a_{\vn},\,b_{\vn})$ for the Kitaev model. We will consider the fermion bilinear
\bea
{\bar{\chi}}_\vn \chi_\vn & =  \int \frac{d^D k d^D k^\prime}{(4\pi^2)^D} e^{-i(\vk - \vk^\prime)\cdot \vn} & \{ a^\dagger (\vk) a (\vk^\prime) U^\dagger (\vk) \sigma_3 U(\vk^\prime)  +
a^\dagger (\vk) b^\dagger (-\vk^\prime) U^\dagger (\vk) \sigma_3 V(-\vk^\prime)
\nonumber \\
& &  b (-\vk) a (\vk^\prime) V^\dagger (-\vk) \sigma_3 U(\vk^\prime) +  b (-\vk) b^\dagger (-\vk^\prime) V^\dagger (-\vk) \sigma_3 V(-\vk^\prime) \} \,.
\nonumber
\eea
In terms of the two-component momentum space fermion field,\footnote{Implicitly, we are defining $\bar\chi=\chi^\dagger\sigma_3$ in both models. \label{footy2}} these expectation values become
as
\bea
_{in}\langle0| {\bar{\chi}} \chi |0\rangle_{in} & = &\int\frac{d^D k}{(2\pi)^D} ~_{in}\langle0|{\bar{V}}(\vk,t) V(\vk,t)|0\rangle_{in}
\labell{3-11}\\
& = & \int\frac{d^D k}{(2\pi)^D} \left[ - |\partial_t \phi_{in}|^2
+\left[G(\vk)^2 - m(\vk,t)^2\right] |\phi_{in}|^2 + 2 m(\vk,t)~{\rm {Im}}[
\phi_{in}\partial_t \phi_{in}^\star] \right] \,, \nonumber
\eea
where $\phi_{in}$ is the solution given in Eq.~(\ref{3-33}) for a
given protocol, \ie for specific values of the constants $a$ and $b$
in Eq.~\reef{31-1}.\footnote{Recall that $D=1$ for the transverse
field Ising model and $D=2$ for the  Kitaev honeycomb model.} A
measure of the excitation of the system is given by the difference
between this quantity measured in the quench and its adiabatic value
\ben \langle{\bar{\chi}} \chi \rangle_\mt{diff} \equiv~~
_{in}\langle0| {\bar{\chi}} \chi |0\rangle_{in} - \op_\mt{adia}\,.
\label{3-12} \een The adiabatic value is obtained by replacing the
exact solution by the lowest order adiabatic solution, \ben
\op_\mt{adia}  = \int\frac{d^D k}{(2\pi)^D} \left[ - |\partial_t
\phi_{adia}|^2 +[(G(\vk))^2 - (m(\vk,t))^2] |\phi_{adia}|^2 + 2
m(\vk,t)~{\rm {Im}}[ \phi_{adia}\partial_t \phi_{adia}^\star]
\right] \,, \label{3-13a} \een where \beqa &&\phi_{adia}   \equiv
\frac{1}{|G(\vk)|}\,\sqrt{\frac{\omega (\vk,t) + m(\vk,t)} {2 \omega
(\vk,t)}}\, {\rm exp}[- i\omega (\vk,t)  t ] \,,
\label{3-13}\\
&&\qquad {\rm with}\qquad \omega (\vk,t)  \equiv  \sqrt{G(\vk)^2 +
m(\vk,t)^2} \,. \nonumber \eeqa and $m(k,t)$ is given by
Eq.~(\ref{3-2}) for a specified time $t$. Substituting this
solution into Eq.~(\ref{3-13a}), the adiabatic value at time $t$
simplifies to \beq \op_{adia}(t) = \int\frac{d^D k}{(2\pi)^D} \frac{
m(k,t)}{\omega(k,t)}\,. \label{oadia} \eeq


\section{Results}\label{sec-4}

In this section, we summarize our results of the calculation of
$\langle{\bar{\chi}} \chi \rangle_\mt{diff}$ in Eq.~(\ref{3-12})
for various quench protocols in the two lattice models described in
section \ref{sec-2}.

\subsection{Transverse Field Ising Model}\label{resice}

The Ising model has an isolated critical point which separates two
massive phases. The most interesting quench protocol in this case is
a $g(t)$ which starts in one of the phases, crosses the critical
point at $g=1$ (at time $t=0$) and ends in the other phase at late
times. Hence with the profile in Eq.~(\ref{31-1}), we choose $a=1$
and consider quenches with various values of $b$.

To begin, we consider small values of $b$ so that throughout the
quench, the model is close to the critical point and we may expect
that the results can be compared to the continuum limit. That is, as
in Eq.~\reef{work}, our profile has the form $g = 1 -\epsilon (t)$
with $\epsilon(t)=-b\tanh(t/\dt)$, \ie $b$ controls the amplitude in
the variation of the dimensionless ``mass" parameter $\epsilon(t)$.
In particular, $\epsilon_{in}=b$ (and $\epsilon_{out}=-b$). Further,
as in \cite{dgm1,dgm2,dgm3,dgm4}, we examine the effect of varying
the quench rate by varying $\dt$. Following the results of
\cite{dgm4}, we can expect to see different scalings in $\langle
{\bar{\psi}} \psi \rangle_\mt{diff}$ for fast and slow quenches. In
passing, we note that the only dimensionful quantity in Hamiltonian
\reef{ham2} is the overall factor of $J$, the nearest neighbor bond
strength, and as noted in footnote \ref{footy3}, this coupling
essentially defines the lattice spacing. Hence $J\,\dt$ is the
natural dimensionless quantity with which to discuss the different
quench rates. Further, let us note that in the lattice models, $\op$
is a dimensionless quantity which does not scale with $J$, \eg  one
finds that the factors of $J$ cancel out in eq.~\reef{oadia}.

\begin{figure}[h!]
\setlength{\abovecaptionskip}{0 pt}
\centering
\includegraphics[scale=0.9]{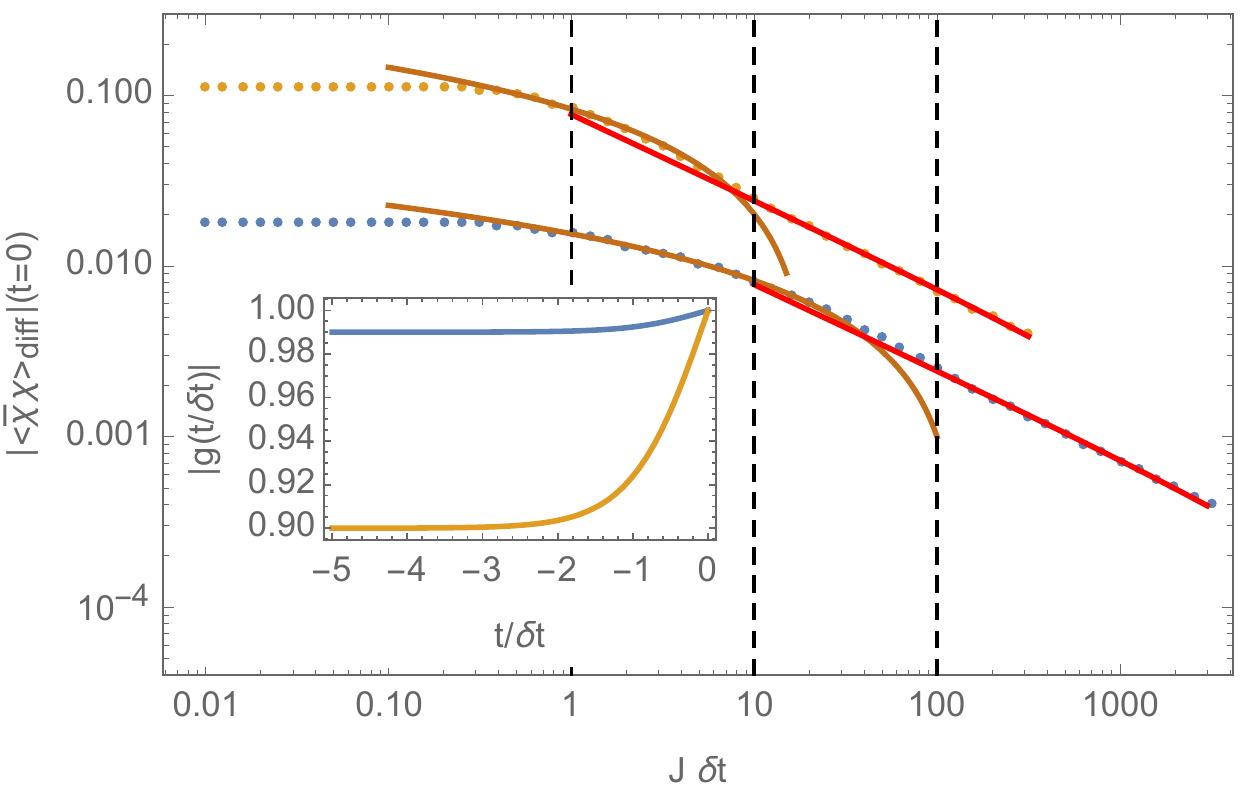}
\caption{The response $\langle {\bar{\chi}} \chi \rangle_\mt{diff}$
at fixed $t=0$ as a function of $J\dt$ for the (one-dimensional)
transverse field Ising model. The time-dependences of the coupling
$g(t)$ are given in the inset. The orange curves are the log fits in
the fast regime, $P \log J\delta t + Q$, and the red lines are the
fits to the Kibble-Zurek scaling in the slow regime, $P' (J\delta
t)^{-1/2} + Q'$. The dotted lines indicate expected cross-overs at
$J\dt\sim 1$ and $J\dt\sim 1/\epsilon_{in}$. The blue dots are the
response for an initial gap of $\epsilon_{in}=0.01$ and the best
fits give $P=-0.00313$, $Q=0.0154$, $P'=0.0248$ and $Q'=5.90 \times
10^{-5}$. For the yellow dots, $\epsilon_{in}=0.1$ and the best fits
give $P=-0.0273$, $Q=0.0830$, $P'=0.0777$ and $Q'=5.10 \times
10^{-4}$. } \label{fig_ising}
\end{figure}

Figure \ref{fig_ising} shows the response $\langle {\bar{\chi}} \chi \rangle_\mt{diff}$ at $t=0$ as a function of the inverse quench rate $J\delta t$.
The two cases shown in the figure begin at $t = -\infty$ with  $\epsilon_{in}=0.01$ and $0.1$, and the plots show several clear features:
\begin{enumerate}
\item First, for small $J\dt$ of order one or less, the response saturates as a function of the quench rate. We can think of this as the ``instantaneous quench" regime, where the quench rate is of the same order as the lattice spacing. We will discuss a comparison with the results of an instantaneous quench for the Ising model, as well as the Kitaev model, in section \ref{sec-sudden}. We will further discuss the saturation point for small amplitudes later in this section and provide more details in Appendix \ref{app-sat}.
\item For $J\dt$ roughly between $1$ and $1/\epsilon_{in}$, the quench time scale is much larger than the lattice scale (and so the results should be comparable to the continuum theory), but dimensionless combination $\epsilon_{in}\,J\dt$ is small. This is the ``fast quench" regime, as defined in \cite{dgm1,dgm2,dgm3,dgm4} for the continuum theory. The conformal dimension of the operator $\bar\chi\chi(x)$ is $\Delta=1$ (which matches one half of the spacetime dimension), and so the continuum result \reef{1-6} suggests that there should be no leading power law dependence on $\dt$. Instead, in this regime, the curves in the figure are well fit with a dependence of the form $P\log(\dt) + Q$, which is exactly what is expected from the continuum calculations. An explicit derivation of the logarithmic dependence in this regime is given in Appendix \ref{app-kubo}.
\item Finally, for $J \dt > 1/\epsilon_{in}$, the response can be fit with $P'\dt^{-1/2} + Q'$ where $Q'$ is essentially zero --- see the figure caption. This result is consistent with Kibble-Zurek scaling of the continuum theory, discussed in \cite{dgm4}. The Kibble-Zurek time is $t_{KZ} \sim \dt^{1/2}$ and then Eq.~(\ref{1-1}) predicts a $\dt^{-1/2}$ scaling since $\Delta = 1$ in this case.
\end{enumerate}
Hence for quenches which only make small excursions from the
critical point, our results agree with those expected for the
continuum theory \cite{dgm1,dgm2,dgm3,dgm4}. This behaviour can be
anticipated because these ``small-amplitude" quenches are largely
only exciting very low energy or long wavelength modes which are
described well by the continuum theory. Of course, the saturation of
$\op_\mt{diff}$ observed in the very fast regime with $J\dt<1$ is
not a feature found in the continuum theory.\footnote{However, as
discussed in footnote \ref{footy78}, this saturation can be emulated
by considering nonlocal operators, \eg $\langle\bar\chi(\delta\vec
x,t)\,\chi(0,t)\rangle$ \cite{dgm3}. The expectation value of these
operators saturates in the regime $\dt\ll |\delta\vec x|$, \ie in
the regime where modes with wavelengths much shorter than
$|\delta\vec x|$ are being excited.} One's naive intuition about
this regime may be that the quench is exciting short wavelength
modes where the nonlinearities of the lattice model become apparent and so the results depart from anything observed in the continuum theory. However, we will see below that this intuition is not quite correct.

One might expect that the response will saturate when the quench is
exciting all possible modes in the lattice theory, \ie when
excitations are being created in all modes. To understand this
point, we begin by substituting the profile \reef{31-1} (with $a=1$
and $b=\e_{in}$) into the dispersion relation (\ref{green}) for the
Ising model and we find
\begin{eqnarray}
E_k^2 = 4 J^2 \left[ 4 \sin^2(k/2) [1+ \epsilon_{in} \tanh(t/\dt) ] + \epsilon_{in}^2 \tanh^2(t/\dt) \right]\,.
\end{eqnarray}
Now we can ask when a mode at a particular wave-number $k$ is going
to be excited. According to the Landau criterion \reef{1-1-2}, this
mode will remain adiabatic throughout the quench (and in particular,
at $t=0$) if \beq
 \frac{1}{E_{k}^2}\left|\frac{dE_{k}}{dt}\right|_{t=0} = \frac{|\e_{in}|}{8 J \dt\, \sin (k/2)} \lesssim 1\,.
\label{gopher} \eeq On the other hand, if the above expression
exceeds one, then this mode with momentum $k$ is excited by the
quench. Now it is clear that there will always be excited modes in
the  neighborhood of $k=0$. The above discussion may now suggest
that saturation will be achieved when the violation of the above
criterion \reef{gopher} extends out to $k=\pi$. However, it turns
out that for small amplitude quenches, only a narrow band of momenta
near $k=0$ contribute significantly to the expectation value
\reef{3-12}. In particular, we show in appendix \ref{app-sat} that
$\langle{\bar{\chi}}\chi \rangle_\mt{diff}$ only receives
significant contributions from modes with $0\le k< k_{max}$ with
$k_{\max} = c\,|\e_{in}|$ where $c$ is some order one number. Hence
the violation of the criterion \reef{gopher} need only extend to
$k=k_{max}$ in order to produce saturation with small amplitudes.
Hence it is straightforward to see from Eq.~\reef{gopher} that the
small amplitude quenches will saturate for  $J \dt \leq 1/(4c)$ or
more simply $J \dt \lesssim 1$.\footnote{This result requires that
the factor $c$ is independent of $J\dt$, which is shown to be
correct in appendix \ref{app-sat}.} Hence we have an explanation of
the saturation behaviour observed for the small amplitude quenches
shown in figure \ref{fig_ising}. To contrast with the following, we
emphasize that the point where saturation sets in is independent of
the initial amplitude here.

\begin{figure}[h!]
\setlength{\abovecaptionskip}{0 pt}
\centering
\includegraphics[scale=0.9]{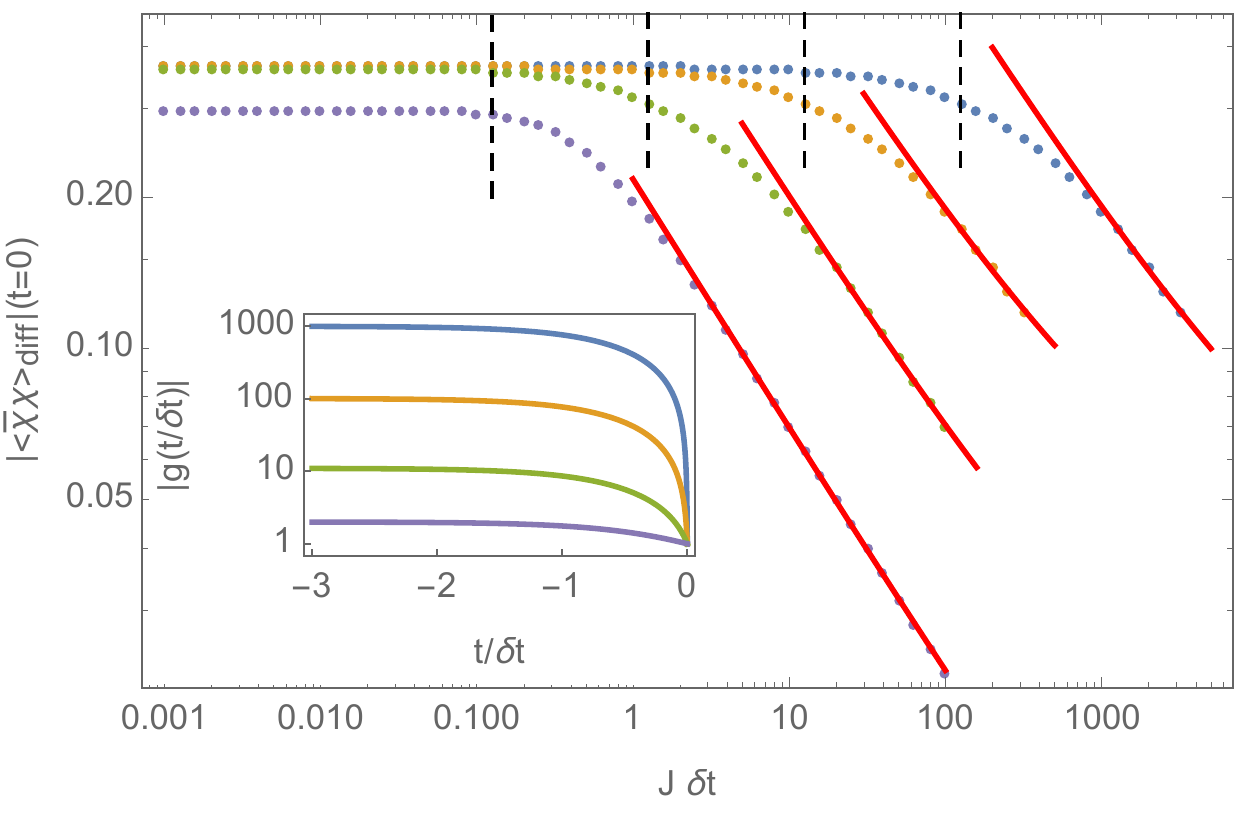}
\caption{The response $\langle {\bar{\chi}} \chi \rangle_\mt{diff}$
at fixed $t=0$ as a function of $J\dt$ for the (one-dimensional)
transverse field Ising model, for large initial amplitudes. The
time-dependences of the coupling $g(t)$ are given in the inset. The
red lines are the fits to the Kibble-Zurek scaling, $P (J\delta
t)^{-1/2} + Q$, that starts at $J\dt\sim |\e_{in}|$. For
$\e_{in}=-1$, $P= 0.216$ and $Q=0.00122$; for $\e_{in}=-10$,
$P=0.604$ and $Q=0.00984$; $\e_{in}=-100$, $P=1.595$ and $Q=0.0299$;
and for $\e_{in}=-1000$, $P=5.247$ and $Q=0.0255$. The black dashed
lines show the expected saturation point $J \dt = |\e_{in}|/8$, in
each case.} \label{fig_ising_large}
\end{figure}

Now we can also examine the scaling behaviour of $\op_\mt{diff}$ for
``large-amplitude'' quenches of the Ising model, \ie with
$|\epsilon_{in}|\gtrsim 1$. In this regime,\footnote{Note we are
choosing $\epsilon_{in}=b$ to be {\it negative} in these
large-amplitude quenches. With this choice, we avoid the other
critical point at $g=-1$ before measuring the expectation value at
$t=0$ --- see footnote \ref{footy}.} we expect that the quenches are
probing the nonlinear regime of the lattice dispersion relation
\reef{green}. The response as measured by the expectation value of
$\bar{\chi}\chi$ is shown for a family of four such quenches in
Figure \ref{fig_ising_large}. These large-amplitude quenches exhibit
three distinct features, which contrast with the behaviour found for
the small-amplitude quenches: {\it a}) there is no longer a fast
quench regime; {\it b}) the Kibble-Zurek scaling regime begins at $J
\dt \sim |\e_{in}|$ (rather than $1/|\e_{in}|$ for small
amplitudes); and {\it c})  the response saturates to the
instantaneous quench value for $J \dt \lesssim |\e_{in}|/8$ (instead
of the previous $J\dt \lesssim 1$). We now discuss each of these in
somewhat more detail.

The three features above are related because an essential
characteristic of these large amplitude quenches is a smooth
transition between the KZ and instantaneous quench regimes, without
an intermediate scaling regime.\footnote{One might try to fit the
plots near the saturation point with some kind of log behaviour.
However, these fits are not reliable, so there is no clear evidence
for a logarithmic behaviour in the narrow window between $J \dt \sim
|\e_{in}|/8$ and $J \dt \sim |\e_{in}|$.}. In the KZ regime, we
still see scaling compatible with $\Delta=1$ expectation, \ie
$\vev{\bar{\chi}\chi}_{\text{diff}} \sim \dt^{-1/2}$, while the
response is saturated in the instantaneous regime. Recall that the
origin of the fast quench scaling \reef{1-6}, is that the quench
activates a growing number of short wavelength modes as $\dt$
shrinks, but further that these new modes are organized as in a CFT
\cite{dgm1,dgm2}. In the transition between the KZ and instantaneous
quench regimes, the large amplitude quenches are already exciting
modes in the nonlinear regime of the Ising dispersion relation
\reef{green}, \ie the quench is probing modes with wavelengths
comparable to the lattice spacing. Hence the latter condition is not
achieved in the large amplitude quenches and we should not expect to
see a fast quench scaling regime.

The second important feature was that the KZ scaling sets in at a
quench rate which grows with the amplitude, rather than decreasing
as in the small amplitude quenches. That is, we observe that the KZ
scaling begins when $J \dt \sim |\e_{in}|$ (instead of
$1/|\e_{in}|$) in Figure \ref{fig_ising_large}. This (naively)
surprising behaviour can be understood by looking carefully at the
conditions required for KZ scaling, as already discussed in the
introduction --- see the discussion around Eqs.~\reef{1-1-4} and
\reef{1-1-5}. We repeat the salient points here: The first condition
for our quench protocol \reef{31-1} was that adiabaticity should
breakdown when the profile is in the linear regime. That is, we must
have $t_{KZ}<\dt$, which in turn yields $J \dt > 1/|\e_{in}|$, when
expressed in terms of Eq.~(\ref{1-1-4}). The second condition for
KZ scaling to hold was that the system has to be close to critical
at $t=t_{KZ}$, \ie $E_{KZ}<\Lambda_\mt{UV}$. Interpreting $J$ as the
inverse lattice spacing (see footnote \ref{footy3}),  this condition
expressed as Eq.~(\ref{1-1-5}) yields $J\dt > |\e_{in}|$. Of
course, when $|\e_{in}|$ is small,  the first condition is stronger
and we recover the behaviour observed in the small amplitude
quenches. However, for the large amplitude quenches, it is the
second restriction that determines the onset of KZ scaling, in
agreement with the results found in Figure \ref{fig_ising_large}.

The third and final feature observed in figure
\ref{fig_ising_large} is that the saturation point, where the
instantaneous quench regime begins, also grows with increasing
amplitude in the large amplitude quenches. In fact, we observe that
the expectation value saturates into the instantaneous quench value
for roughly $J \dt \lesssim |\e_{in}|/8$, instead of $J\dt \lesssim
1$ as observed for the small amplitude quenches. This behaviour can
be understood by the same reasoning used above in discussing the
small amplitude quenches. The key difference is that for large
amplitude quenches, the response \reef{3-12} does, in fact, receive
significant contributions from all modes, \ie $0\le k\lesssim \pi$
--- see appendix \ref{app-sat} for details. Hence we should ask when
is the highest mode going to be excited\footnote{Strictly
speaking, the contribution of the $k=\pi$ mode to the expectation
value $\langle{\bar{\chi}} \chi \rangle_\mt{diff}$ is zero for any
amplitude. However, this same argument holds for any large $k$. In
the case of large amplitudes, these modes have significant
contributions to $\langle{\bar{\chi}} \chi \rangle_\mt{diff}$ and so
the following argument applies. See Appendix \ref{app-sat} for more
details.} and substituting $k=\pi$  into Eq.~\reef{gopher}, we find
\beq \frac{1}{E_\pi^2}\,\left|\frac{dE_\pi}{dt}\right|_{t=0} =
\frac{|\e_{in}|}{8 J \dt}  \,. \label{4-1-1} \eeq Therefore all of
the modes will be excited when the above expression is bigger than
one, \ie for $J \dt \lesssim |\e_{in}|/8$, and hence we expect that
the expectation value will saturate in this regime for large
amplitude quenches.

Finally, we should note that the quenches were studied here by
measuring response exactly at the critical point. That is, we
evaluated the expectation value precisely at $t=0$. One can also
evaluate the response at some finite time $\tau \equiv t/\dt$. At
least for small amplitudes, the response of the continuum theory
should provide a guide \cite{dgm1,dgm2,dgm3,dgm4}. Hence we expect
that for large enough $J \dt$, the KZ scaling regime should give way
to an adiabatic regime. In fact, we expect to see this transition
around $J \dt \sim 1/(\e_{in} \tau^2)$. As shown in \cite{dgm2}, the
adiabatic expansion for relativistic theories will be a series in
$1/\dt^{2}$ and hence to leading order, we expect that after
subtracting the zeroth-order term for large $J \dt$,
$\vev{\bar{\chi}\chi}_{\text{diff}} \sim \dt^{-2}$. We have verified
that the Ising model indeed produces this behaviour. However, it is
technically challenging to separate clearly all four regimes
(instantaneous, fast, KZ and adiabatic) for a fixed initial
amplitude and finite time.

\subsection{Kitaev Honeycomb Model} \label{honey22}

Recall that to simplify our discussion of quenches in the Kitaev
model \reef{2-12}, we restricted our attention to the subspace
within the full space of couplings where $J_1=J_2=J>0$ and
$J_3=-2J\,g$. With these restrictions,  the Hamiltonian reduces to
that given in Eq.~\reef{2-12a}. The critical region where the
energy gap vanishes in this space of couplings reduces to the
one-dimensional line segment $|g|\le1$, with $g=\cos\bk$ and
$k_1=k_2=\bk$ as in Eq.~\reef{2-15}. Further, as discussed in
detail in section \ref{honey}, there are three classes of critical
models: 1) interior points with $1<|g|<0$; 2) edge points with
$g=\pm1$; and 3) ``interior" edge points with $g=0$; as well as the
gapped phases with $|g|>1$.

As described in section \ref{sec-3}, we quench the system with $g(t)
= a + b \tanh(t/\dt)$. In the following, we will always examine the
response $\op_\mt{diff}$, as defined in Eq.~\reef{3-12}, at $t =
0$. Clearly, there is a wide variety of different quenches depending
on the choice of the parameters, $a$ and $b$. In particular, as we
describe below, the results depend crucially on the phase in which
the quench begins and on the phase at $t=0$ where we measure the
response, \ie the scaling of the responses depends on $g(t\to
-\infty)=a-b$ and on $g(t=0)=a$. Of course, $\op_\mt{diff}(t=0)$
will not depend on the profile of $g(t)$ at latter times $t > 0$.

Given the three different types of critical theories, there are
certainly a wide variety of critical quenches which one might choose
to explore. In the following, we will focus on three protocols: a)
`gapped-to-edge' quenches which begin in the gapped phase and are
measured at the edge critical point; b) `gapped-to-interior'
quenches which begin in the gapped phase and are measured at an
interior point at some finite distance into the critical region; and
c) `interior-to-interior' quenches where the entire protocol only
passes through interior critical points. Clearly this selection is
not exhaustive and only provides a preliminary study of the critical
quench dynamics of the Kitaev model --- see section \ref{sec-5} for
a discussion of other possible protocols.

One feature common to all of the different quenches is that for
$J\dt\lesssim1$, the response saturates as a function of the quench
rate. As described for the Ising model above, we can think of this
as the ``instantaneous quench" regime, where the quench rate is of
the same order as the lattice spacing. In the discussion of the
individual protocols below, we focus on the scaling of the response
for the regime $J\dt>1$. We return to consider the instantaneous
quench regime in section \ref{sec-sudden}.

\subsubsection{Gapped-to-edge} \label{gapedge}

Here, we consider quenches which start in the gapped phase with
$g(-\infty)<-1$ and pass to the edge point with $g(0)=-1$. That is,
we consider profiles \reef{31-1} with $a=-1$ and $b>0$ (and hence
$a-b<-1$).\footnote{Of course, the results for quenching to the edge
point at $g=+1$ are identical. One needs to simply flip the sign of
all of these parameters, \ie $g,a,b\to-g,-a,-b$.} In this case, $b$
sets the scale of the gap in the initial phase. Further note that
even though the system continues into the region of interior
critical points for $t>0$, these protocols are identical to a quench
from a gapped phase which crosses an isolated critical point, since
a measurement at $t=0$ does not care about the values of the
coupling for $t > 0$.

\begin{figure}[htbp]
\setlength{\abovecaptionskip}{0 pt}
\centering
\includegraphics[scale=0.9]{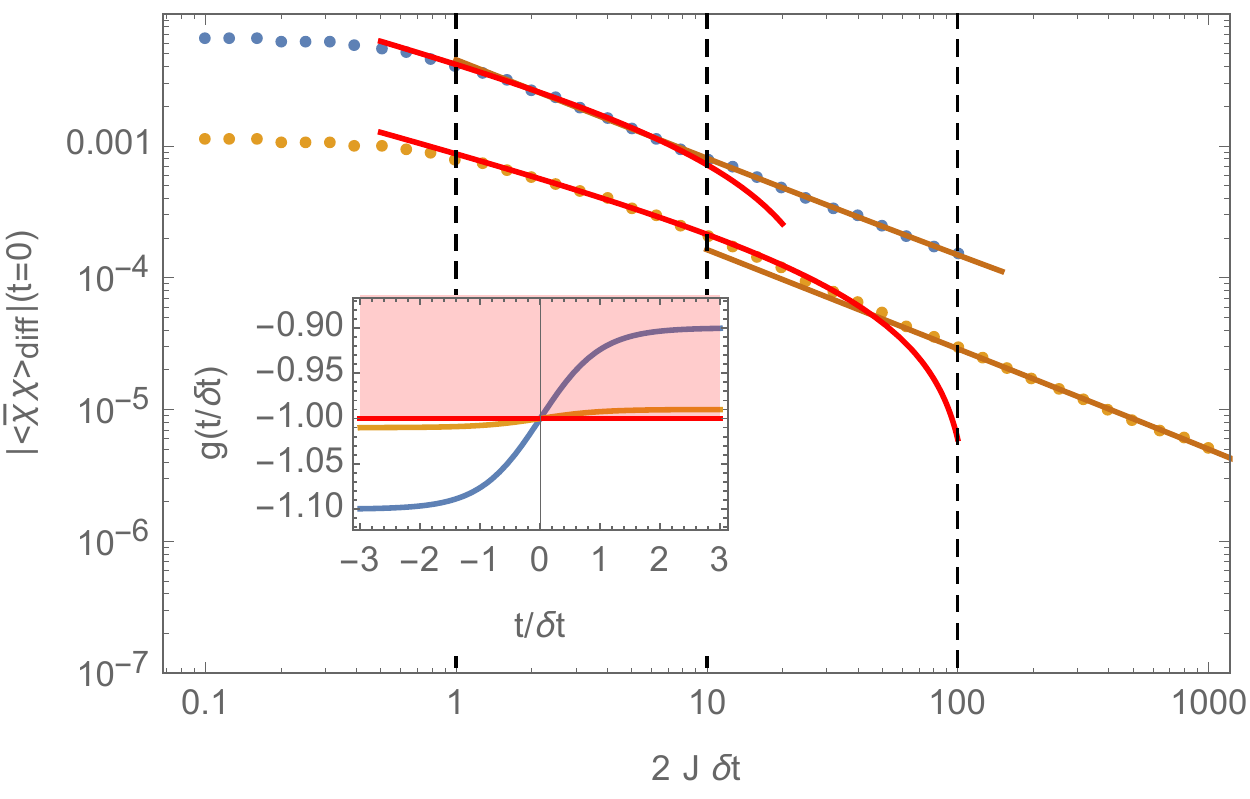}
\caption{ Plot of $\op_\mt{diff}(t=0)$ as a function of $2 J\dt$ for the gapped-to-edge quench shown in the inset with $g(t) = -1 + b \tanh t/\dt$, with $b=0.1$(blue),$0.01$(yellow). The red curves are the best fit for the slow regime by a function $P+Q (2 J \dt)^{-3/4}$, where $P=-0.00448$(blue),$-0.000917$ (yellow) and $Q=-6.97\times 10^{-6}$(blue), $1.617\times 10^{-7} $(yellow). In the fast regime the brown curves indicate the best fit of a function $P + Q (2 J \dt)^{-1/2}$, with $P=-0.00499$(blue), $-9.56 \times 10^{-4} $ (yellow) and $Q=0.000860$(blue), $8.96 \times 10^{-5} $ (yellow).
  }\label{fig-kitaev-edge}
\end{figure}

Figure \ref{fig-kitaev-edge} shows the response for a quench with a
small amplitude, \ie $|b|\ll1$. We see that there are two distinct
scaling behaviours for $\langle{\bar{\chi}} \chi \rangle_\mt{diff}
(t=0)$:  with exponent $-1/2$ in the fast quench regime with
$1\lesssim J\dt\lesssim 1/|b|$, and with the exponent $-3/4$ in the
slow quench regime with $ J\dt\gtrsim 1/|b|$.  Let us note that the
scaling exponent $-3/4$ for slow quenches was also observed in
earlier work \cite{hikichi}, where the variation of the coupling was
taken to be always linear in time and the quench was started at $t =
-\infty$.

Again for small amplitudes, we expect that the quench is described
well by the anisotropic continuum theory, given in
Eq.~\reef{2-23-4}. Recall from the discussion below
Eq.~\reef{2-23-4}, that the dimension of the operator
$\bar\psi\psi$ is $\Delta=3/2$. However, the anisotropy of the
theory is important to identify the fast scaling dimension. In
particular, while $t$ and $y$ scale as regular coordinates with
mass dimension --1, the dimension of the $x$ coordinate was --1/2.
Hence the effective spacetime dimension in various formulae is
$d_\mt{eff}=5/2$, rather than $d=3$. For example, the dimension of
$\hJ m$, the coupling conjugate to $\bar\psi\psi$, is
$d_\mt{eff}-\Delta=1$ and {\it not} $d-\Delta=3/2$. Hence using
$d_\mt{eff}$ in Eq.~\reef{1-6}, we find the scaling\footnote{Note
that $\bar\chi\chi$ is a dimensionless quantity, and in accord with
footnote \reef{footy3}, we are canceling powers of $J$ and the
lattice spacing $a$ in converting the first expression to the
second, \ie we set $Ja\sim1$.} \beq \langle\bar\psi\psi\rangle \sim
\frac{\hJ\, m}{\dt^{1/2}} \quad\implies \quad
\langle\bar\chi\chi\rangle \sim \frac{ b}{(J\dt)^{1/2}} \,,
\label{fastE} \eeq for fast quenches, in agreement with the results
noted above and shown in figure \ref{fig-kitaev-edge}. In the slow
quench regime, the Kibble-Zurek time is given by $ t_{KZ} =
(\dt/\hJ\,m)^{1/2}$,  since in Eq.~(\ref{1-2}) $\nu=1$ and $E_0 =
\hJ m$, and so the response \reef{1-1} becomes \beq
\langle\bar\psi\psi\rangle \sim
\frac{1}{t_{KZ}^{\,\Delta}}=\left(\frac{\hJ\, m}{\dt}\right)^{3/4}
\quad\implies \quad \langle\bar\chi\chi\rangle \sim
\left(\frac{b}{J\dt}\right)^{3/4}\,. \label{slowE} \eeq Again this
reproduces precisely the exponent $-3/4$ found in our numerical
results.\footnote{One can also try to match the $b$ dependence of
$\langle\bar\chi\chi\rangle$ in Eqs.~\reef{fastE} and \reef{slowE}
with our numerical results. While the agreement is promising for
small values of $b$, we expect that it is not exact because we are
working with bare lattice quantities in our numerical calculations.}

\begin{figure}[h!]
\setlength{\abovecaptionskip}{0 pt}
\centering
\includegraphics[scale=0.9]{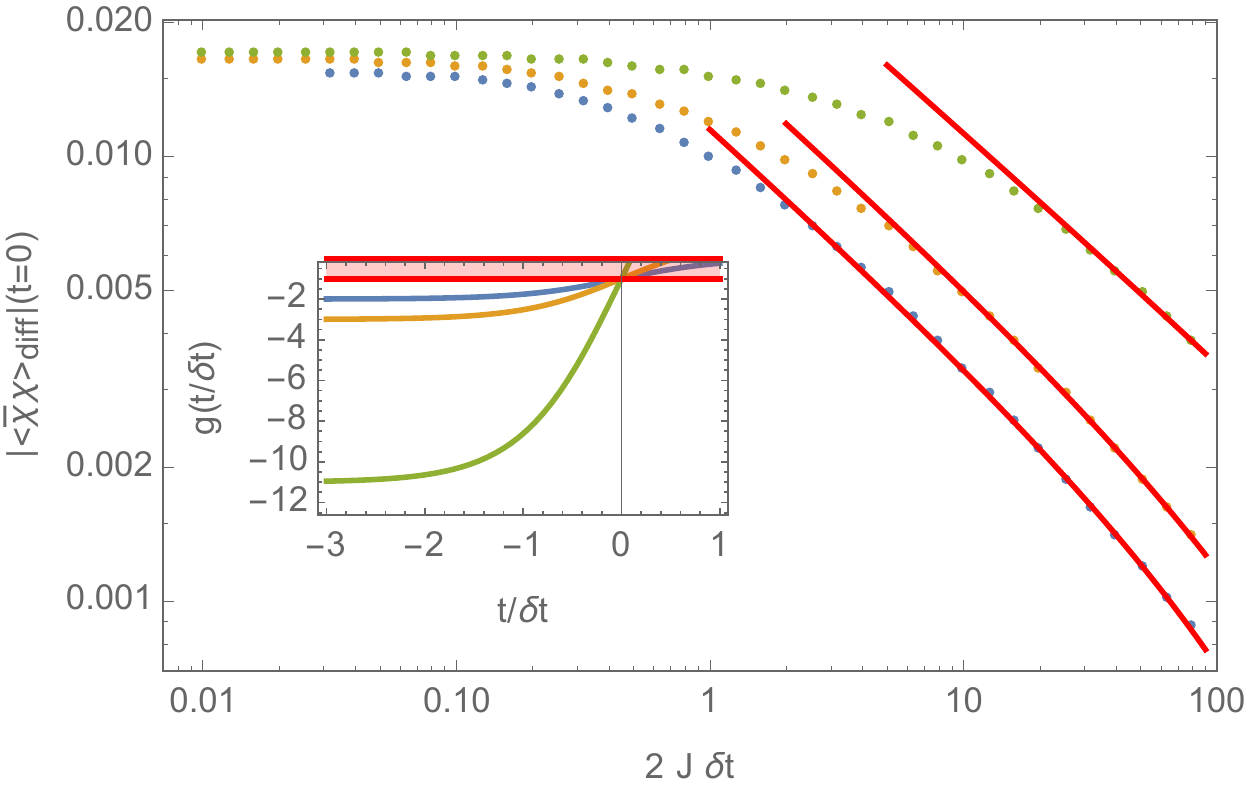}
\caption{ Plot of $\op_\mt{diff}(t=0)$ as a function of $2 J\dt$ for the gapped-to-edge quench shown in the inset with $g(t) = -1 + b \tanh t/\dt$, with $b=1$(blue), $2$(yellow) and $10$ (green). The red curves are the best fit by a function $P+Q (2 J \dt)^{-1/2}$, where $P=-0.0120$(blue),$-0.0176$ (yellow), $-0.0363$ (green) and $Q=4.845\times 10^{-4}$(blue), $5.807\times 10^{-4} $(yellow), $2.322\times 10^{-4} $(green). Note also that the saturation point for small $\dt$ increases as $b$ increases and that there is no fast scaling in the large-amplitude regime.
  }\label{fig-kitaev-edge-large}
\end{figure}

Of course, we can also extend these quenches to large amplitudes,
and as shown in figure \ref{fig-kitaev-edge-large}, the behaviour is
somewhat different. First, as expected, we do not observe any fast
scaling in this case. As discussed for the Ising model above, in the
transition between the instantaneous and slow quench regimes, the
large amplitude quenches are probing modes with wavelengths
comparable to the lattice spacing and hence they cannot be described
by an effective UV CFT.  Hence the fast quench scaling \reef{1-6} is
not produced in these large amplitude quenches. However, there does
appear to be a {\textit{slow}} scaling regime. The exponent,
however, changes continuously as we increase the amplitude from
$\op_\mt{diff} \sim (J\dt)^{-3/4}$ to $\op_\mt{diff} \sim
(J\dt)^{-1/2}$.
Of course, it will be interesting to develop an analytical
understanding of this change in scaling behaviour between the small
and the large amplitude quenches. This new result does not
contradict the results of \cite{hikichi}: in the regime where we see
this scaling, the coupling is proportional to $t$ with a coefficient
which is not quite small, whereas the results of \cite{hikichi}
refer to a regime where this coefficient is very small in units of
$1/J$. Finally, in parallel to what happens in the Ising case when
the amplitude is large, we see that the saturation point for small
$J\dt$ increases roughly linearly as we increase $b$.

As a final note, let us add that in the next section, we will see that for small amplitudes, the exponent for both the fast and slow quench regimes begins to change as
soon as we continue the quench into the gapless phase.

\subsubsection{Gapped-to-interior} \label{gapint}

Next, we consider quenches where we start in the gapped phase with
$g(-\infty)<-1$, pass beyond the edge point and measure the response
at an interior point with $0>g(0)>-1$. That is, we are studying
quench profiles \reef{31-1} with $a-b<-1$ and $0>a>-1$ (and $b>0$).
In this case, there are many different quench protocols that can be
studied and that yield different responses. To concisely go through
them, it will be convenient to define two new parameters: $\delta
g_\mt{in} \equiv b - a -1\ (>0)$, which sets the scale of the gap in
the initial phase, \ie measures the initial distance of $g$ to the
critical region, and $\delta g_\mt{fin} \equiv 1+ a \ (>0)$, which
measures the final distance of $g$ inside the interior critical
region, \ie, the distance of the point at which we measure the
response from  the edge point. There will be three clearly distinct
behaviours depending on whether $\delta g_\mt{fin}$ is much smaller
than, greater than or the same order as $\delta g_\mt{in}$.

\begin{figure}[h!]
\setlength{\abovecaptionskip}{0 pt}
\centering
\includegraphics[scale=0.9]{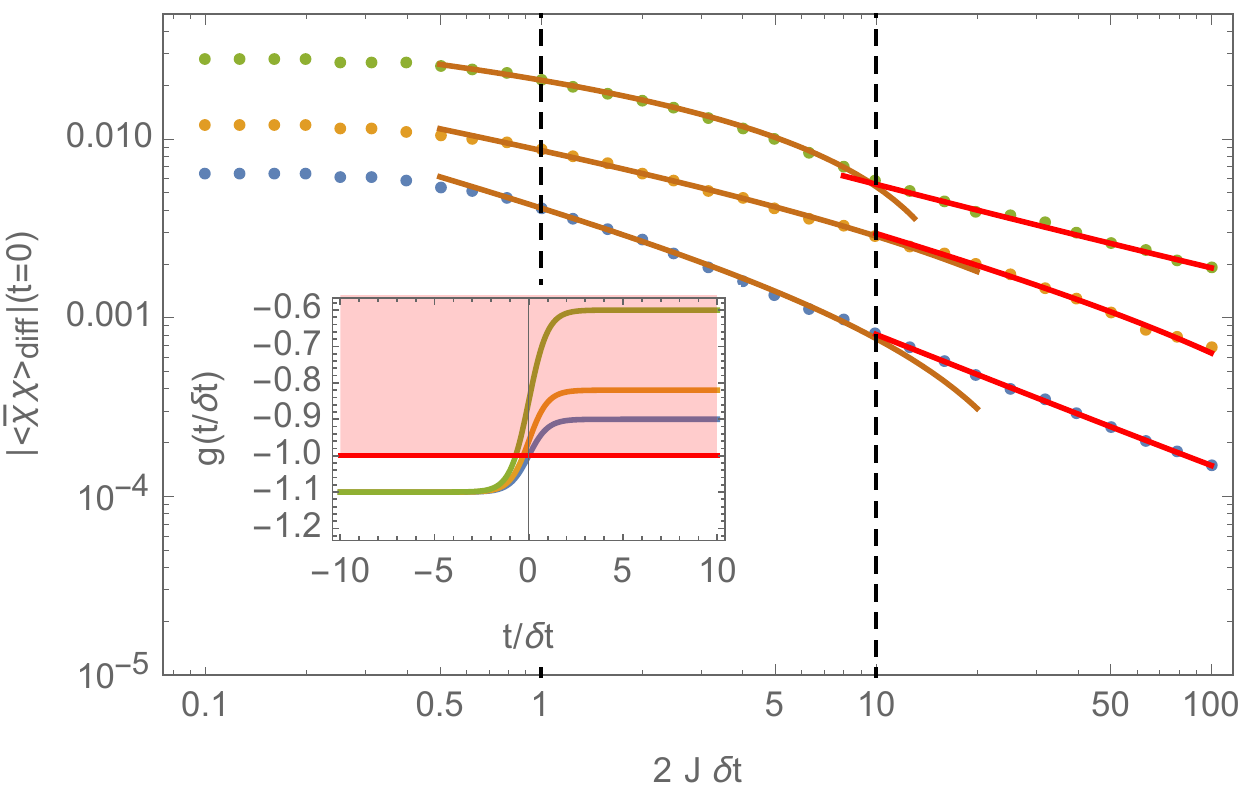}
\caption{ Plot of $\op_\mt{diff}$ as a function of $2 J\dt$ for the
gapped-to-gapless quenches shown in the inset. The brown and red
curves are the best fits in the fast and slow quench regimes,
respectively.  As an example, we show here the transition between 3
different scalings. In the inset, the three quenches protocols are
shown. All of them start at a distance of 0.1 from the critical
region. The blue protocol goes to the edge at $t=0$,
$g(t/\dt)=-1+0.1 \tanh(t/\dt)$; the yellow protocol, just enters the
critical region at $t=0$, $g(t/\dt) = - 0.96 + 0.14 \tanh(t/\dt)$;
and the green protocol is well inside the critical area at $t=0$
(compared to the initial amplitude), $g(t/\dt) = - 0.85 + 0.25
\tanh(t/\dt)$. As in the previous subsection, the scalings of the
blue dots correspond to the gapped-to-edge quench: in the fast
regime it goes as $P + Q (2 J \dt)^{-1/2}$ (brown curve) with
$P=-0.00482$ and $Q=0.000750$; in the slow regime, it behaves as $P
+ Q (2 J \dt)^{-3/4}$ (red curve) with $P=-0.00448$ and
$Q=-6.97\times 10^{-6}$. Now, when you measure inside the critical
area, the slow scaling changes instantaneously to $-1/2$, while the
fast quench continually changes from $-1/2$ to a logarithmic
scaling. Thus, the two regimes in the yellow quench are given by $P
+ Q (2 J \dt)^{-\a}$ (brown, fast), with
$P=-0.00963,Q=0.000884,\a=0.406$ and $P + Q (2 J \dt)^{-1/2}$ (red,
slow), with $P=-0.0107$,$Q=0.000439$. Finally, the green curve shows
the prototypical example of a gapped-to-interior quench with best
fits $P+ Q \log(2 J \dt)$ (brown, fast), with $P=0.00690$,
$Q=-0.0213$ and $P + Q (2 J \dt)^{-1/2}$ (red, slow), with
$P=-0.0170$ and $Q=-0.000200$. }\label{fig-kitaev-inside-3}
\end{figure}

Let us start by considering the case in which $\delta g_\mt{in} >
\delta g_\mt{fin}$. In the previous section, we already analyzed the
case where $\delta g_\mt{fin}=0$, which corresponds to quenching to
the edge point. In that case with small amplitudes, we found that
the expectation values scale as $\dt^{-1/2}$ in the fast regime, and
$\dt^{-3/4}$ in the slow one. Now we hold $\delta g_\mt{in}$ fixed
and slowly increase $\delta g_\mt{fin}$ away from zero. We observe
different scaling behaviours depending on whether the quench is slow
or fast compared to $\delta g_{in}$, as shown in figure
\ref{fig-kitaev-inside-3} where the initial amplitude is fixed to
$\delta g_{in} = 0.1$. Note that the scaling exponent in the slow
quench regime, \ie $2 J \dt > 1/\delta g_{in}$, immediately changes
from $-3/4$ to $-1/2$ as soon as $\delta g_\mt{fin}>0$. The latter
exponent corresponds to the KZ scaling of a (1+1)-dimensional
fermionic mass quench. The situation is different for the fast
quench regime. In this case, the scaling behaviour varies
continuously from $\op_\mt{diff} \sim (J\dt)^{-1/2}$ at $\delta
g_\mt{fin}=0$ to a logarithmic scaling when $\delta
g_\mt{fin}\sim\delta g_\mt{in}$.

The three examples shown in figure \ref{fig-kitaev-inside-3} were
chosen to show the evolution of the scaling behaviour described
above: The blue dots show the example where $\delta g_\mt{in}=0.1$
and $\delta g_\mt{fin}=0$, and hence these are quenches to the edge,
as in the previous section, with a scaling exponent ${-1/2}$ in the
fast regime and ${-3/4}$ in the slow regime. The yellow dots show an
example where $\delta g_\mt{in}=0.1$ and $\delta g_\mt{fin}=0.04$.
In this case, the slow quench regime already scales with an exponent
$-1/2$ while the fast quench regime has an intermediate scaling
(between $-1/2$ and logarithmic) with an exponent of roughly $-0.4$.
The green dots correspond to quenches with $\delta g_\mt{fin} = 0.15
\gtrsim \delta g_\mt{in}=0.1$.  Here, the fast quenches have already
settled to a logarithmic scaling while the slow quenches again
exhibit the scaling exponent $-1/2$.

\begin{figure}[h!]
        \centering
        \subfigure[Quenches Protocols]{
                \includegraphics[scale=0.5]{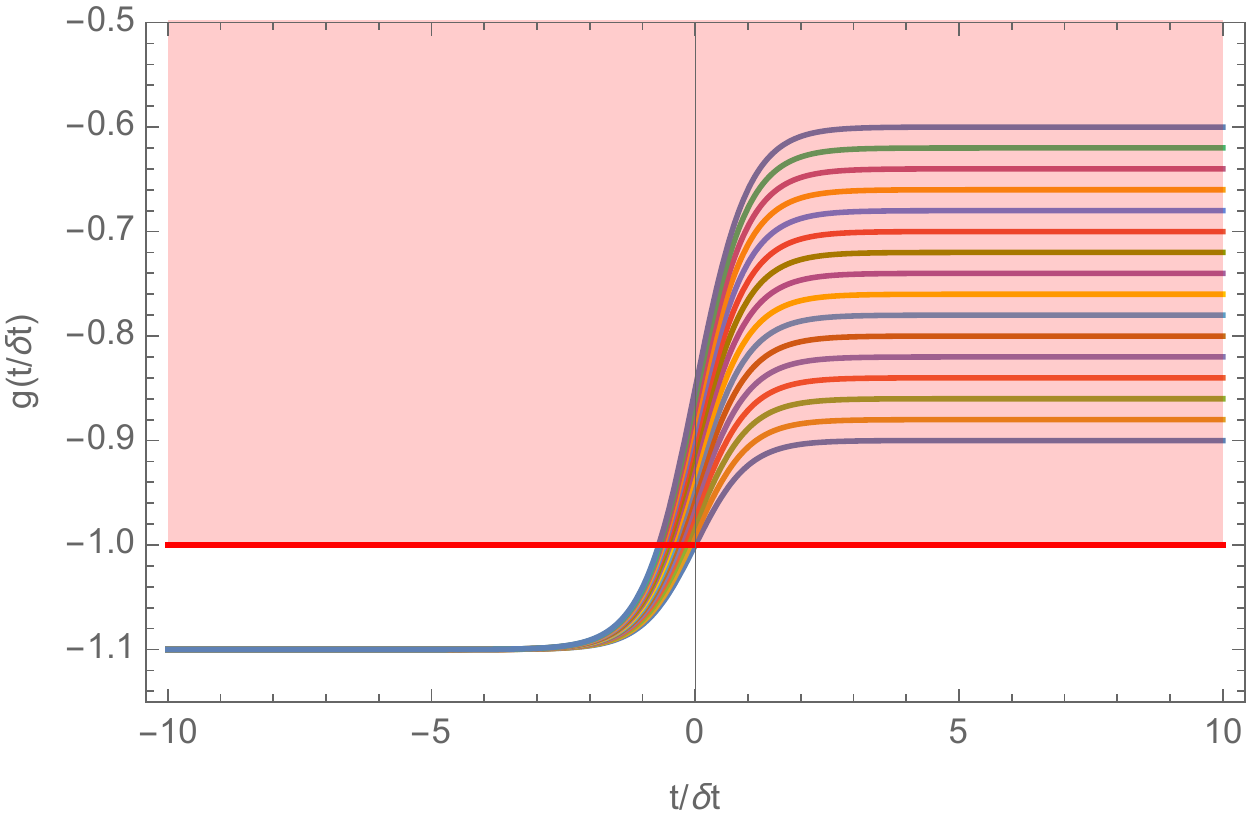} \label{t_quarter}}
         \subfigure[Resulting scaling in the fast regime]{
                \includegraphics[scale=0.5]{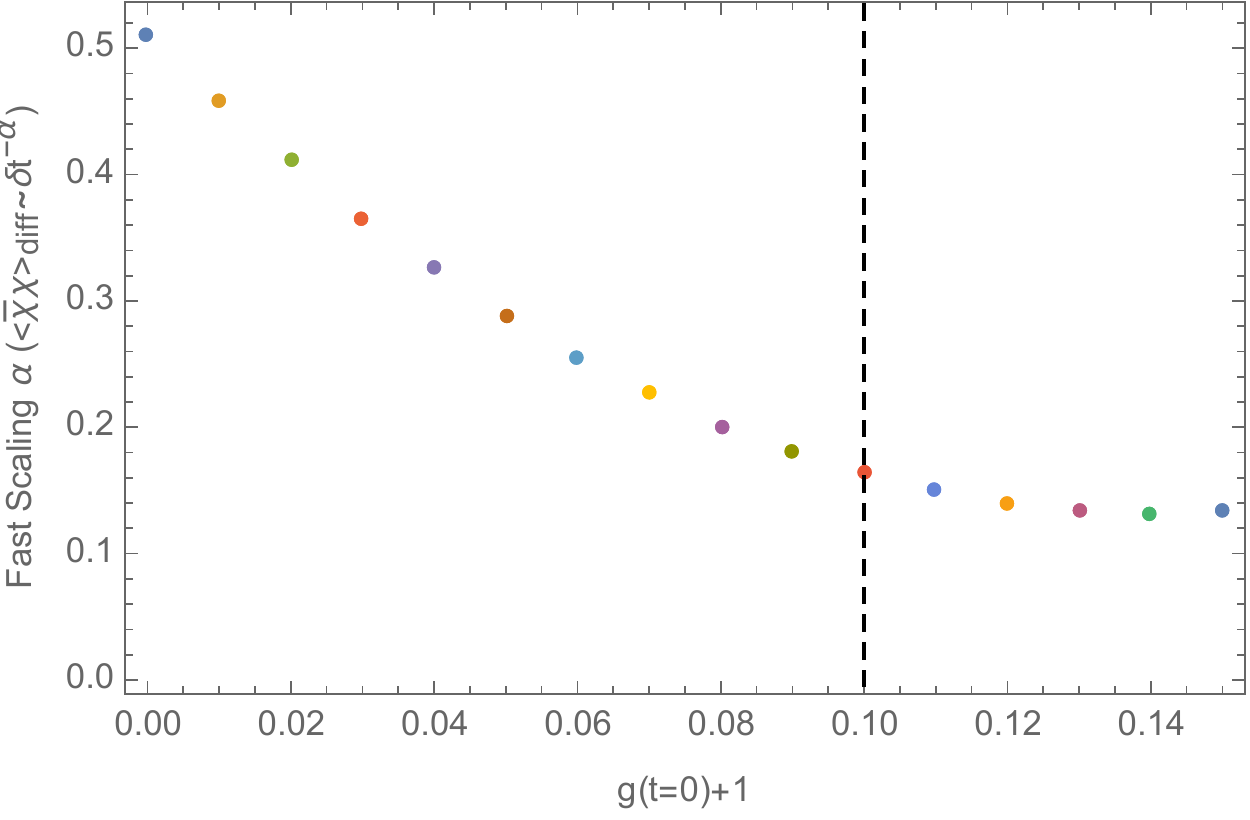} \label{t_half}}
                 \caption{Resulting scaling in the fast regime as a function of how deep inside the critical region it is measured. For the quench to the edge, it is expected a fast scaling of $\dt^{-1/2}$ (see last subsection). That scaling slowly decays until it stabilizes when the distance inside the critical region is of the order of the initial amplitude (in this case, 0.1). At that stage (shown in dashed black line), the scaling of the expectation value becomes logarithmic, that in figure (b) is presented as a ``small" power-law behaviour.}
\label{fig_scaling2}
\end{figure}

In figure \ref{fig_scaling2}, we explore the transition between the
edge scaling to logarithmic scaling in the fast quench regime. We
analyze different quench protocols, all with fixed $\delta g_\mt{in}
= 0.1$ and with $\delta g_\mt{fin}$ varying from $0$ to $0.15$ ---
see figure \ref{fig_scaling2}a. In figure \ref{fig_scaling2}b, we
show the corresponding scaling exponent in the fast quench regime
fit with $\op_\mt{diff} \sim (J\dt)^{-\alpha}$. We see that the
exponent $\alpha$ begins at 1/2 and smoothly decreases until some
value just below $0.1$ when $\delta g_\mt{fin} \sim 0.1$ At this
point, $\alpha$ saturates for large $\delta g_\mt{fin}$. In fact,
the scaling of the expectation value has become logarithmic, but it
turns out that this is indistinguishable from power-law scaling with
the small exponents shown in the figure.

In summary, the (small amplitude) quenches from the gapped phase to
the interior of the critical region exhibit two scaling regimes: The
slow quench regime with $2 J \dt > 1/\delta g_\mt{in}$ where
$\op_\mt{diff}$ always scales as $(J\dt)^{-1/2}$ for any $\delta
g_\mt{fin}>0$, which is distinct from $(J\dt)^{-3/4}$ scaling
observed with $\delta g_\mt{fin}=0$. We might add that this behaviour
would be the KZ scaling for a (1+1)-dimensional fermionic theory. In
the fast quench regime with $2 J \dt < 1/\delta g_\mt{in}$, the
scaling behaviour makes a smooth transition from the quench-to-edge
scaling of $(J\dt)^{-1/2}$ to a logarithmic scaling, that would be
expected for a (1+1)-dimensional fermionic mass quench. This
transition occurs over the range $0<\delta g_\mt{fin}\lesssim \delta
g_\mt{in}$.

\begin{figure}[h!]
\setlength{\abovecaptionskip}{0 pt}
\centering
\includegraphics[scale=0.9]{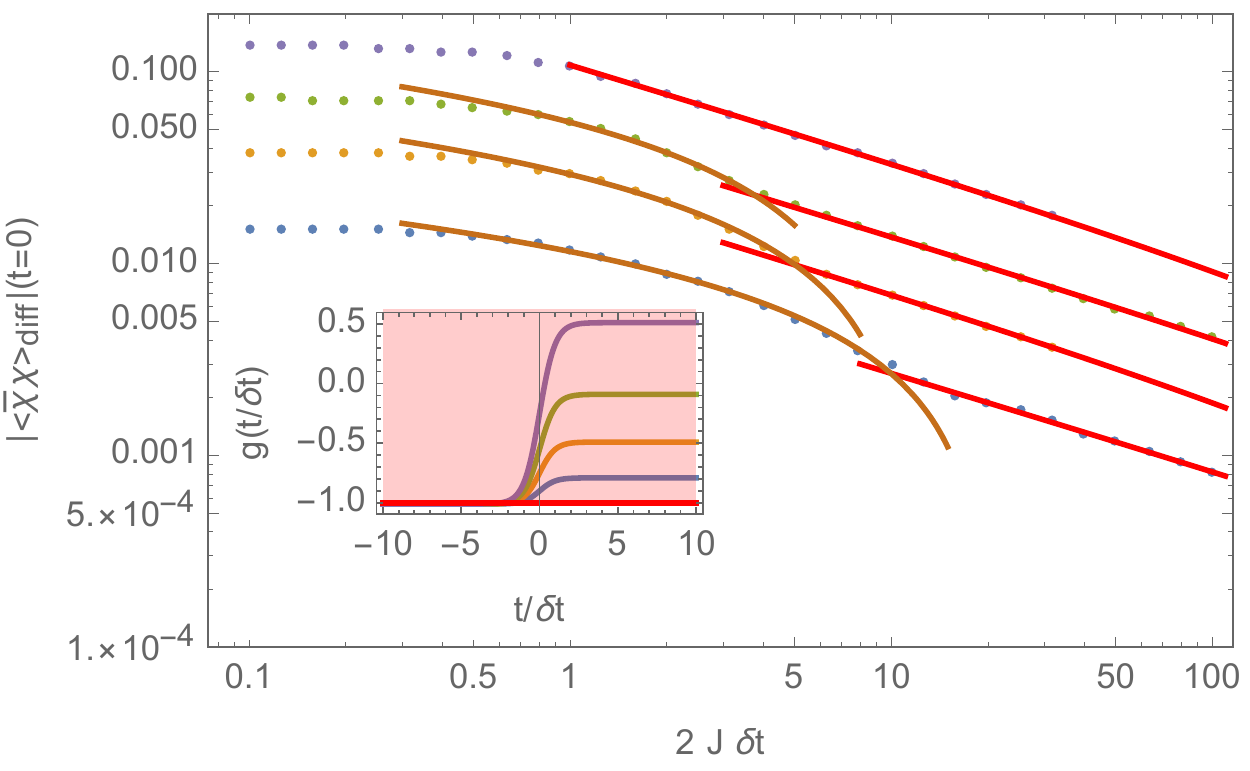}
\caption{Plot of $\op_\mt{diff}$ as a function of $2 J\dt$ for the
large-amplitude gapped-to-gapless quenches shown in the inset. The
initial distance to the critical region is 0.01. The brown and red
curves are the best fits in the fast and slow quench regimes,
respectively. In all cases they correspond to fits by $P_1 + Q_1
\log(2 J \dt)$ for the fast and $P_2 + Q_2 (2 J \dt)^{-1/2}$ for the
slow regime. $P_1= 0.00385$ (blue), $0.0119$ (yellow), $0.0240$
(green); $Q_1=-0.0115$ (blue),$-0.0291 $ (yellow), $-0.0544$
(green); $P_2=-0.00857$ (blue),$-0.0229 $ (yellow), $-0.0448 $
(green), $-0.110$ (violet); $Q_2=0.0000391$(blue), $0.000421$
(yellow), $0.000443$ (green), $0.00194$ (violet). The crossover
between fast and slow regimes happens at a $ J \dt$ that is
inversely proportional to the amplitude of the quench, making the
fast regime to effectively disappear when the amplitude is large
enough (like for the violet dots).}\label{fig-kitaev-inside-4}
\end{figure}

Now it is natural to ask whether the same scaling holds for large
amplitude quenches inside the critical region, \ie for quenches with
$\delta g_\mt{fin}\gtrsim \delta g_\mt{in}$. These large amplitude
quenches are explored in figure \ref{fig-kitaev-inside-4}. As the
whole region on interior critical points is traversed with $\delta
g_\mt{fin}=1$ to analyze larger amplitudes, we need to start with
smaller $\delta g_\mt{in}$. For the quenches shown in figure
\ref{fig-kitaev-inside-4}, we fix $\delta g_\mt{in} = 0.01$ and vary
the final amplitude $\delta g_\mt{fin} = 0.1, 0.25, 0.45, 0.75$, all
of which satisfy $\delta g_\mt{fin} \gg \delta g_\mt{in}$. In this
large amplitude regime, the corresponding fast and slow quench
scalings remain the same as above, \ie $-1/2$ and logarithmic,
respectively. However, now the transition between the two scaling
regimes is no longer at $2 J \dt \sim 1/\delta g_\mt{in} = 100$,
rather we find the transition at $2 J \dt \sim \delta g_\mt{fin}$.
In fact, we see that as $\delta g_\mt{fin}$ increases and becomes of
order one, the fast quench scaling regime shrinks more and more
until is no longer observable, as shown with the violet dots in
figure \ref{fig-kitaev-inside-4} which predominantly scale with the
slow quench scaling.

Another alternative approach to large amplitude quenches is to
consider quenches where we fix $\delta g_\mt{fin}$ but we increase
the gapped amplitude $\delta g_\mt{in}$. We followed this approach
with $\delta g_\mt{fin}=0.1$ fixed and varying $\delta
g_\mt{in}=0.1,1,10,100$. When both $\delta g$'s are comparable, we
found the same result as before; namely, a logarithmic fast scaling
regime and a power-law slow regime with exponent $-1/2$, separated
at a scale $\delta g_\mt{in}^{-1} \sim \delta g_\mt{fin}^{-1}$. As
we increased $\delta g_\mt{in}$, the logarithmic scaling remains but
the exponent of the power-law starts decreasing, until around
$\delta g_\mt{in}=10$, where we can only observe a logarithmic
behaviour. This is a rather surprising behaviour, given that in every
other case the fast scaling was the regime which disappeared for
large amplitudes. One possibility is that there is indeed still a
slow quench regime (which sets in at some scale given by a
combination of $\delta g_\mt{in}$ and $\delta g_\mt{fin}$) that will
only appear for large enough $J \dt$.\footnote{Unfortunately, our
numerical analysis does not allow to consider large enough $J \dt$'s
to check whether this hypothesis holds or not.}

We do not have a good understanding of the various scaling behaviours
described above. The scaling exponent of $-1/2$ in the slow quench
regime has been observed previously for quenches linear in time for
this model \cite{sengupta}. These linear quenches began at $t =
-\infty$ and the response is measured at $t = +\infty$. With these
simple protocols, the response can be examined analytically and the
$1/(J\dt)^{1/2}$ scaling stems from the fact that the excitation
probability predominantly depends on one of the directions in
momentum space. Recall that the critical models \reef{2-19} for the
interior points are not really anisotropic and so we expect that
this result must be related to the fact that the quenched operator
itself is anisotropic. That is, from the critical Hamiltonian in
Eq.~\reef{2-19a}, we see that the quenched operator corresponds
to\footnote{Here we are using Dirac matrix notation, \ie $\gamma_x$
is the Dirac matrix associated with the $x$ direction.}
$\bar\psi\gamma_x\psi$. Hence we should think of $m(t)$ as the
$x$-component of a vector coupling and accordingly, the roles of
$p_x$ and $p_y$ are clearly distinguished in the dispersion relation
\reef{wild} --- see further discussion in section \ref{sec-5}. Of course, for the protocols used in
this paper, our numerical results exhibit a behaviour similar to that
in the linear quenches, as long as the gapless interior region is
traversed, irrespective of the starting point.

For slow quenches, the change of the exponent from 3/4 to 1/2 as soon as $\delta g_{in} \neq 0$ may be understood as follows: Consider first the line $k_1=k_2 = \bk$, so that $G(k)=0$. In this case the equations for $\chi_1$ and $\chi_2$, as defined in equation (\ref{3-1-1}), decouple. The solutions of the Dirac equation can be readily written down
\bea
\chi_1 (\bk) & = & A_1\, {\rm exp} [ -i \alpha(\bk,t)] \nonumber \,, \\
\chi_2 (\bk) & = & A_2\, {\rm exp} [ i \alpha(\bk,t)] \,,
\label{422-1}
\eea
where $A_1,A_2$ are integration constants and
\ben
\alpha (\bk,t) = -\int^t m(\bk,t^\prime) dt^\prime = -4J[(a-\cos \bk)t+b\dt \log (\cosh t/\dt)] \,.
\label{422-2}
\een
To determine the ``in" solution, we need to examine the behavior at $t \rightarrow -\infty$,
\ben
\alpha(\bk,t) \rightarrow -4J(a-b-\cos\bk)t \,.
\een
Since we are considering quenches which start from the gapped phase we have $(a-b) < -1$, so that $(a-b-\cos\bk) < 0$. Therefore for the positive energy solution we must set $A_2=0$, while for the negative energy solution we must set $A_1=0$,
\begin{eqnarray}
U(\bk,t) = A_1 \left( \begin{array}{c}
 e^{-i\alpha(\bk,t)}\\
0
\end{array} \right) \nonumber \,, \\
V(\vk,t) = A_2\left( \begin{array}{c}
0 \\
e^{i\alpha(\bk,t)}
\end{array} \right)  \,.
\label{422-3}
\end{eqnarray}
Substituting these in the mode expansion for the operator $\chi$ as in eq.~(\ref{3-8}) and imposing the anticommutation relations then determines the integration constants $A_1=A_2=1$ upto a phase.

Therefore for these $k_1=k_2=\bk$ modes, we have
\ben
_{in}\langle0| {\bar{\chi}} \chi |0\rangle_{in} = {\bar{V}}(\vk,t) V(\vk,t) = -1 \,,
\label{422-4}
\een
which is {\em independent of $\vk$ and time}.

Consider now the response which we measure, viz. the quantity 
$\langle{\bar{\chi}} \chi \rangle_\mt{diff}$ defined in eq.~(\ref{3-12}). The adiabatic modes at time $t=0$ for $k_1=k_2=\bk$ are
\bea
\chi_1^{adia} (\bk) & = & A_1\, {\rm exp} [ i\, m(\bk,0) t] \nonumber \,,\\
\chi_2^{adia} (\bk) & = & A_2\, {\rm exp} [ -i\, m(\bk,0) t] \,,
\label{422-5}
\eea
Therefore the negative frequency adiabatic modes have $\chi_1 =0$ when $m(\bk,0) <0$ while they have $\chi_2=0$ when $m(\bk,0) >0$. Therefore we have
\bea
_{in}\langle0| {\bar{\chi}} (\vk) \chi (\vk) |0\rangle_{in} & = & -1~~~~(m(\bk,0) < 0) \nonumber \,,\\
& = & +1 ~~~~(m(\bk,0) > 0)  \,.
\label{422-6}
\eea
which subsequently means that
\bea
\langle{\bar{\chi}}(\vk) \chi (\vk) \rangle_\mt{diff} & = &\ 0 \qquad\ (m(\bk,0) < 0) \nonumber \,,\\
& = & -2 \qquad(m(\bk,0) > 0) \,.
\label{422-7}
\eea

Let us now consider quench protocols ``gapped to edge". These have $a=-1$ and $b > 0$. This means $m (\vk,0) = -4J(\cos \vk -a) = -8J \cos^2 (\vk/2) \leq 0$ for these quenches, so that $\langle{\bar{\chi}}(\vk) \chi (\vk) \rangle_\mt{diff} =0$. On the other hand for quench protocols ``gapped to interior", we need $a =-1+2\eta$ where $\eta >0$, so that $m(\vk,0) = -8J (\cos^2 \bk/2 - \eta)$. This can be positive for $\vk$ such that $\cos^2 \bk/2 < \eta$.

We therefore conclude that along the $k_1=k_2=\bk$ the contribution to the response, $\langle{\bar{\chi}}(\vk) \chi (\vk) \rangle_\mt{diff}$ is always independent of $\vk$. This vanishes for all $\vk$ for gapped to edge quenches, whereas for gapped to interior quenches there is a range of $\vk$ for which this has the maximal value $-2$.

Consider now the response for generic $k_1$ and $k_2$. The components $\chi_1, \chi_2$ satisfy the equation
\ben
\left[ \partial_t^2 \pm  i \partial_t m(\vk,t) + [ G(\vk)^2 + m(\vk,t)^2 ] \right] \chi_{1.2}= 0\,.
\label{422-8}
\een
For slow quenches the response is substantial for times when the quench profile can be approximated by a profile which is linear in time. For these times, eq. (\ref{422-8}) becomes
\ben
\left[ \partial_t^2 \pm i 4Jb/\dt + [ G(\vk)^2 + m(\vk,t)^2 ] \right] \chi_{1,2} = 0 \,.
\label{422-9}
\een
Rescaling $t \rightarrow t^\prime = t /\sqrt{\dt}$ it is clear that solutions have a functional form
\ben
\chi_{1,2}= F (G(\vk)\sqrt{\dt}, m(\vk,t)\sqrt{\dt}, t /\sqrt{\dt}) \,.
\een
This, in turn, implies that the quantity $\langle{\bar{\chi}}(\vk) \chi (\vk) \rangle_\mt{diff}$ also has this functional form. 

Since we have shown that this quantity is independent of momenta when $G(k)=0$, the simplest form of this function at time $t=0$ is
\ben
\langle{\bar{\chi}}(k) \chi (k) \rangle_\mt{diff} = c_1\, F_1(G(\vk)\sqrt{\dt}) + c_2 \,(G(k)\sqrt{\dt})^\alpha \,F_2(G(\vk)\sqrt{\dt}, m(\vk,t)\sqrt{\dt}) + \cdots \,,
\label{422-11}
\een
where $F_{1,2}$ are some functions and $\alpha$ is some positive real exponent. The ellipsis indicates higher orders in $G(\vk)\sqrt{\dt}$.

For gapped-to-edge quenches, we have also shown that $\langle{\bar{\chi}}(\vk) \chi (\vk) \rangle_\mt{diff}$ vanishes for all $\vk$. Therefore for such quenches, the first term must be absent, or we should set $c_1=0$. On the other hand, for gapped-to-interior quenches, $\langle{\bar{\chi}}(\vk) \chi (\vk) \rangle_\mt{diff}$ is nonvanishing for some range of momenta. Therefore for such quenches the first term is generically nonvanishing.

Since we are considering slow quenches, most of the contribution will come from the region in momentum space where $G(k)$ is small. In this region we can approximate
\ben
G(k) \sim 2 k_- \cos (k_+) \,,
\een
where $k_\pm = \frac{1}{2}(k_1 \pm k_2)$. Thus, for gapped-to-interior quenches, we have
\ben
\langle{\bar{\chi}} \chi \rangle_\mt{diff} = \int dk_+ dk_- F_1(\cos k_+ (k_- \sqrt{\dt})) \sim (\dt)^{-1/2} \,.
\een
On the other hand, for gapped-to-edge quenches, most of the contribution should come from the single gapless point which has
$k_+=\pi, k_-=0$. Expanding around this point, $k_+=\pi + \delta k_+$  and $k_-=\delta k_-$ where both $\delta k_\pm$ are small we have
\ben
G(k) \sim -2\delta k_   -~~~~~ m(k,0) \sim (\delta k_+)^2 + (\delta k_-)^2 \,.
\een
Therefore, we have
\ben
\langle{\bar{\chi}} \chi \rangle_\mt{diff} = \int d\delta k_+ d \delta k_- (\delta k_- \sqrt{\dt})^\alpha F_2 (\delta k_-\sqrt{\dt}, (\delta k_+^2 + \delta k_-^2)\sqrt{\dt}) \,.
\een
To extract the leading $\dt$ dependence we rescale $\delta k_- \rightarrow \delta k_-\sqrt{\dt}$ and $\delta k_+ \rightarrow \delta k_+ (\dt)^{1/4}$ to get
\ben
\langle{\bar{\chi}} \chi \rangle_\mt{diff} \sim (\dt)^{-3/4} \,.
\een

\subsubsection{Interior-to-interior} \label{intint}

In this section, we consider quenches where both $g(-\infty)$ and
$g(0)$ are at interior critical points --- and in fact, where the
entire protocol from $t=-\infty$ to 0 passes only through interior
points. That is, we consider profiles \reef{31-1} with $-1<a<0$ and
$-1<a-b<0$ (as well as $b>0$).

\begin{figure}[h!]
\setlength{\abovecaptionskip}{0 pt}
\centering
\includegraphics[scale=0.9]{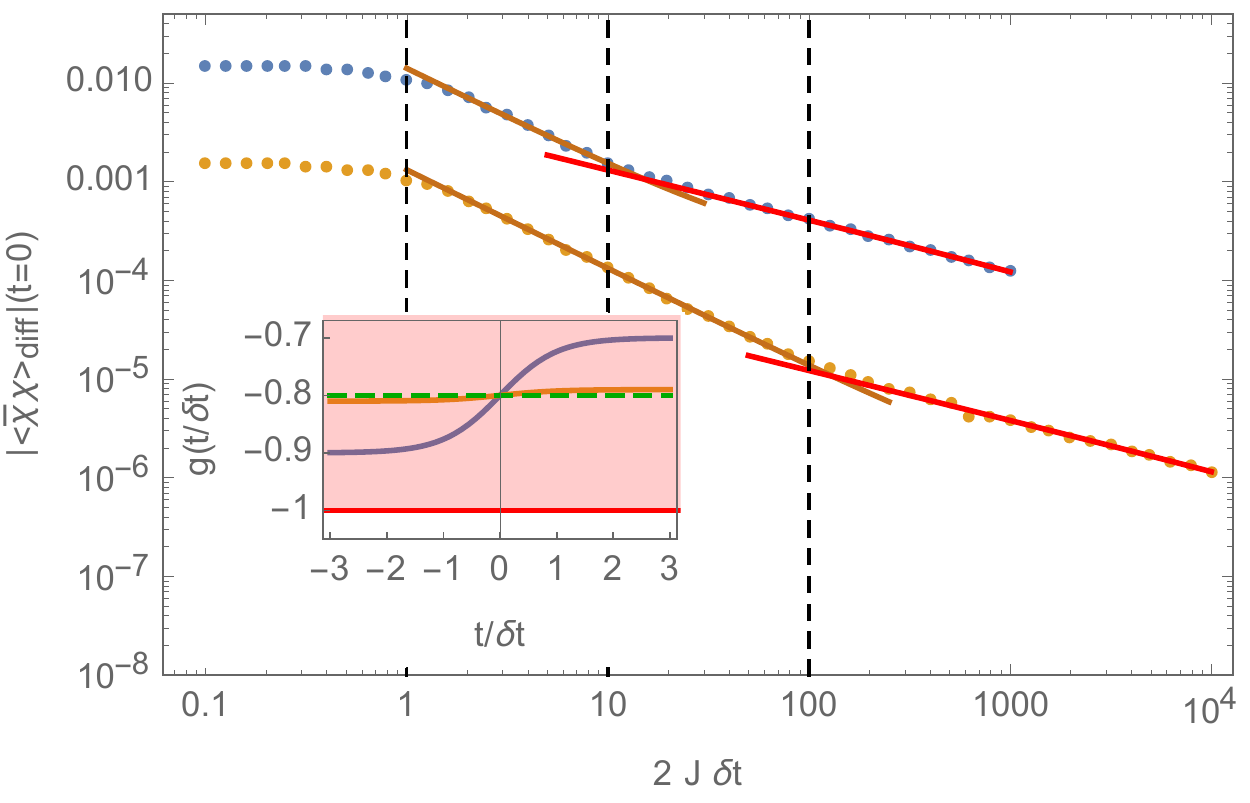}
\caption{Plot of $\op_\mt{diff}$ as a function of $2 J\dt$ for the interior-to-interior quench. The quench protocols are shown in the inset with $g(t/\dt) = -0.8 + b \tanh t/\dt$, where $b=0.1$ (blue), $0.01$ (yellow). The green dashed line shows the value at which we are computing, $g(t=0)=-0.8$. The shaded red area is critical. The red fits are for functions $P (2 J\dt)^{-1/2}+Q$ with $P=-0.00416, 1.22 \times 10^{-4} ,Q=1.01 \times 10^{-5}, 7.60 \times 10^{-8}$, respectively. The brown curves show the best fit for functions $P (2 J \dt)^{-1}+ Q$ with $P=0.0139, -0.00131,Q=1.42 \times 10^{-4}, 6.04\times 10^{-7}$.}\label{fig-kitaev-gapless}
\end{figure}

Typical results are shown in figure \ref{fig-kitaev-gapless} where
we consider two different protocols one with $b= 0.1$ and the other
with $b=0.01$. One somewhat surprising feature is that we, in fact,
observe two different scaling regimes, separated at a scale of $J
\dt$ of the order of the inverse of the amplitude, \ie $J\dt \sim
1/b$, even though the quench only travels across interior critical
points at all times. The two scaling regimes are characterized by
distinct scaling exponents. For $1<J\dt<1/b$, we found that
expectation values scale as $1/(J\dt)$, which is consistent with the
fast scaling of a (2+1)-dimensional fermionic mass quench. On the
other hand, in the slow regime where $J\dt>1/b$, we find
$\vev{\bar{\chi}\chi}_\mt{diff} \sim (J\dt)^{-1/2}$, which matches
the slow quench scaling found for gapped-to-interior quenches in the
previous section.

Of course, since the amplitudes considered in Figure
\ref{fig-kitaev-gapless} are still small,\footnote{In the case of
interior-to-interior quenches, the amplitude cannot be too large as
the critical region is parametrized by $|g|<1$. So at most, the
amplitude can be of order one for these quenches.} it would be very
interesting to develop an understanding of these scalings in terms
of a continuum theory described in section \ref{honey}. In
particular, there is a striking fact to understand, namely, why is
there a slow quench regime at all? Since the quench only passes
through interior points where the model is gapless, there is no
intrinsic energy scale with which to compare the quench rate, \ie
the quenches are never in an adiabatic regime, no matter how large
$J\dt$ becomes. We checked this last fact by computing expectation
values at a finite time but never observing an adiabatic
evolution.\footnote{As shown in \cite{dgm1,dgm2,dgm3,dgm4}, the
adiabatic evolution in these cases can be computed in a series
expansion in $\dt$ that will be characterized by terms proportional
to inverse $\textit{even}$ powers of $\dt$. } However, the numerical
computations still show that there is a slow scaling regime for
these quenches, which therefore cannot be adiabatic and somehow the
scaling exponent matches the KZ scaling for a mass quench in
$(1+1)$-dimensional fermionic theory! We also reiterate that this
behaviour also matches the slow quench scaling found for
gapped-to-interior quenches in the previous section. There it was
suggested that this unusual scaling was associated with the
anisotropy of the quenched operator.

One observation, which provides a step towards a possible
explanation, is the following: In \cite{dgm1,dgm2}, it was shown
that the fast quench scaling behaviour essentially follows from
linear response theory. In our case, the dimensionless parameter
which controls the renormalized perturbation theory is $Jb\,\dt$.
Thus the linear response would be valid when $Jb\,\dt$ is small, and
indeed in this regime, we get the expected result from the continuum
theory. That is, we find the fast quench scaling, \ie $1/(J\dt)$,
for a mass quench in a (2+1)-dimensional fermionic theory.  On the
other hand, when $Jb\,\dt > 1$, the linear response calculation is
no longer valid, and the arguments which lead to the fast quench
scaling do not hold any more. Indeed the change of the scaling
exponent changes precisely around $Jb\,\dt \sim 1$. While this does
not explain the new scaling $(J\dt)^{-1/2}$ found beyond this point,
it does indicate that we should expect that the quenches are
entering a new regime here.

\subsection{Instantaneous quench limit}\label{sec-sudden}

Irrespective of the particular model under consideration, when the
quench rate is faster than the lattice scale, \ie $J\dt\lesssim1$,
we expect that our results should agree with that of an
instantaneous quench in which the coupling is switched abruptly from
$g(t = -\infty)$ to $g(t=0)$ at the time of measurement. In
particular, the system has no time to respond to the change in the
coupling and so to a good approximation, we have \beq {\rm
instantaneous\ quench:}\ \ \
\op_\mt{diff}=\op_\mt{adia}\big|_{t=-\infty} -
\op_\mt{adia}\big|_{t=0}\,. \label{instantx} \eeq Hence given the
expression for $\op_{adia}(t)$ in Eq.~(\ref{oadia}), it is
straightforward to calculate $\op_\mt{diff}$ for this regime.

\begin{figure}[h]
\setlength{\abovecaptionskip}{0 pt}
\centering
\includegraphics[scale=0.4]{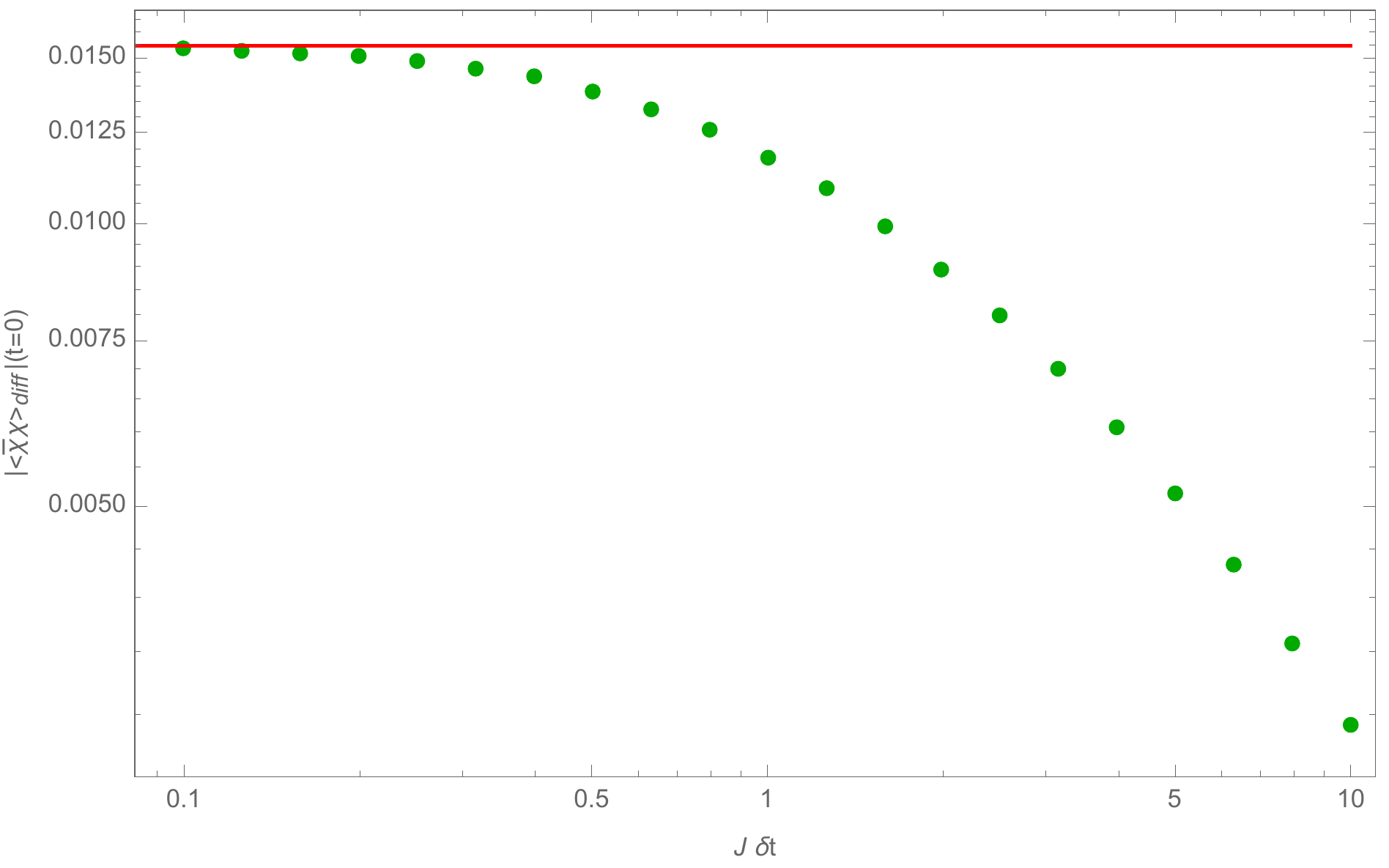}
\caption{ Plot of $\op_\mt{diff}$ at $t=0$ as a function of $J\dt$
for the gapped-interior quench shown in the inset. At the beginning
of the quench $g(t=-\infty) = 1.01$ and we measure at $g(t=0) =
0.9$. The red line is the result \reef{instantx} for instantaneous
quenches evaluated using Eq.~(\ref{4-8}), which in this case yields
$0.0155$. }\label{fig-kitaev-sudden}
\end{figure}

For the Kitaev model, Eq.~(\ref{oadia}) yields \beq
\op_\mt{adia}(t) =  - \int_0^\pi \int_0^\pi \frac{dk_1
dk_2}{(2\pi)^2}\, \frac{ \cos k_1 + \cos k_2 -2g(t)}{\sqrt{ (  \cos
k_1 + \cos k_2-2g(t) )^2 + (\sin k_1 - \sin k_2)^2}}\,.\label{4-8}
\eeq Now as an example, figure \ref{fig-kitaev-sudden} shows the
result for $\op_\mt{diff}$ at $t=0$ for a quench protocol which
starts in the gapped phase with $g(t=-\infty)=-1.01$, and ends in
the interior of the critical region with $g(t=0)=-0.9$. The figure
also shows the instantaneous quench result \reef{instantx} evaluated
with Eq.~\reef{4-8}. The figure clearly shows that in the limit
$J\dt \rightarrow 0$, the exact numerical results smoothly approach
the instant quench value \reef{instantx}. Similarly, we have
verified that the saturation value of $\op_\mt{diff}$ agrees with
Eq.~\reef{instantx} in the Ising model.

\subsubsection*{Beyond $t=0$}
\begin{figure}[h]
\setlength{\abovecaptionskip}{0 pt}
\centering
\includegraphics[scale=0.6]{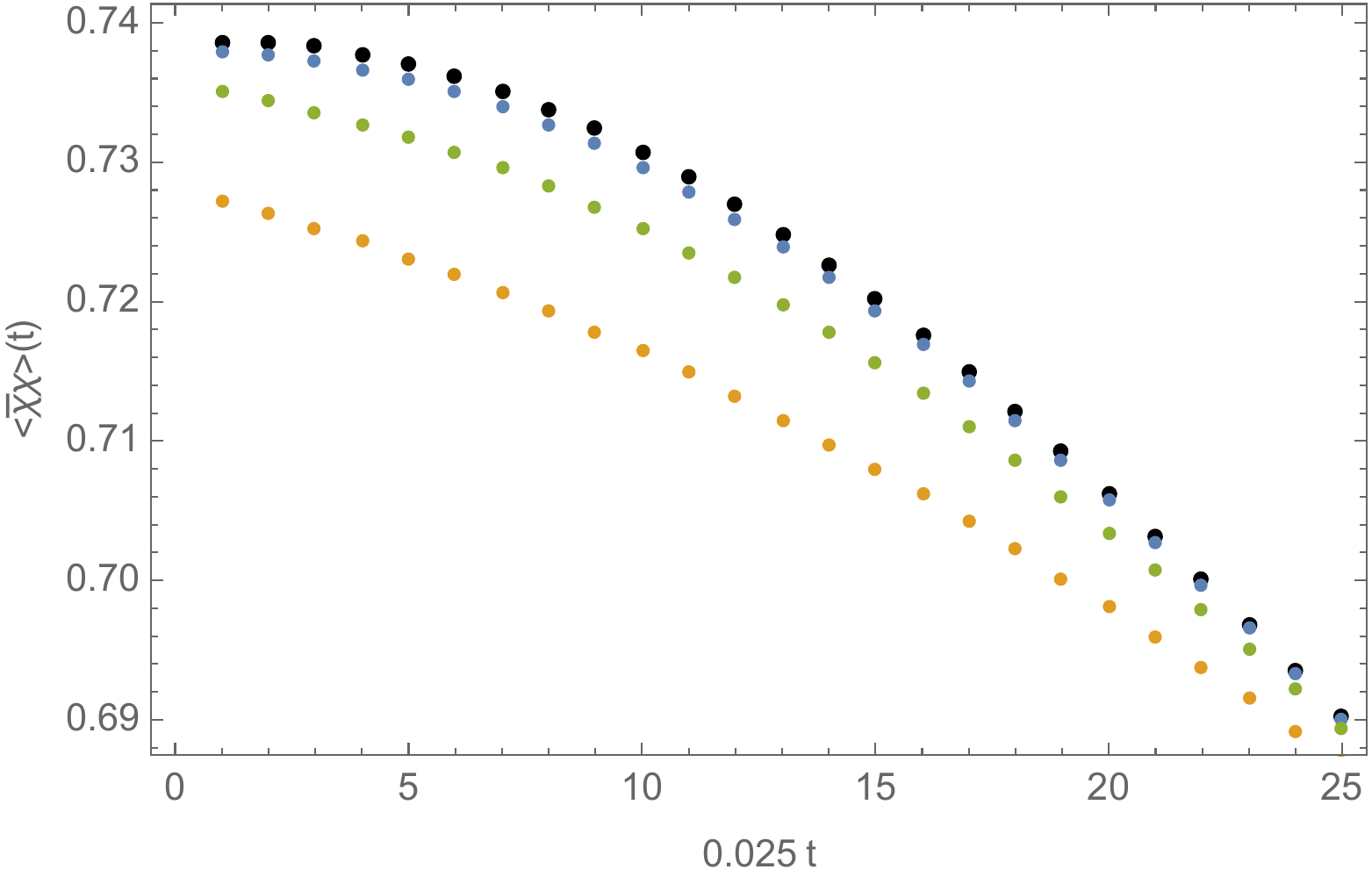}
\caption{Comparison of the instantaneous quench  answer (black dots)
in Eq.~(\ref{4-1}) with the time-dependent solution for $J\dt=0.5$
(orange dots), $J\dt= 0.25$ (green dots) and $J\dt=0.1$ (blue
dots).} \label{comparison}
\end{figure}

The exact result for an instantaneous quench in the Ising model from a coupling $g=g_0$ to $g=g_1$
was calculated in \cite{barouch},
\bea \op_\mt{inst}(t) &=&
\int_0^\pi \frac{dk}{\pi\,\epsilon_0(k)\, \epsilon_1^2(k)}\, \Big[
(g_0-g_1) \sin^2 k ~\cos(2\epsilon_1 t)  \nonumber\\
&&\qquad-(\cos k - g_1) \big( (\cos k - g_0)(\cos k - g_1) +\sin^2 k
\big) \Big] \,, \label{4-1} \eea where \ben \epsilon_i = \Big(\sin^2
k + (g_i - \cos k)^2\Big)^{1/2}\,. \een Hence in the Ising model, we
can compare our full solution $\op(t)$ for quench rates faster than
the lattice scale with this instantaneous quench answer, for all
times $t > 0$.  In Figure (\ref{comparison}), we compare the time
dependence of $_{in}\langle0| {\bar{\chi}} \chi |0\rangle_{in}$ at
small values of $\dt$ with the exact instant quench result
Eq.~(\ref{4-1}) and we see that the agreement gets better as $J\dt$
decreases.

\section{Discussion}
\label{sec-5}

In this paper, we have studied critical quench dynamics in the
transverse field Ising model \reef{ham2} on one-dimensional chain
and in the Kitaev honeycomb model \reef{2-12} in two dimensions. We
studied an exactly solvable quench protocol which asymptotes to
finite values of the coupling at early and late times, and focused
on the response of the operator by which we carried out the quench.
The exact solutions in terms of free fermions  were used to study
the scaling of the response with the quench rate. Our answers have
been obtained in the thermodynamic limit with an infinite number of
sites and thus are free from finite size effects. Our results are
summarized in Table \ref{table33}.
\begin{table}[htbp]
\centering
\begin{tabular}{|c|c|c|c|c|}
\hline
\multicolumn{2}{|c|}{Theory} & Slow & Fast & Transition  \\ \hline \hline
\multirow{2}{*}{\begin{tabular}{c} Continuum \\ Free Fermion  \end{tabular}} & $d=1+1$ & 1/2 & $\log$ & $m_0^{-1}$  \\
& $d=2+1$ & 1 & 1 & $m_0^{-1}$ \\
 \hline
 \hline
 \multirow{2}{*}{\begin{tabular}{c} Transverse \\ Ising Model \end{tabular}} & small amplitude; $\e_{in} \ll 1$ & 1/2 & $\log$ & $\e_{in}^{-1}$ \\
& large amplitude; $\e_{in} \gg 1$ & 1/2 & none & $\e_{in}$ \\
 \hline
 \hline
 \multirow{8}{*}{\begin{tabular}{c} Kitaev \\ Honeycomb \\Model \end{tabular}} &  gapped-to-edge &&&\\
 & small amplitude; $b \ll 1$  & 3/4 & 1/2 & $|b|^{-1}$ \\
 & large amplitude; $b \gg 1$ & 1/2 & none & $|b|\ $ \\ \cline{2-5}
 & gapped-to-interior &&&\\
& $\delta g_\mt{in}\gg \delta g_\mt{fin}$ & 1/2 & $\frac{1}{2} > \a > 0$ & $\delta g_\mt{in}^{-1}$ \\
& $\delta g_\mt{in}\sim \delta g_\mt{fin}$ & 1/2 & $\log$ & $\delta g_\mt{in}^{-1}$ \\
& $\delta g_\mt{in}\ll \delta g_\mt{fin}$ & 1/2 & $\log$ & $\delta g_\mt{fin}^{-1}$ \\ \cline{2-5}
& interior-to-interior & 1/2 & 1 & $|b|^{-1}$ \\
\hline
\end{tabular}
\caption{A summary of our results for the scaling behaviour in
critical quenches of the transverse field Ising model \reef{ham2}
and the Kitaev honeycomb model \reef{2-12}. We have also included
the analogous results for a free fermion in two and three
dimensions, for comparison. The columns `Slow' and `Fast' indicate
the scaling exponent in the response, \ie  $\langle {\bar{\chi}}
\chi \rangle_\mt{diff}\sim (J\dt)^{-\alpha}$, for the slow and fast
quench regimes --- `log' indicates a logarithmic scaling was found
and `none' indicates the fast scaling regime disappears. The column
`Transition' indicates the approximate value of $J\dt$ when the
scaling makes a transition between the two scaling regimes, or the
minimum value for the slow scaling when there is no fast scaling
regime. For the continuum free fermion, we are  indicating the value
of $\dt$ at this transition.} \label{table33}
\end{table}

The results for the Ising model, discussed in section \ref{resice},
are the ones which are best understood. The Ising model has an
isolated critical point. Our quench protocol takes us through this
point and we measure the response at the moment (chosen to be $t=0$)
when the quench hits the critical point. For small amplitude
quenches (with $\epsilon_{in}=b\ll1$), there are three different
regimes, as shown in figure \ref{fig_ising}: 1) for $J\dt\lesssim1$,
the response saturates and we are in the instantaneous quench
regime; 2) for $1\lesssim J\dt\lesssim 1/\epsilon_{in}$, the
response scales logarithmically with $J\dt$, as expected from the
continuum description of the fast quench regime; and 3) for $J \dt >
1/\epsilon_{in}$, the response scales as $1/(J\dt)^{1/2}$, as
expected for Kibble-Zurek scaling in continuum. Hence for quenches
which only make small excursions from the critical point, our
results agree with those expected for the continuum theory
\cite{dgm1,dgm2,dgm3,dgm4}. However, we can also consider large
amplitude quenches (with $\epsilon_{in}=b\gtrsim1$) and the scaling
behaviour changes in this regime. In particular, the fast quench
scaling regime disappears and rather there is a smooth crossover
between the instantaneous and slow quench regimes. Further, the slow
scaling behaviour matches the KZ scaling found above, but the
transition into this regime occurs roughly when $J \dt
\sim\epsilon_{in}$.

As described in section \ref{resice}, for the Ising model, we have a
good theoretical understanding of the response in all of the
different situations described above. Perhaps one of the most
interesting results here is that there is a fast scaling regime for
small amplitude quenches. That is, our analysis of the Ising model
confirms that even with a finite lattice spacing, certain quench
protocols produce the fast scaling behaviour originally discovered in
the study of continuum field theories \cite{numer,fastQ,dgm1,dgm2}
--- see further discussion below.

The Kitaev model is distinguished by having an extended region of
couplings for which the theory is gapless. In section
\reef{honey22}, we considered a variety of different quench
protocols with different starting points and measuring the response
at different points in the critical region (again, chosen to be time
$t=0$), and the results are summarized in Table \ref{table33}.

The simplest case to consider is the gapped-to-edge quench,
described in section \ref{gapedge}, where the quench of the Kitaev
model starts in the gapped phase and at $t=0$, the system is at the
edge of the gapless region. The situation here is very similar to
that of quenching across an isolated critical point, as in the Ising
model. Indeed for small amplitude quenches, we observe three
distinct scaling regimes: instantaneous, fast and slow regimes,
which can be understood in terms of the continuum model
\reef{2-23-4}. However, we must add that the latter is an
anisotropic theory and the scaling behaviour does not match the
scaling of a conventional fermionic field, which is also shown in
Table \ref{table33}. Further, for large amplitude quenches, the fast
scaling regime disappears. One difference in the Kitaev case is that
for the large amplitude quenches, the scaling exponent in the slow
regime is different from that in the small amplitude quenches.

We began to explore the extended critical region of the Kitaev model
with the gapped-to-interior and interior-to-interior quenches, which
are discussed in sections \ref{gapint} and \ref{intint}. In both
cases, the response is measured when the system is at an interior
critical point, while in the first family, the quench starts in the
gapped phase and in the second, the system is initially at an
interior point. For all of these quenches, we again observe three
distinct scaling regimes: slow, fast and instantaneous, with smooth
transitions between them. The scaling behaviour in the fast regime
depends on the details of the quench, as shown in Table
\ref{table33}. However, for all of these quenches which traverse a
finite part of the critical region, all exhibit the same scaling
exponent in the slow quench regime. Namely, $\langle {\bar{\chi}}
\chi \rangle_\mt{diff}\sim (J\dt)^{-1/2}$, which corresponds to the
Kibble Zurek scaling of a relativistic fermion in one lower
dimension. In fact, our preliminary results indicate this slow
scaling behaviour extends to any quenches which  traverse some
distance in the interior region, \eg quenches beginning at an
interior point and ending at an edge point or vice versa.

This slow quench behaviour has been observed earlier for quenches of
the Kitaev model with linear protocols which start at $t = -\infty$,
and the measurement is performed at $t= \infty$ \cite{sengupta}. In
this case, the equivalent Landau-Zener problem has a simple solution
and for slow quench rates, the excitation probability predominantly
depends on $k_y$ alone. We do not have a clear explanation for how
this behaviour emerges with our quench protocols at the moment.
However, examining the integrand $\op(\vec k)$ in an expansion
around the critical modes, we find that this quantity depends
primarily on $k_y=k_1-k_2$ and is almost independent of
$k_x=k_1+k_2$. Hence this momentum dependence reflects the same
behaviour found for the excitation probability for the linear
quenches \cite{sengupta}. Further, as
commented in section \ref{gapint}, while the critical models
\reef{2-19} for the interior points are not anisotropic, these
quenches are inherently anisotropic because the quenched operator
corresponds to $\bar\chi\gamma_x\chi$. Accordingly, we should think
of coupling which we are varying in the quenches as the
$x$-component of a vector. We expect that this anisotropy will play
a central role in the explanation of the unusual scaling found in
both the slow and fast quench regimes.


As commented above, one of the most interesting results here is that
there is a fast scaling regime in many of our lattice quenches. For
small amplitudes, our numerical results here match to the
expectations of a continuum analysis for the Ising model and for the
gapped-to-edge quenches in the Kitaev model. While we presently lack
a theoretical understanding, it also appears that a fast scaling
regime arises for quenches of the Kitaev model which traverse a
finite distance across the critical region, \eg for the
gapped-to-interior and interior-to-interior quenches.\footnote{Our
preliminary results that a fast scaling regime also appears in
interior-to-edge and edge-to-interior quenches of the Kitaev model.}
Therefore our results indicate that for systems with a finite
lattice spacing, there is quite generally a regime where the quench
rates lie between the inverse lattice spacing and the physical mass
scales, and where the fast scaling behaviour found previously only in
continuum field theories holds. This opens up the interesting
possibility that such scaling can be indeed observed in experiments.

In the cases where we can match the theoretical and numerical
analysis for the fast scaling regime, the dimensionless couplings
are small and the change in the couplings are small as well. It is
only in this case that the crossover from the fast to the slow
quench happens when $\dt$ is of the order of the inverse physical
mass scale. This is expected, since small dimensionless couplings
correspond to finite physical mass scales and the continuum fast
quench scalings are expected when the quench rate is fast compared
to the physical scales but slow compared to the UV cutoff scale.
Indeed, in the case of the Ising model, when $\Delta g(t) \sim
\cO(1/|m_{in}|)$ and in the case of the Kitaev quench from gapped to
the edge for $\Delta J_3(t) \sim \cO(1/|m_{in}|)\sim \cO(1/ J)$ we
still have three regimes, but now the crossover between the fast and
the slow regime is no longer at $\dt \sim 1/|m_{in}|$. On the other
hand the slow quench behaviour is insensitive to this, since this is
the regime where the physical mass scale is much higher than the
scale set by the quench rate.


There are a wide variety of different avenues to follow in extending our study of critical quench dynamics in lattice models. In particular, the three protocols introduced in section \ref{honey22} to study the Kitaev model do not form an exhaustive list of the possibly interesting quench protocols in this lattice model. As alluded to in the above discussion,  we have made some preliminary studies of interior-to-edge and edge-to-interior quenches. One feature that seems to extend to these protocols is the slow quench scaling: $\langle {\bar{\chi}} \chi \rangle_\mt{diff}\sim (J\dt)^{-1/2}$. It also appears that there is a fast scaling regime separating the instantaneous and slow quench regimes. As noted in section \ref{honey}, there is a distinct ``interior'' edge point with $g=0$. It would be interesting to probe this new critical theory with new quench protocols.\footnote{Unfortunately, we found interior-to-``interior" edge quenches to be problematic, \ie producing reliable numerics for these quenches seems challenging.}

From the quantum information perspective, it will be
important to understand quench dynamics on the open chain Cluster-Ising model
which has non-trivial symmetry protected edge-states --- see appendix \ref{app-cluster}. In that context, it will be interesting to study the response of the string order parameter \cite{cluster-ising}, in a similar manner in terms of free fermions.

As we commented above, the quenches in the Kitaev model which
traverse the critical region are inherently anisotropic. While the
critical theories corresponding to the interior critical points are
not anisotropic, we are quenching the system with an anisotropic
operator, \ie $\chi^\dagger\sigma_3\chi \sim \bar\chi\gamma_x\chi$
(in the continuum langauge). We also measure the response as the
expectation value of this same operator. Hence it would be
interesting to see if similar scaling laws hold in these quenches
for other operators like $\chi^\dagger\sigma_1\chi$,
$\chi^\dagger\sigma_2\chi$ or $\chi^\dagger\chi$. Undoubtedly, this
would give us new insights into the unusual scaling behaviour found
for, \eg the gapped-to-interior and interior-to-interior quenches.

Of course, we do not have a good theoretical understanding of many
results for the quenches in the Kitaev model, particularly, for
quenches that traverse a finite distance in the critical region.
Certainly, this situation should be improved. We might note that
this is required even for the small amplitude quenches with the
gapped-to-edge protocols, where we heuristically applied
Eq.~\reef{1-6} with an effective spacetime dimension to predict the
scaling exponent should be --1/2. While the fast quench scaling is
well understood in relativistic theories \cite{dgm1,dgm2}, it is
interesting to confirm that these ideas properly extend to
non-relativisitic theories, and to semi-Dirac point appearing at the
edge of critical region. Similarly, as discussed above, our quenches
involving moving across the interior critical region are not really
mass quenches, \ie $m$ is actually the $x$-component of a vector
coupling. Hence it would also be interesting to extend the
discussions in \cite{dgm1,dgm2} to smooth fast quenches involving
anisotropic operators.

\section*{Acknowledgements} We would like to thank Vladimir Gritsev and Ganpathy Murthy for valuable discussions.
The work of DD is partially supported by DOE contract DE-SC-0009919.
The work of SRD is partially supported by the National Science Foundation grant NSF-PHY-1521045. Research at Perimeter Institute is supported by the Government of Canada through the Department of Innovation, Science and Economic Development and by the Province of Ontario
through the Ministry of Research \& Innovation. RCM and DAG were also supported
by an NSERC Discovery grant. RCM is also supported by research funding
from the Canadian Institute for Advanced Research and from the Simons Foundation through ``It from Qubit" Collaboration. DAG is supported by Nederlandse Organisatie voor Wetenschappelijk Onderzoek (NWO) via a Vidi grant. The work of DAG is part of the Delta ITP consortium, a program of the NWO that is funded by the Dutch Ministry of Education, Culture and Science (OCW). SRD would like to thank Galileo Galilei Institute for Theoretical Physics, the Indian Association for Cultivation of Science and Tata Institute of Fundamental Research for hospitality during the completion of this paper.

\appendix

\section{Fast Scaling from CFT}\label{app-kubo}

We imagine doing a global quantum quench in $d$ spacetime dimensions
by the Hamiltonian, $ H= H_{CFT } -i\delta \lambda \int  F(t/\dt)\,
\cO_{\Delta}$, where the profile $F(t/\dt)$ is non-zero for $t \in
(-\dt, \dt)$. We then calculate $\vev{ \cO_\Delta(t) }$  for $t \ll
\dt$. If we choose the scale $\dt$ such that it is much smaller than
the gap, then we can use the Kubo formula about the CFT to obtain,
\beq \delta \vev{ \cO_\Delta(t,0) } =-i \delta \lambda
\int_{-\dt}^{t} dt' F(t'/\dt) \int_{-(t-t')}^{t-t'} d^{d-1}x \vev{
\left[ \cO_\Delta(t,0) , \cO_\Delta(t',x)\right] }_\mt{CFT} +
\cdots\,. \label{kubo} \eeq The Lorentzian unequal time commutator
in the CFT involves crossing of a branch cut --- see section 3.4 of
\cite{Hartman:2015lfa}. The result is \beq \vev{\left[
\cO_\Delta(t,0) , \cO_\Delta(t',x) \right] }_\mt{CFT} = \frac{2i
\sin \pi \Delta  }{ \left( (t-t')^2 - x^2\right)^\Delta}.
\label{comm} \eeq for timelike separations, zero otherwise. This is
reflected in the range of the $x$ integrals. In polar coordinates,
for the spatial integral, we have \beq \Omega_{d-1} \int_0^{t-t'} dr
\frac{r^{d-2}}{ \left( (t-t')^2 - r^2 \right)^\Delta}  =
\Omega_{d-1}\,\frac{ \Gamma\left( \frac{d+1}{2} \right) \,\Gamma
\left( 1- \Delta\right) }{\Gamma\left( \frac{  d+1 - 2 \Delta}{2}
\right)\, \left( d-1\right)} \left( t-t'\right)^{ d - 2\Delta -1
}\,, \eeq where $\Omega_{d-1}=2\pi^{d/2}/\Gamma(d/2)$ is the volume
of a unit $(d-1)$-sphere. Thus Eq.~(\ref{kubo}) yields \beq
\delta\vev{ \cO_\Delta(t,0) } = \frac{ \delta \lambda\,
\pi^{\frac{d+1}{2}} }{\Gamma\left( \frac{ d+1 - 2 \Delta}{2}
\right)\, \Gamma( \Delta )} \int_{-\dt}^{t} dt' \frac{F(t'/\dt)
}{\left( t-t'\right)^{2\Delta -d +1} }+\cdots \, . \eeq

Now if we choose a simple impulse profile with $F(x)=1$ in the range
where it is nonvanishing, then \beq \delta\vev{ \cO_\Delta(t,0) } =
\frac{ \delta \lambda\, \pi^{\frac{d+1}{2}} }{\Gamma\left( \frac{
d+1 - 2 \Delta}{2} \right)\, \Gamma( \Delta )}\, \int_{-\dt}^t dt'
\frac{1 }{\left( t-t'\right)^{2\Delta -d +1 } }+\cdots \, . \eeq
When $d\neq 2\Delta$, the integral evaluates to \beq \delta\vev{
\cO_\Delta(t,0) } = \frac{\delta \lambda\,\pi^{\frac{d+1}{2}}
}{\Gamma(\Delta)\, \Gamma\left(\frac{d+1}{2} - \Delta
\right)\,(d-2\Delta) }\, \left( t + \dt \right)^{d-2\Delta} +
\cdots\,. \eeq In our discussion of quenches in the main text, the
observable was measured at $t \rightarrow 0$. Thus the leading order
scaling with $\dt$ is given by $\vev{ \cO_{\Delta}(0,0) } \sim
\dt^{d-2\Delta}$, as noted earlier in Eq.~(\ref{1-6}). When
$d=2\Delta$, the upper limit of the $t'$ integral is divergent and
we regulate this by shifting the upper limit of the integral, $t
\rightarrow t + \epsilon$. The leading behaviour is then logarithmic,
\beq \delta\vev{ \cO_{\Delta}(t,0) } = \lim_{\epsilon \rightarrow
0}\, \frac{\delta \lambda\,\pi^{\frac{d+1}{2}} }{\Gamma(\Delta)\,
\Gamma\left(\frac{d+1}{2} - \Delta \right) } \, \log \left( \frac{t
+ \dt}{\epsilon} \right) + \cdots\,. \label{goog} \eeq

For the example of the ($1+1$)-dimensional Ising model in the
continuum limit, we have a massless fermion and $\cO = \bar \psi
\psi$. Hence, we have $\Delta = 1$ and $d =2$ and Eq.~\reef{goog}
becomes \beq \delta\vev{ \bar \psi \psi }(t=0) =\lim_{\epsilon
\rightarrow 0}\,  2 \pi \delta \lambda \,\log \left( { \dt
}/{\epsilon} \right) + \cdots\, . \eeq

\section{Saturation in the Transverse Field Ising model} \label{app-sat}

One feature which distinguishes the lattice quenches from their counterparts in a continuum field theory is that for small enough $\dt$, the expectation value of the quenched operator saturates in the lattice quenches. While this is expected since the lattice provides a natural cutoff given by the lattice spacing (which is hidden in the interaction strength $J$ --- see footnote \ref{footy3}), the details of this saturation are important to understand the different regimes which we are analyzing in this paper.

In this Appendix, we will provide the details for understanding why
the point at which saturation sets in is qualitatively different for
small and large amplitudes in the Transverse Field Ising model. This
difference was observed in the results in section \ref{resice} and
already discussed there. The main result is that for small
amplitudes the expectation value saturates to the instantaneous
answer at a scale of $J \dt \sim 1$, independent of the amplitude
$\e_{in}$, while for large amplitudes the saturation occurs for a
value of $J \dt$ proportional to the amplitude $|\e_{in}|$. We
expect that an analogous description will hold for quenches in the
Kitaev model, which exhibits similar behaviour as described in
section \ref{honey22}.

To understand the above difference, it is important to carefully
account for the contributions of the various momentum modes in the
two different cases. For that, we analyze the
integrand in
\beq \langle{\bar{\chi}} \chi \rangle_\mt{diff}(t=0)
\equiv \int \frac{dk}{2\pi}\, X(k) \label{tall} \eeq where \beq X(k)
=    \left[- |\partial_t \phi_{in}|^2 +[(G(\vk))^2 - (m(\vk,t))^2]
|\phi_{in}|^2 + 2 m(\vk,t)~{\rm {Im}}[ \phi_{in}\partial_t
\phi_{in}^\star]  - \frac{ m(k,t)}{\omega(k,t)}\right]_{t=0}
\label{tall2}
\eeq
which comes from combining Eqs.~(\ref{3-11}) and
(\ref{oadia}) in Eq.~\reef{3-12}. Recall that all of the needed
definitions are given in section \ref{sec-3}. We note that for $k=0
\, {\rm and}\, \pi$, $G(k)=0$ --- see Eqs.~\ref{3-4a} and
\ref{3-4b}; consequently, the instantaneous energy levels at these
momenta are decoupled. This in turn ensures that the probability of
excitation, for any quench rate and amplitude for these modes, is
either $0$ (for $k=\pi$ when the instantaneous levels do not cross)
or $1$ (for $k=0$ when there is exact level crossing). Consequently,
$X(k=\pi)=0$ and $X(k=0)=1$.

\begin{figure}[h!]
        \centering
        \subfigure[Small Amplitude ($|\e_{in}|=0.1$)]{
                \includegraphics[scale=0.45]{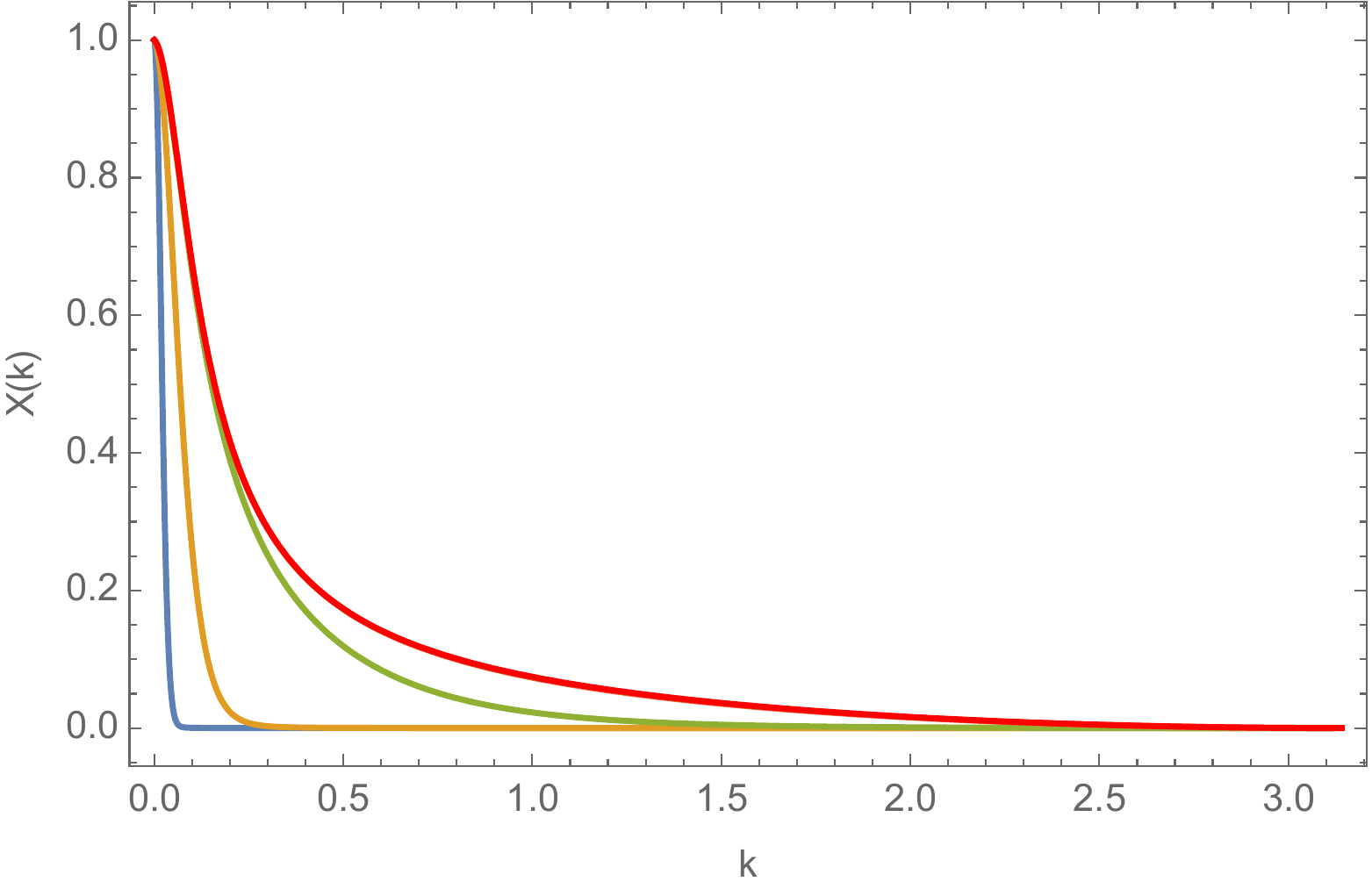} \label{bsmall}}
         \subfigure[Large Amplitude ($|\e_{in}|=100$)]{
                \includegraphics[scale=0.45]{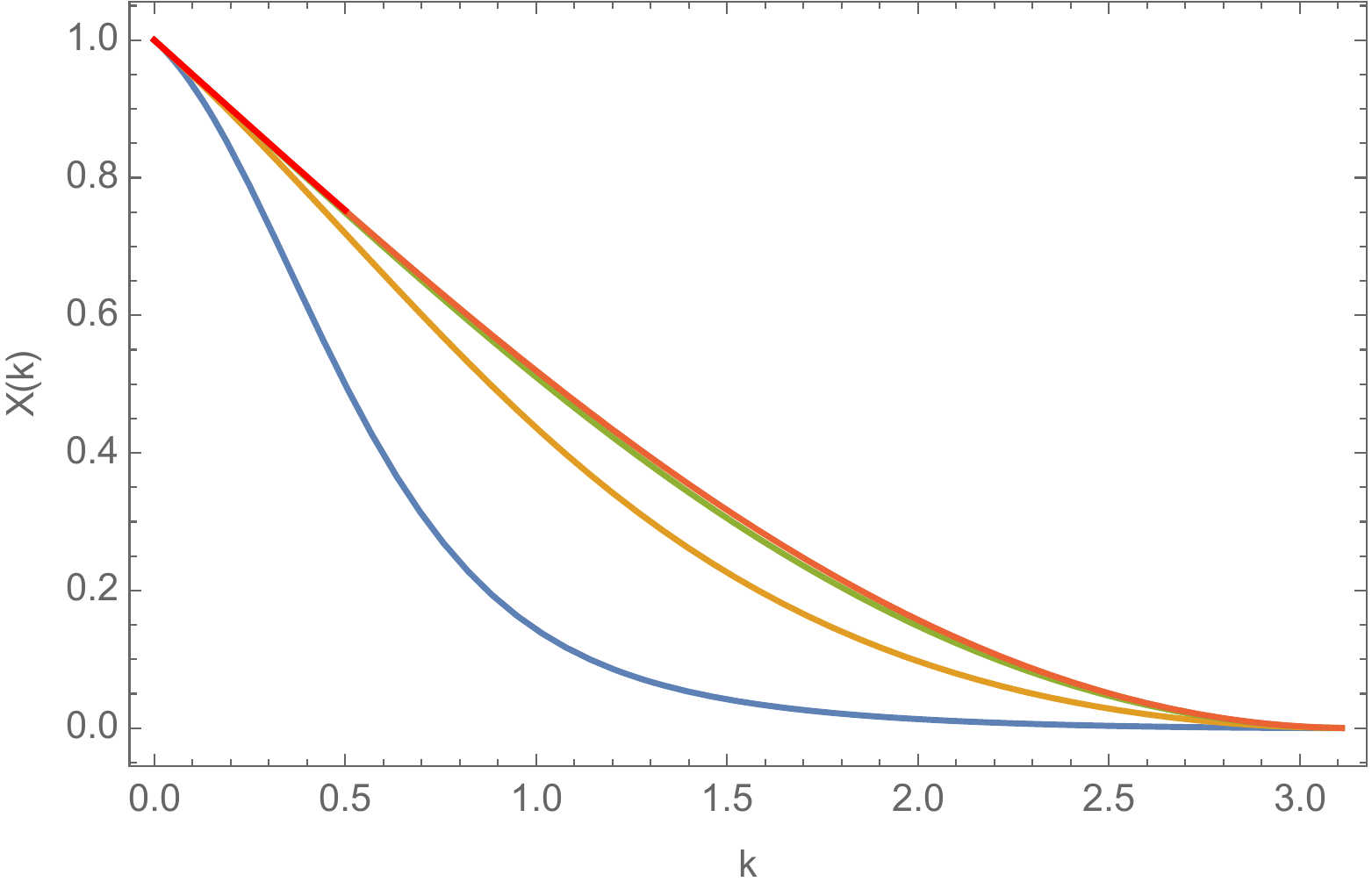} \label{bbig}}
                 \caption{Momentum mode contribution to the expectation value as a function of the amplitude and the quench rate. Different curves correspond to different quench rates: $J\dt=100$ is blue; $J\dt=10$ is yellow; $J\dt=1$ is green; and $J\dt=0.1$ overlaps in both figures with the red curve that corresponds to the instantaneous answer \reef{instant8}. The different rate of decay in each figure is clear, showing that only small momenta contribute in the small amplitude case while high modes start contributing for large amplitudes.}
\label{fig_app1}
\end{figure}

Now from plotting this integrand, it is straightforward to see that
the behaviour is qualitatively different depending on whether the
amplitude of the quench is small or large. In figure \ref{fig_app1},
we show this for two different cases, $|\e_{in}|=0.1$ and
$|\e_{in}|=100$ for different values of $\dt$. In the case of small
amplitudes, the integrand is rapidly decaying and hence only low
momenta contribute to the expectation value; while for large
amplitudes, the integrand does not decay rapidly and so large
momenta also give important contributions to the expectation value.

In figure \ref{fig_app1}, we also compare these profiles with the
integrand for an `instantaneous' quench, as in Eq.~\reef{instantx}.
From Eq.~\reef{oadia}, this instantaneous integrand takes the
simple form
\beqa
X_{inst}(k) &=&  \frac{ m(k,t=-\infty)}{\omega(k,t=-\infty)}    - \frac{ m(k,t=0)}{\omega(k,t=0)} \labell{instant8}\\
&=& \frac{2\sin^2\left(k/2\right)-\e_{in}}{\left[2(2-\e_{in})\sin^2\left(k/2\right)+\e_{in}^2\right]^{1/2}}-\sin\left(k/2\right)\,,
\nonumber
\eeqa
where we have combined the various definitions in section \ref{sec-3} to produce the final expression. Figure \ref{fig_app1} shows that for either large or small amplitudes, the integrand $X(k)$ quickly approaches $X_{inst}(k)$ as $J\dt\to0$. This will become a key fact in analyzing the modes contribution for small amplitudes, as we will see below. In passing, we also note that with $\e_{in}<0$, $X_{inst}(k=\pi)=0$ and $X_{inst}(k=0)=1$, as was argued must be the case on general grounds above.

\

\begin{figure}[h!]
        \centering
        \subfigure[$J\dt=0.1$]{
                \includegraphics[scale=0.3]{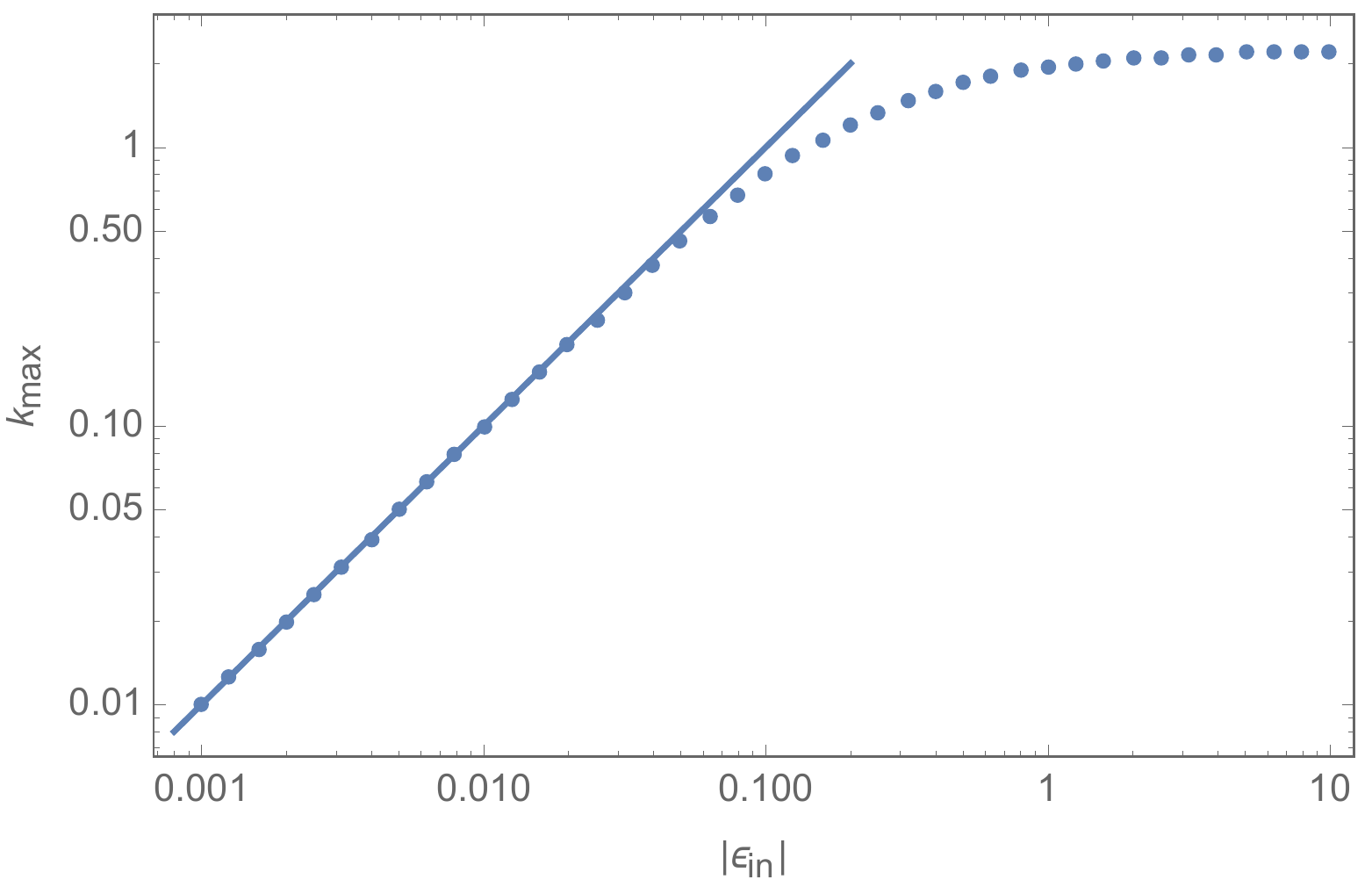} }
         \subfigure[$J\dt=1$]{
                \includegraphics[scale=0.3]{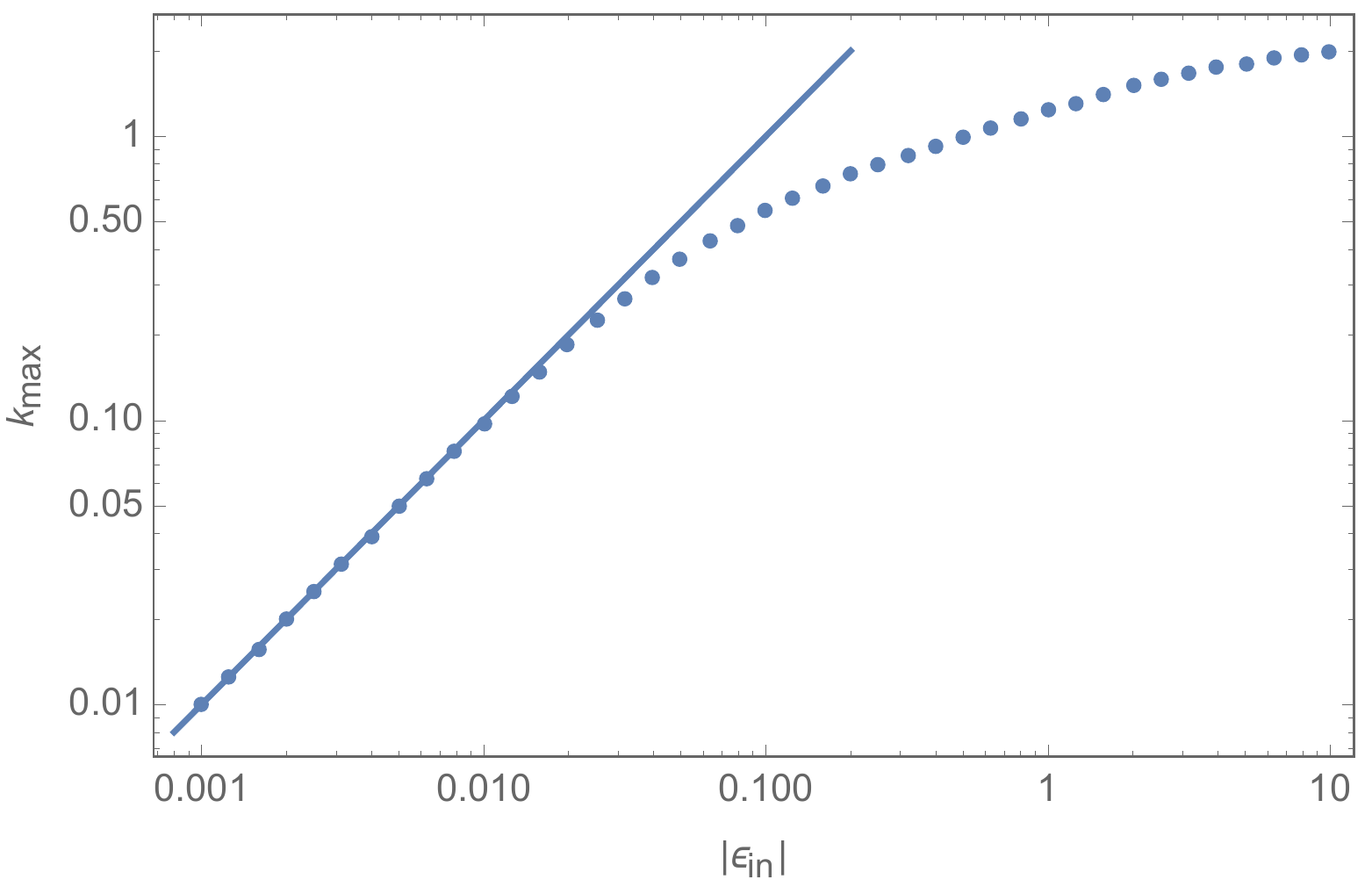} }
         \subfigure[$J\dt=10$]{
                \includegraphics[scale=0.3]{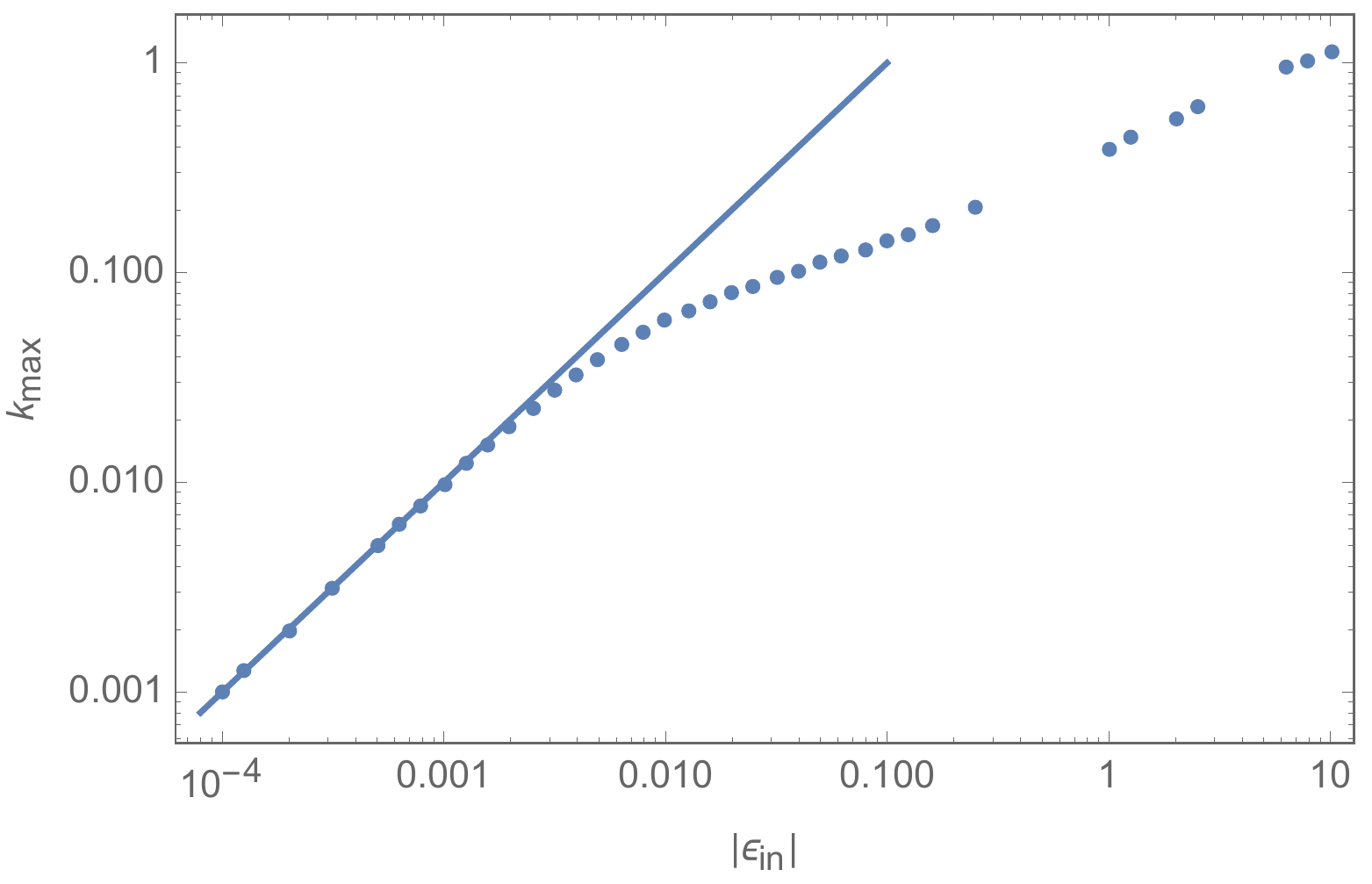} }
                 \caption{Numerical evaluation of $k_{max}$ as a function of $|\e_{in}|$ for different values of $J\dt$. In the three cases, it is possible to observe a linear dependence of $k_{max}$ on $|\e_{in}|$ for small enough $|\e_{in}|$. This is verified by the fit with a function $k_{max}= c\, |\e_{in}|$ (solid blue line). Further, we find the coefficient $c$ is the same in all three plots, \ie it is independent of $J\dt$ but depends, of course, on the choice of the threshold, being $X(k_{max})=0.1$ in the present case.}
\label{fig_app2}
\end{figure}

Let us first analyze the small amplitude case in more detail. We first determine which modes are contributing significantly to the expectation value. Then since we are interested in understanding when the expectation value saturates,  we will ask for which values of $J \dt$ are all of these modes excited by the quench (as first discussed in section \ref{resice}), and how this varies as we vary the initial amplitude. We approached these questions with a combination of numerical and analytic analyses. In figure \ref{fig_app2}, we study the decay of the integrand as a function of $|\e_{in}|$. In particular, we find numerically the momentum $k_{max}$ where $X(k)=0.1$, \ie where the integrand decays to 10 percent of its maximum value. This gives us an estimate of the range of momenta for which the corresponding modes are contributing significantly to the expectation value \reef{tall}. The result in figure (\ref{fig_app2}) is robust: for any $J\dt$, there is a region for small enough $|\e_{in}|$ where $k_{max} \propto |\e_{in}|$ and moreover, the coefficient of proportionality is independent of $J \dt$. However, the coefficient obviously depends on the chosen threshold, \ie $X(k_{max})=0.1$ in the present case.

We can also provide an analytic calculation which supports this same conclusion. First, using Eqs.~\reef{3-33} and \reef{3-7}, we can expand the expression in Eq.~\reef{tall2} for $X(k)$ to find
\begin{equation}
X(k) = X_{inst} (k) + O((J \dt)^2)  \,,
\label{job0}
\end{equation}
where $X_{inst}$ is given by Eq.~(\ref{instant8}). Of course, this result is in agreement with the observation that $X(k)$ quickly approaches $X_{inst} (k)$, made from the plots in  figure \ref{fig_app1}.
However, this result can be refined since if we carefully examine the expressions in
Eqs.~\reef{3-33} and \reef{3-7}, it is possible to see that every contribution of $J \dt$ to the integrand \reef{tall2} is accompanied either by a factor of the amplitude $\e_{in}$ or a momentum factor, $\sin (k/2)$. It will serve our purposes below to expand $X(k)$ simultaneously for small $x\equiv \e_{in}\,J\dt$ and $y\equiv \sin (k/2)\,J\dt$. With these variables, we find that the previous expansion is replaced by
\begin{equation}
X(k) = X_{inst} (k) + O(y^2,xy)  \,.
\label{job}
\end{equation}
where implicitly we have assumed that $x\sim y$ in our expansion.\footnote{While our explicit calculations show that there is no $O(x^2)$ correction here, we expect that terms involving only a power of $x$ may appear at higher orders in the expansion.} Of course, we may ensure that $|x|,|y|\ll1$ by simply taking $J\dt\ll1$ and with this choice, we recover Eq.~\reef{job0}. However, the above expansion indicates for any $J \dt \ll 1/|\e_{in}|, 1/k$, the integrand is well approximated by the instantaneous integrand. Hence for small amplitudes, as considered here, we may still consider $J \dt \sim 1$ as long as the relevant momenta are also small. This will now let us self-consistently prove that only small momenta contribute to the expectation value \reef{tall} for small amplitude quenches.

Above we argued that in the limit of small momenta and small amplitude, $X(k)$ reduces to the instantaneous integrand. However, it remains to expand the expression in eq.~\reef{instant8} when taking $k,|\e_{in}|\ll1$ while keeping
$k/|\e_{in}|\sim 1$. This expansion yields
\beq
X_{inst}(k) = -\frac{\e_{in}}{\left[k^2+\e_{in}^2\right]^{1/2}}-\frac{k}{2} +\frac{(2k^2+\e_{in}^2) k^2}{4\left[k^2+\e_{in}^2\right]^{3/2}}+ O(k^2)\,,
\label{instant91}
\eeq
where the $O(k^2)$ is used above in a sense where the third term is $O(k)$.

Now, to determine the significant contributions, we want to see when the leading term in eq.~\reef{instant91} reaches a particular threshold $\gamma$ in this limit. Hence we set
\beq
\frac{|\e_{in}|}{\left[k_{max}^2+\e_{in}^2\right]^{1/2}} =\gamma
\quad\implies\quad
k_{max}= \frac{\sqrt{1-\gamma^2}}\gamma\,|\e_{in}|\,,
\label{instant92}
\eeq
where we assumed that $\e_{in}<0$ while $\gamma>0$. We note that this $O(1)$ term for $X_{inst}(k)$ in eq.~\reef{instant91} has a long tail. However, let us explicitly evaluate the pre-factor for $\gamma=0.1,\,
0.2,\,0.5$, and we find $c \equiv \sqrt{1-\gamma^2}/\gamma=9.95,\, 4.90,\,
1.73$, respectively. So even with $\gamma=0.1$, we have a consistent solution for $k_{max}$. That is, our expansion assumed that the relevant $k$'s were small and now we find that the maximum contributing momentum is proportional to $|\e_{in}|$ and so it is indeed small for the small amplitudes considered here. Moreover, as both the amplitudes and momenta are small, this result is still valid for $J \dt \sim 1$, which will be needed below for our final result on the saturation point. It is also important that the pre-factor $c$ in eq.~\reef{instant92} is independent of $J \dt$.

Hence both our numerical and analytic calculations suggest that for small amplitude quenches, the modes which contribute significantly to the expectation value \reef{tall} lie in a narrow band: $0\le k\le k_{max}$ with $k_{\max} = c\,|\e_{in}|$ where $c$ is some order one number (which is independent of $J\dt$). Now as discussed in section \ref{resice}, we can expect that the expectation value saturates when all of these modes are in fact excited by the quench. In particular then, we must confirm that the last mode at $k=k_{max}$ is excited according to the Landau criterion \reef{gopher}. Substituting in $k_{\max} = c\,|\e_{in}|$ and keeping in mind that we are considering small amplitude quenches, \ie $|\e_{in}|\ll1$, we find
\beq
{\rm small\ amplitude:}\qquad
\frac{1}{E_{k_{max}}^2}\,\left|\frac{dE_{k_{max}}}{dt}\right|_{t=0} = \frac{1}{4 c\,J \dt} \,,
\eeq
which should then be larger then one to excite all of the modes contributing to $\langle{\bar{\chi}} \chi \rangle_\mt{diff}(t=0)$. That is, we expect saturation for  $J \dt \leq 1/(4c)$ or more simply $J \dt \lesssim 1$. As stressed in the main text, this result shows that for the small amplitude quenches, the point where saturation sets in is independent of the initial amplitude. Note also that while the argument that the maximum momentum contribution is proportional to the initial amplitude is valid for larger $J \dt$ (provided the amplitude is small enough), the Landau criterion in this case will show that it is still possible to keep exciting these modes and then we will see no saturation for larger $J \dt$.

\

The situation is qualitatively  different for large amplitude
quenches.  As shown in figure \ref{fig_app1}, in this case, the
profile of the integrand \reef{tall2} is much broader and hence
almost every mode makes a significant contribution to the
expectation value. Hence we essentially have, $k_{max} \simeq \pi$.
Of course, the mode $k$ exactly at $\pi$ is not contributing since
as noted above, $X(k=\pi) =0$ in every quench, \ie for any amplitude
or quench rate. However, other modes near $\pi$ will contribute
significantly to the expectation value.\footnote{In fact, it is sufficient to choose any finite value for $k_{max}$ (which is independent of $|\e_{in}|$) and one will reach the same conclusion. The only change is an additional factor of $\sin(k_{max}/2)$ from Eq.~\reef{gopher} so that saturation occurs for $J \dt \lesssim |\e_{in}|/(8 \sin(k_{max}/2))$.}  It is straightforward to
evaluate the Landau criterion \reef{gopher} for $k_{max} \simeq
\pi$, and as in eq.~\reef{4-1-1}, we obtain
\begin{eqnarray}
{\rm large\ amplitude:}\qquad
\frac{1}{E_{k_{max}}^2}\,\left|\frac{dE_{k_{max}}}{dt}\right|_{t=0} = \frac{|\e_{in}|}{8 J \dt } \,.
\end{eqnarray}
As before, all the modes contributing to the expectation value \reef{tall} will be excited when the above expression is bigger than one. Hence we should expect saturation for $J \dt \lesssim |\e_{in}|/8$. This is the result reported in section \ref{resice}, and hence the saturation point is proportional to the initial amplitude for large amplitude quenches.
While all of the above analysis is particular to the Transverse Field Ising model, let us re-iterate that we expect an analogous description will hold for the results for the quenches in the Kitaev model described in section \ref{honey22}.

\section{Cluster-Ising model} \label{app-cluster}

The Hamiltonian of the Cluster-Ising model \cite{cluster-ising} on a
one-dimensional bipartite lattice is \beq H_\mt{CI} = - \sum_{j=1}^N
\tau^{(1)}(j-1) \tau^{(3)}(j) \tau^{(1)}(j+1) + \lambda(t)
\sum_{j=1}^N \tau^{(2)}(j) \tau^{(2)}(j+1)\,. \label{clump} \eeq
where we have allowed for a time-dependent Ising coupling to
consider quenches and $\tau^{(1,2,3)}$ denote Pauli spin operators.

For $\lambda = 0$, the ground state is protected by a $\IZ_2 \times
\IZ_2$ symmetry generated by $\prod_{i \in even} \tau_i^z \times
\prod_{i' \in odd} \tau_{i'}^z$. This short-ranged entangled state
has topological order (no long range order) on a open chain,
realizing projective representations of the symmetry group via edge
states. At $\lambda = 1$, the model has a quantum critical point and
for $\lambda >1$, the system is  antiferromagnetic with long range
order. At the quantum phase transition, the critical exponents are
$\nu = z = 1$, $\beta = 3/8$ and $\alpha = 0$. We show below that on
a closed chain with periodic boundary conditions, the model can be
mapped to ($1+1$)-dimensional free fermions as in the Ising case.

Introducing Jordan-Wigner fermions,
$$
c(j) = \prod_{m=1}^{j-1} \tau^{(3)}(m) \tau^-(j).
$$
where $\tau_j^\pm = ( \tau^{(1)}(j) \pm i \tau^{(2)}(j) )/2$, the
Hamiltonian \reef{clump} becomes \bea \label{ham}H_\mt{CI} &=&
\sum_{l=1}^N \left( c(l-1)^\dagger - c(l-1)^\nd \right)\left(
c^\dagger(l+1) + c(l+1)^\nd \right) \nonumber\\
&& + \lambda \sum_{l=1}^N \left( c^\dagger(l) +
c^\nd(l)\right)\left(c^\dagger(l+1) - c^\nd(l+1) \right)\,. \eea
Next we apply the Fourier transform \beq b(k) = \frac{1}{\sqrt{N}}
\sum_{j=1}^N c(j)\, e^{ -  i k j}\,. \eeq with
$$
 k  = \pm 2\pi/(2N), \pm 6\pi/(2N)\dots ,\pm 2\pi(N-1)/(2N)\,. $$
The spin Hamiltonian \reef{ham} can now be written as,
\beq
H_\mt{CI} =\sum_{k>0}
\chi^\dagger(k)
{\cal H}_k
\chi^\nd(k) \,,
\eeq
where \beq {\cal H}_k =  ( \cos 2k - \lambda \cos k ) \sigma_3  +(
\sin 2 k + \lambda \sin k ) \sigma_2 \,, \label{cim-ham1} \eeq
$\sigma_i$ denotes Pauli matrices in particle-hole space of fermions
and $\chi(q)$ is a two-component Majorana fermion defined by
\[ \chi(q) = \left( \begin{array}{c}
b(q) \\
b^\dagger (-q)
\end{array} \right)\,.
\]

To bring this Hamiltionian into a standard Dirac form, we carry out
a few unitary transformations. Let us first write the Hamiltonian
(\ref{cim-ham1}) as \beq \label{B-5} {\cal H}_k  = ( \alpha_k
+\beta_k \lambda  )\, \sigma_3  +( \gamma_{1k}+ \gamma_{2k} \lambda
)\, \sigma_2\,. \eeq where $\alpha_k  = \cos 2k$,  $\beta_k  = -
\cos  k$, $\gamma_{1k} = \sin 2 k $ and $ \gamma_{2k} = \sin  k $.
First we do a global rotation by the unitary, $e^{i \sigma_3 \pi/4
}$, which leaves $\sigma_3 $ invariant and rotates $\sigma_2 $ into
$\sigma_1$. Next, let us rewrite the Hamiltonian in terms of new
Pauli matrices, $\tau_i$ as \beq \label{B-6} \tilde { \cal H}_k =
\Lambda_{1k} ( \lambda(t) - t_{1k} ) \tau_3  + \Lambda_{2k}
\tau_1\,. \eeq We have from Eqs.~(\ref{B-5}) and \reef{B-6}, \bea
\Lambda_{1k} \tau_3  &=& \beta_k \sigma_3 + \gamma_{2k} \sigma_1\,, \nonumber \\
\Lambda_{2k} \tau_1 &=& \sigma_3 ( \alpha_k + t_{1k} \beta_k) + \sigma_1 ( \gamma_{1k} + t_{1k} \gamma_{2k} )\,.
\nonumber
\eea
This gives
\bea
\Lambda_{1k}^2 &=& \beta_k^2 + \gamma_{2k}^2
\,,\label{B-7}  \\
\Lambda_{2k}^2 &=&  ( \alpha_k + t_{1k} \beta_k)^2 + ( \gamma_{1k} +
t_{1k} \gamma_{2k} )^2\,. \label{B-8} \eea
From the canonical
condition, $\{ \tau_1 , \tau_3 \}_+ = 0$, we find
\beq \beta_k(
\alpha_k + t_{1k} \beta_k ) + \gamma_{2k} ( \gamma_{1k} + t_{1k}
\gamma_{2k} ) = 0\,. \label{B-9} \eeq
Using Eqs.~(\ref{B-7}),
\reef{B-8} and \reef{B-9}, the Hamiltonian can be brought into the
form
 \beq
\tilde { \cal H}_k =  \left(  \lambda(t) - \cos 3k  \right)\,\tau_3  +
\sin 3k\, \tau_1\,.
\label{cim-ham}
\eeq
Note that this expression \reef{cim-ham}  is
the same as the lattice Hamiltonian \reef{2-5} for the Ising model with the
replacement of $ k \rightarrow -3k$, and there is a periodicity $ k \rightarrow k + \frac{2\pi}{3}$. This means that we can rewrite the
theory in terms of {\em three flavors} of Majorana fermions, each
living on a chain of size $N/3$.

The eigenvalues of the Hamiltonian \reef{cim-ham} are
\begin{equation}
E_k = \pm\sqrt{1 + \lambda(t)^2 - 2\lambda(t) \cos 3 k} \,.
\end{equation}
Thus we see at $\lambda = 1$ the gap closes for $k \rightarrow 0$, $2\pi/3$ and $4\pi/3$.\footnote{As in the Ising model, there is another critical point at $\lambda=-1$ with $k\to \pi/3,$ $\pi$ or $5\pi/3$.}
Expanding around this critical point, we can see that the theory is
described by three massive Majorana fermions in the
continuum limit. With the lattice spacing $a$, we introduce dimensionful momenta $p$, and a dimensionful mass $m$,
\ben
p=\frac{3k}a\qquad {\rm and}\qquad  m(t) = \frac{\lambda
(t)-1}{a} \,,    \een
 the Hamiltonian becomes in the
$a\rightarrow 0 $ limit
 \ben H_\mt{CI}^\mt{cont} = \sum_{i=1}^3\int
\frac{dp}{2\pi}\, \psi^{i\dagger}(p)\left[ m(t)\, \tau_3 + p\,
\tau_1 \right] \psi^i(p)\,. \label{LLLLL} \een which is equivalent
to three identical copies of Eq.~\ref{2-5b}, describing the
continuum theory for the Ising model.

\end{document}